\documentclass[fleqn,usenatbib]{mnras}

\usepackage{newtxtext,newtxmath}

\usepackage[T1]{fontenc}

\DeclareRobustCommand{\VAN}[3]{#2}
\let\VANthebibliography\thebibliography
\def\thebibliography{\DeclareRobustCommand{\VAN}[3]{##3}\VANthebibliography}

\usepackage{graphicx}	
\usepackage{amsmath}	

\usepackage{orcidlink}  

\defcitealias{brown_assembly_2024}{B24} 

\title[ICL Assembly I.]{A Consistent Comparison of Intracluster Light Assembly in Simulations\newline I. Redshift Evolution and Progenitor Galaxies}

\author[Brown et al.]{
Harley J. Brown\orcidlink{0009-0004-7521-8204}$^{1}$\thanks{E-mail: Harley.Brown@nottingham.ac.uk},
Garreth Martin\orcidlink{0000-0003-2939-8668}$^{1}$,
Frazer R. Pearce\orcidlink{0000-0002-2383-9250}$^{1}$,
Yannick M. Bah\'{e}\orcidlink{0000-0002-3196-5126}$^{1,2}$,
Madeline Butler\orcidlink{0009-0004-9273-4630}$^{1}$,
\newauthor~Weiguang Cui\orcidlink{0000-0002-2113-4863}$^{3,4,5}$,
Nina A. Hatch\orcidlink{0000-0001-5600-0534}$^{1}$,
and Alexander Knebe\orcidlink{0000-0003-4066-8307}$^{3,4,6}$
\\
$^{1}$School of Physics \& Astronomy, University of Nottingham, University Park, Nottingham NG7 2RD, UK\\
$^{2}$Laboratoire d'Astrophysique, \'{E}cole Polytechnique F\'{e}d\'{e}rale de Lausanne (EPFL), Observatoire de Sauverny, 1290 Versoix, Switzerland\\
$^{3}$Departamento de Física Te\'orica, M\'odulo 15, Facultad de Ciencias, Universidad Aut\'onoma de Madrid, 28049 Madrid, Spain\\
$^{4}$Centro de Investigaci\'on Avanzada en F\'isica Fundamental (CIAFF), Facultad de Ciencias, Universidad Aut\'onoma de Madrid, 28049 Madrid, Spain\\
$^{5}$Institute for Astronomy, Royal Observatory, Edinburgh EH9 3HJ, UK\\
$^{6}$International Centre for Radio Astronomy Research, University of Western Australia, 35 Stirling Highway, Crawley, Western Australia 6009, Australia\\
}

\date{Accepted XXX. Received YYY; in original form ZZZ}

\pubyear{\the\year{}}

\begin{document}
\label{firstpage}
\pagerange{\pageref{firstpage}--\pageref{lastpage}}
\maketitle

\begin{abstract}
The tidal stripping of satellite galaxies and the stellar detritus ejected during galaxy mergers builds up a diffuse stellar component in galaxy clusters known as the intracluster light (ICL). 
We investigate ICL assembly in cluster-mass haloes ($M_{178c}\sim10^{14}-10^{15}$\,M\textsubscript{\sun}) using four different hydrodynamical simulations (\textsc{Horizon-AGN}, \textsc{TNG100}, \textsc{The Three Hundred Gizmo-Simba 7K}, and \textsc{Hydrangea}) under a homogenized ICL identification framework. 
For our fiducial ICL definition 
we obtain broadly consistent $z\approx0$ ICL stellar mass fractions  
($\sim0.1-0.2$) and, by tracking the progenitors of $z\approx0$ clusters back to $z\gtrsim2$, find no significant evolution in average ICL 
mass fractions.
Alternative approaches for distinguishing the ICL from the central galaxy
show the absolute ICL fraction to be highly sensitive to adopted definition,
but we never find any significant inter-simulation discrepancies when implementing a consistent  
methodology to identify the ICL. 
Whether the average ICL mass fraction falls with increasing redshift or does not evolve is determined by the ICL definition adopted. 
By tracing $z\approx0$ ICL stars back to their progenitor galaxies, we find that lower-mass satellites typically make slightly larger ICL contributions relative to their mass in every considered simulation, 
but which galaxies 
make the dominant contribution
to the ICL is primarily controlled by the infalling satellite mass function.
Most ICL stars sourced from satellite galaxies are therefore expected to originate from galaxies with infall stellar masses above $\sim10^{10}$\,M\textsubscript{\sun} and largely within $10^{10.5}-10^{11.5}$\,M\textsubscript{\sun}. 
\end{abstract}




\section{Introduction}\label{Intro}

The light produced by the diffuse ensemble of stars that permeate the intergalactic space in a galaxy cluster 
-- not belonging to
any particular galaxy but still bound to the cluster potential as a whole --
is termed \textit{intracluster light} (ICL;
see \citealt{Mihos_review_2016}, \citealt{Contini_review_2021}, \citealt{Montes_review_2022} for relevant reviews). The existence of this diffuse stellar component was first theorised and observed by \citet{zwicky_masses_1937, zwicky_coma_1951}. 
Though much of the ICL in a typical cluster will be centrally concentrated -- composing an extended diffuse stellar envelope around the 
brightest cluster galaxy 
(BCG;  
e.g. \citealt{presotto_intracluster_2014}, \citealt{ellien_euclid_2025}, \citealt{englert_intracluster_2025})
-- this low surface-brightness cluster component ($\mu_{r} \gtrsim 26\,\textrm{mag arcsec}^{-2}$; \citealt{brough_preparing_2024}) can extend to the far fringes of a cluster, 
with observational studies tracing the ICL out to $\gtrsim1$\,Mpc (\citealt{gonzalez_discovery_2021},
\citealt{ zhang_dark_2024}, \citealt{jimenez-teja_deep_2025}). 

Typically over 10 per cent of the stellar mass in a cluster may be associated with the ICL, though with substantial scatter between individual clusters (e.g. \citealt{ragusa_does_2023}, 
\citealt{montenegro-taborda_stellar_2025}, 
\citealt{mayes_coevolution_2025}).
Significant scatter also occurs between different definitions and methodologies for extracting the ICL (e.g. \citealt{cui_characterizing_2014}, \citealt{Tang_mockICLims_2018}, \citealt{kluge_photometric_2021}, \citealt{brough_preparing_2024}).
Likewise unresolved is the question of how this typical ICL mass fraction evolves with redshift,
though a mounting body of both observational (e.g. \citealt{ko_evidence_2018}, \citealt{joo_intracluster_2023},
but see also e.g. \citealt{canepa_dependence_2025}) and theoretical work (e.g. 
\citealt{joo_tracing_2024}, \citealt{contini_moreimptime_2024}, \citealt{kimmig_intra-cluster_2025}, but see also e.g. \citealt{rudick_quantity_2011}) suggests 
this diffuse stellar component should remain significant 
to $z\sim1$ (and plausibly to $z\gtrsim2$; e.g. \citealt{werner_intracluster_2023}, 
\citealt{kimmig_intra-cluster_2025}, \citealt{joo_mature_2026}). 

As hierarchically formed structures, galaxy clusters are thought to grow through a protracted procession of accretion and merger events (\citealt{press_formation_1974}, \citealt{white_core_1978}) and so the ICL 
-- as the stellar debris generated by these and related processes -- is expected to encode a record of each cluster's merger history as well as its past and present dynamical state (e.g. \citealt{golden-marx_hierarchical_2025}, \citealt{kimmig_intra-cluster_2025}, \citealt{gassis_tracing_2025}, and references therein). 
Given that the ICL can span the entire cluster halo, with observational detections made even beyond the splashback radius  \citep{gonzalez_discovery_2021},
and that these intergalactic stars can 
be considered 
collisionless particles with motions governed only by the overall cluster potential, 
it has also been proposed that the ICL could be used as a visible tracer of the total mass distribution in galaxy clusters 
(e.g. \citealt{Montes_lumtrace_2019}, 
\citealt{asensio_intra-cluster_2025}, \citealt{Fernandez_IclDmHaloShape_2026}; 
see also \citealt{butler_intracluster_2025}).
However, exploiting observations of the ICL for these purposes requires 
both a detailed understanding of the processes that contribute stars to the ICL  and the objects from which these constituent stars are sourced (\citealt{martin_intracluster_2026}). 

The currently prevailing paradigm is that the bulk of ICL stars should originate from cluster satellite galaxies, chiefly through a combination of tidal stripping (e.g. \citealt{willman_origin_2004}, \citealt{rudick_tidal_2009}, \citealt{martin_stellar_2024}) and ejection by violent relaxation processes during galaxy mergers (chiefly between the BCG and massive satellites; e.g. \citealt{murante_importance_2007}, \citealt{groenewald_close_2017}, \citealt{contini_different_2018}). There is an emerging consensus that the main progenitor objects from which these ICL stars are drawn should be massive 
galaxies (stellar mass, $M_{*} \gtrsim10^{10.5}$\,M\textsubscript{\sun}). A developing body of theoretical work supports this perspective (e.g. \citealt{contini_formation_2014, contini_theoretical_2019}, 
\citealt{chun_formation_2023, 
chun_formation_2024}, \citealt{ahvazi_progenitors_2024}, \citealt{martin_stellar_2024}, \citealt{brown_assembly_2024}, \citealt{mayes_coevolution_2025}, \citealt{bilata-woldeyes_tracing_2025}; but see also e.g. \citealt{tang_importance_2021, Tang_ICLfromSats_2023}). The alternative, that ICL stars are predominately sourced from less massive satellites, 
requires in some combination either significant modifications to the galaxy stellar mass function or a considerable suppression of stripping rates at higher satellite masses compared to current expectations (\citealt{Montes_review_2022}, \citealt{martin_stellar_2024}, \citealt{brown_assembly_2024}, \citealt{martin_intracluster_2026}). 

Observational studies remain more divided. Though many agree with the prevailing view from theoretical studies, that the ICL should be largely sourced from more massive galaxies (e.g. \citealt{montes_intracluster_2014, montes_intracluster_2018},  \citealt{demaio_origin_2015, demaio_lost_2018}, \citealt{montes_buildup_2021}), some instead suggest
that less massive galaxies ($M_{*}\lesssim10^{10}$\,M\textsubscript{\sun}) are the main progenitors of ICL stars,
on the basis of e.g. observed ICL colours and metallicities 
(e.g. \citealt{melnick_intergalactic_2012}, \citealt{morishita_characterizing_2017}, \citealt{gu_spectroscopic_2020}, \citealt{ellien_euclid_2025}). 
However, among this latter 
body of work 
there is an acknowledged degeneracy which might ease this tension:
declining stellar metallicity with increasing galactocentric radius
(e.g. \citealt{baes_metallicity_2007}, \citealt{gonzalez_delgado_califa_2015}, \citealt{iza_distribution_2025}, \citealt{tau_age_2025}), allowing a population of stars tidally stripped 
from the outskirts of a more massive galaxy to have a typical metallicity mimicking that 
characteristic of a less massive 
galaxy (\citealt{gu_spectroscopic_2020}, \citealt{ellien_euclid_2025}).

Numerous prior studies have already investigated the ICL in individual hydrodynamical simulations, but the findings of these studies can significantly diverge concerning various key ICL properties and statistics such as the typical ICL stellar mass fraction (e.g. see table~1 in \citealt{kluge_photometric_2021}). Some disagreement is generally expected between different simulations, from (for example) the differing numerical methodologies implemented to model gravity and baryonic physics (\citealt{somerville_physical_2015}, \citealt{sembolini_nifty_2016, sembolini_nifty_2016-1}, \citealt{vogelsberger_cosmological_2020}, \citealt{crain_hydrodynamical_2023}) 
as well as the simulation resolution achieved (\citealt{martin_stellar_2024}, \citealt{lovell_numerical_2025}, \citealt{chiang_universal_2026}). Conceivably even the broad simulation strategy adopted, such as uniform-resolution 
vs \textit{``zoom-in''} simulation, may have an effect (e.g. \citealt{bagla_comments_2005}, \citealt{power_impact_2006}). However, an understanding of these essential differences is obscured by these studies also generally employing bespoke and often incompatible methods to identify the ICL. In order that any differences can be understood and reconciled separately from the impact of differing ICL extraction methodologies, comparison studies employing a uniform approach to identifying the ICL are necessary. 

Here, and in a forthcoming paper (Brown et al. in prep.), we investigate the assembly of the ICL 
in a number of simulated galaxy clusters drawn from four state-of-the-art hydrodynamical simulations 
($z\approx0$ halo masses $\in[10^{14},10^{15}]$\,M\textsubscript{\sun}; $10-16$ clusters per simulation) which leverage a diverse selection of 
simulation practices and numerical methods, 
upon which we have imposed a uniform ICL definition. 
In this paper we consider the division of stellar mass between galaxies and the ICL in these simulated clusters. 
We do this both at $z\approx0$ and in the progenitors of these structures back to $z\sim3$, and probe how our findings are influenced by the specific ICL definition implemented. 
We also investigate the relative contributions to the ICL from satellite galaxies
with differing stellar masses preceding cluster infall
and so assess the main progenitor galaxies of ICL stars, extending the work described in \citet[][hereafter \citetalias{brown_assembly_2024}]{brown_assembly_2024}. We leave 
a comprehensive evaluation of the ICL contribution from assembly channels 
beyond the liberation of satellite galaxy stars 
as well as an investigation into various 
observed ICL trends with cluster-centric radius and an interrogation of plausible origins for these trends to the forthcoming companion paper. 

This paper is organized as follows. 
In Section~\ref{Methods} 
we introduce and briefly detail the four simulations (\ref{Methods-Simulations}), our fiducial homogenized ICL definition (\ref{Methods-IclDef}), how we associate $z\approx0$ ICL stars to progenitor galaxies (\ref{Methods-TrackingGalaxies}), and the cluster samples we employ (\ref{Methods-ClusterSamples}).  
In Section~\ref{Results-StellarMassDistribution} we present our results for the division of stellar mass between galaxies and the ICL at $z\approx0$ (\ref{Results-StellarMassDistribution-z0}), and how this has evolved since $z\sim3$ (\ref{Results-StellarMassDistribution-IclEvo}), as well as probe how our findings might be influenced by the imposed BCG-ICL separation (\ref{Results-StellarMassDistribution-AltDefIclEvo}).
In Section~\ref{Results2} we investigate the ICL contribution from a cluster's own satellite galaxies (\ref{Results2-SummarisedOrigChannels})
as well as the differing contributions made to the ICL by satellite galaxies with different masses (\ref{Results2-IclContribution}).
In Section~\ref{Discussion} we discuss the physical origins of noted trends (\ref{Discussion-flibTrends}), apparent inter-simulation discrepancies (\ref{Discussion-SimDiffs}), as well as important caveats and limitations of our analysis (\ref{Discussion-Caveats}). 
We conclude by summarising the main results of our investigation in Section~\ref{Conclusions}. 
For the following analysis, we adopt the respective cosmologies of the simulations being considered; 
distances should be assumed proper (not comoving);
and presented quantities that depend on the dimensionless Hubble parameter have been evaluated using the value from the corresponding cosmology and do not contain additional factors of ``little $h$'' unless explicitly indicated otherwise. 


\section{Methods}\label{Methods}

\subsection{Simulations}\label{Methods-Simulations}

\begin{table*}
    \centering
        \caption{Key methods and properties of the four cosmological hydrodynamical simulations used in this work. Spatial resolution refers to either the gravitational softening length or minimum grid cell size used for force calculation, as appropriate for the hydrodynamics code employed.}
        \begin{tabular}{l|cccc}
            \hline
            & \textsc{Horizon-AGN} & \textsc{TNG100} & \textsc{The Three Hundred} & \textsc{Hydrangea} \\
            &  &  & \textsc{Gizmo-Simba 7K} & \\
            \hline
            Hydrodynamics & Adaptive mesh refinement & Moving-mesh & Meshless finite-mass & Smoothed particle hydrodynamics \\
            Hydrocode & \textsc{Ramses} & \textsc{Arepo} & \textsc{Gizmo} & \textsc{Gadget-3}\\
            Subgrid model & \citet{dubois_dancing_2014} & TNG & \textsc{Simba-C} & EAGLE [variant AGN-dT9]\\
            \hline
            Simulation strategy & Uniform-resolution & Uniform-resolution & Cluster zoom-in & Cluster zoom-in \\
            Simulation side-length & $100\,h^{-1}\,\text{cMpc}$ & $75\,h^{-1}\,\text{cMpc}$ & - & - \\
            Zoom-in region size at $z=0$& - & - & $15\,h^{-1}\,$Mpc & $10\times r_{200c}$\\
            \hline
            $m_\text{DM}$ [M\textsubscript{\sun}]& $8\times10^{7}$ & $7.5\times10^{6}$ & $2\times10^{8}$ & $9.8\times10^{6}$\\
            $m_{*}$ [M\textsubscript{\sun}]& $3\times10^{6}$ & $1\times10^{6}$ & $4\times10^{7}$ & $1\times10^{6}$\\
            Spatial resolution [kpc]& $1.0$ [min. cell size] & $0.5\,h^{-1}$ & $1.25\,h^{-1}$ & 0.7 \\
            \hline
            Cosmology& WMAP7 & Planck2015 & Planck2013 & Planck2013\\
            $h \equiv H_{0}/100\textrm{\,km\,s}^{-1}\,\textrm{Mpc}^{-1}$ & 0.704 & 0.677 & 0.678 & 0.678\\
            \hline
        \end{tabular}
        \label{tab:sim_props}
\end{table*}

We use data from four 
cosmological hydrodynamical simulation projects: 
the \textsc{Horizon-AGN} simulation (hereafter \textsc{Hz-AGN});
the \textsc{TNG100-1} simulation of the \textsc{IllustrisTNG} project (hereafter \textsc{TNG100});
the \textsc{Gizmo-Simba 7K} galaxy cluster simulations of the \textsc{The Three Hundred} project (hereafter \textsc{The300 GS-7K});
and the \textsc{Hydrangea} galaxy cluster simulations. 
We summarise below the properties, models, and methods employed by each of the simulations. Some key properties are also presented for comparison in Table~\ref{tab:sim_props}. Further details can be found in the respective introductory papers of each simulation.

\subsubsection{\textsc{Horizon-AGN}}\label{Methods-Simulations-HzAGN}

The \textsc{Horizon-AGN} simulation\footnote{\url{https://www.horizon-simulation.org/}} (\citealt{dubois_dancing_2014}, \citealt{dubois_horizon-agn_2016}, \citealt{kaviraj_horizon-agn_2017}) is a uniform-resolution cosmological-volume hydrodynamical simulation employing the Adaptive Mesh Refinement (AMR) Eulerian hydrodynamics code \textsc{Ramses} \citep{teyssier_cosmological_2002}, with a cubic simulation volume of side-length $100\,h^{-1}\,$Mpc (comoving). 
A $\Lambda$CDM cosmology consistent with the 7-year Wilkinson Microwave Anisotropy Probe data (WMAP7; \citealt{komatsu_seven-year_2011}) is adopted with $(\Omega_{m,0}, \Omega_{\Lambda,0}, \Omega_{b,0}, \sigma_{8}, n_{s}, h) = (0.272, 0.728, 0.045, 0.810, 0.967, 0.704)$. 

The simulation follows $1024^3$ dark matter (DM) particles (mass, $m_{\textrm{DM}}\sim8 \times 10^{7}\,$M\textsubscript{\sun}) and an initially uniform $1024^3$ cell gas grid (initial gas resolution  $m_{\textrm{gas,ini}}\sim1\times10^{7}\,$M\textsubscript{\sun}) 
dynamically refined according to a quasi-Lagrangian criterion (based on either the total baryonic or DM mass in a cell exceeding eight times the mass of a DM particle) down to a minimum cell size of 
$1$\,kpc after seven levels of refinement.
Star formation occurs beyond a threshold gas density of $n_{\textrm{H}}=0.1\,\textrm{cm}^{-3}$ following the Kennicutt–Schmidt law  (\citealt{kennicutt_jr_global_1998}) with a constant star efficiency of 2 per cent per free-fall time. Stellar particles have a mass of $m_{*}\sim3 \times 10^{6}\,$M\textsubscript{\sun}.
Gas heating by a uniform UV background begins after $z = 10$; H and He cooling with a contribution from metals using a \cite{sutherland_cooling_1993} model can cool gas to $10^{4}\,\textrm{K}$. Continuous stellar feedback that includes momentum, mechanical energy and metals from stellar winds and supernovae is included,
in addition to dual-mode 
AGN feedback switched according to gas accretion rate following the \citet{dubois_self-regulated_2012} model.

\subsubsection{\textsc{TNG100}}\label{Methods-Simulations-TNG}

The \textsc{TNG100} simulation of the \textsc{IllustrisTNG}  project\footnote{\url{https://www.tng-project.org/}} (\citealt{Nelson_2019_TngPublicDataRelease}; see also \citealt{pillepich_first_2018}, \citealt{springel_first_2018}, \citealt{nelson_first_2018}, \citealt{naiman_first_2018}, \citealt{marinacci_first_2018}) is a uniform-resolution cosmological-volume (magneto-)hydrodynamical simulation employing the moving-mesh (magneto-)hydrodynamics code \textsc{Arepo} (\citealt{springel_e_2010}, \citealt{pakmor_magnetohydrodynamics_2011}, \citealt{pakmor_simulations_2013}) with a cubic simulation volume of side-length $75\,h^{-1}\,$Mpc (comoving). 
From the three flagship \textsc{IllustrisTNG} volumes 
(\textsc{TNG50}, \textsc{TNG100}, and \textsc{TNG300}) we adopt the intermediate \textsc{TNG100} simulation for this study as a compromise between resolution and volume.
A $\Lambda$CDM cosmology consistent with the \cite{planck_collaboration_planck_2016} results is assumed with $(\Omega_{m,0}, \Omega_{\Lambda,0}, \Omega_{b,0}, \sigma_{8}, n_{s}, h) = (0.309, 0.691, 0.049, 0.816, 0.967, 0.677)$. 

\textsc{TNG100} follows $1820^3$ DM particles ($m_{\textrm{DM}}\approx7.5\times10^6\,$M\textsubscript{\sun}) and initially has $1820^3$ gas cells (minimum cell radius $14$\,pc; target gas cell mass, $m_{\textrm{gas}}\approx1.4\times10^6\,$M\textsubscript{\sun}). 
Stochastic star formation occurs above a threshold gas density of $n_{\textrm{H}}=0.1\,\textrm{cm}^{-3}$ following the \citet{Springel_SfModel_2003} model at rates that empirically reproduce the Kennicutt-Schmidt law \citep{kennicutt_jr_global_1998}, with the typical stellar particle formation mass approximately matching the target gas cell mass (i.e. $m_{*}\sim1\times10^6\,$M\textsubscript{\sun}). 
The Plummer equivalent gravitational softening of collisionless
particles is $0.5\,h^{-1}\,$kpc 
at $z\leq1$ and $h^{-1}\,$comoving kpc at $z\geq1$; 
adaptive gravitational softening is used for gas, set at $2.5$ times the effective cell radius with an enforced minimum of $0.125\,h^{-1}\,$kpc (comoving). 
Radiative gas cooling (primordial and metal-line) occurs in the presence of a redshift-dependent UV background  switched on at $z=6$ (\citealt{pillepich_TNGref_2018}). 
Stellar feedback with galactic-scale, energy-driven kinetic winds is implemented following the \citet{pillepich_TNGref_2018} model, and dual-mode 
AGN feedback switched according to accretion rate is implemented following the \citet{weinberger_simulating_2017} model. 

\subsubsection{\textsc{The Three Hundred Gizmo-Simba 7K}}\label{Methods-Simulations-300}

The \textsc{The Three Hundred} simulation suite\footnote{\url{https://the300-project.org/}} (\citealt{cui_three_2018}) consists of hydrodynamical zoom-in simulations of the 324 most massive haloes (masses $\in[10^{14.8},10^{15.5}]\,$M\textsubscript{\sun}) drawn from the \textsc{MultiDark Planck 2} cosmological N-body simulation (side length $1\,h^{-1}\,\textrm{Gpc}$; \citealt{klypin_multidark_2016}). In this work we use data from the new \textsc{GIZMO-SIMBA 7K} simulation runs (Cui et al. in prep.), a higher-resolution and otherwise improved follow-up of the original \textsc{GIZMO-SIMBA (3K)} simulations \citep{cui_three_2022}. The meshless finite-mass Lagrangian hydrodynamics solver \textsc{GIZMO} (\citealt{hopkins_new_2015}, \citealt{hopkins_new_2017}) is employed, coupled with the \textsc{SIMBA-C} galaxy formation model (\citealt{hough_simba-c_2023}, \citealt{hough_simba-c_2024}). This fork of the \textsc{SIMBA} model \citep{dave_simba_2019} is differentiated chiefly by adopting the state-of-the-art \textsc{Chem5} cosmic chemical enrichment model (\citealt{kobayashi_origin_2020} and references therein). The simulation regions 
are each centred on a cluster-scale halo, with the high-resolution region extending at least $15\,h^{-1}\,\textrm{Mpc}$ from the cluster centre at $z=0$. 
A $\Lambda$CDM cosmology consistent with the \cite{planck_collaboration_planck_2014} results is assumed with $(\Omega_{m,0}, \Omega_{\Lambda,0}, \Omega_{b,0}, \sigma_{8}, n_{s}, h) = (0.307,0.693,0.048, 0.829, 0.961, 0.678)$. 

In the high-resolution regions, DM particles have $m_\text{DM}\sim2\times10^{8}\,$M\textsubscript{\sun} and gas particles $m_\text{gas}\sim4\times10^{7}\,$M\textsubscript{\sun}. 
Star formation occurs above a threshold gas density of $n_{\textrm{H}}=0.1\,\textrm{cm}^{-3}$ following a \citet{krumholz_comparison_2011} derived model which scales with gas molecular hydrogen content, and stellar particles form with $m_{*}\sim4\times10^{7}\,$M\textsubscript{\sun}.
The gravitational softening length for collisionless particles is $1.25\,h^{-1}$\,kpc at $z\leq1$ and $2.5\,h^{-1}$\,comoving~kpc at $z\geq1$; 
adaptive gravitational softening is used for gas \citep{hopkins_new_2015} with an enforced minimum softening length of $0.1\,h^{-1}$\,kpc at $z\leq1$ and $0.2\,h^{-1}$\,comoving~kpc at $z\geq1$. 
Radiative gas cooling (including metal cooling) and photoionization is incorporated via the \textsc{GRACKLE-3.3} library \citep{smith_grackle_2017}.
\textsc{SIMBA-C} largely follows the dual-mode AGN feedback implementation of \textsc{SIMBA} (as described in \citealt{dave_mufasa_2016, dave_simba_2019}; see \citealt{hough_simba-c_2023} for minor updates) and a recalibration of \textsc{SIMBA}'s star-formation-driven two-phase galactic winds model is also used, 
but the stellar feedback treatment in \textsc{SIMBA-C} has otherwise been substantially overhauled as compared to \textsc{SIMBA} (see \citealt{hough_simba-c_2023} for details). 

\subsubsection{\textsc{Hydrangea}}\label{Methods-Simulations-Hydra}

The \textsc{Hydrangea} simulations (\citealt{bahe_hydrangea_2017}, \citealt{bahe_disruption_2019}) are a suite of 24 hydrodynamical galaxy cluster zoom-in simulations run using a variant of the EAGLE galaxy formation and evolution code (variant ``AGN-dT9''; \citealt{schaye_eagle_2015}) and form part of the \textsc{C-EAGLE} project \citep{barnes_cluster-eagle_2017}. The EAGLE code is a substantial modification of the \textsc{Gadget-3} Smoothed Particle Hydrodynamics (SPH) code \citep{springel_cosmological_2005}. Each of the simulated regions is selected from an N-body simulation (\textsc{MACSIS}; \citealt{barnes_redshift_2017}) with side length $3.2$\,Gpc, and are each centred on a cluster-scale halo 
(mass $\in[10^{14.0},10^{15.4}]\,$M\textsubscript{\sun}; subject to the additional selection criteria of there being 
no more massive halo nearby at $z=0$). The high-resolution regions extend to at least $10\times r_{200c}$\footnote{For this work, $r_{\Delta c}$ is the radius within which the average DM density is $\Delta$ times the cosmic critical density (where $\Delta$ is either 178 or 200), and $M_{\Delta c}$ is the total mass enclosed by $r_{\Delta c}$.} 
from the centre of the central cluster at $z=0$.
A $\Lambda$CDM cosmology consistent with the \cite{planck_collaboration_planck_2014} results is assumed with $(\Omega_{m,0}, \Omega_{\Lambda,0}, \Omega_{b,0}, \sigma_{8}, n_{s}, h) = (0.307,0.693,0.048, 0.829, 0.961, 0.678)$. 

In the high-resolution regions, DM particles have $m_\text{DM}\approx9.8\times10^{6}\,$M\textsubscript{\sun}, and gas particles have an initial mass of $m_\text{gas,ini}\approx1.8\times10^{6}\,$M\textsubscript{\sun}. Star formation occurs above a metallicity-dependent gas density threshold of $n_{\textrm{H}}=(Z/0.002)^{-0.64}\times0.1$\,cm$^{-3}$ \citep{schaye_star_2004} following the pressure implementation of the Kennicutt--Schmidt law \citep{kennicutt_star_1989} from \citet{schaye_relation_2008}, and stellar particles have $m_{*}\sim1\times10^{6}\,$M\textsubscript{\sun}. The Plummer-equivalent gravitational softening length is 0.7\,kpc at $z<2.8$.
Additions atop the base hydrodynamics scheme of \textsc{Gadget-3} include radiative cooling, photoheating, and reionization (using \citealt{wiersma_effect_2009} models); mass and metal enrichment of gas due to stellar outflows (based on \citealt{wiersma_chemical_2009}); and thermal feedback from both star formation and supermassive black holes (\citealt{dalla_vecchia_simulating_2012}, \citealt{rosas-guevara_impact_2015}, \citealt{schaye_eagle_2015}). 
For further detail on the \textsc{Eagle} code and its ``\textsc{ANARCHY}'' hydrodynamics scheme update see \citet{schaye_eagle_2015} and \citet{schaller_eagle_2015}. 

\subsection{Homogenized ICL definition}\label{Methods-IclDef}

\subsubsection{Distinguishing galactic and intergalactic stars}\label{Methods-IclDef-BaseStrucFinders}

The four simulations we employ for this study by default adopt different codes for structure identification. 
\textsc{Hz-AGN} uses the \textsc{AdaptaHOP} structure finder (\citealt{aubert_origin_2004}, \citealt{tweed_building_2009}). \textsc{TNG100} uses \textsc{Subfind} (\citealt{springel_populating_2001}, \citealt{dolag_substructures_2009}). \textsc{The300 GS-7K} uses the \textsc{Amiga Halo Finder} (\textsc{AHF}; \citealt{gill_evolution_2004}, \citealt{knollmann_ahf_2009}). We employ the \textsc{Cantor} structure catalogue (Bah\'{e} et al. in prep.) for \textsc{Hydrangea}. 
We omit detailed general descriptions of each structure finder for concisenesses, referring the interested reader to the introductory papers of each code (see also \citealt{knebe_haloes_2011}, \citealt{onions_subhaloes_2012} for a comparison of the different approaches). 

\textsc{Subfind}, \textsc{AHF}, and \textsc{Cantor} do not innately separate ``galaxies'' from the DM (sub-)structures hosting them. The stars and gas of a galaxy, the DM of its host (sub-)halo, as well as any circumgalactic medium or diffuse stellar envelope within that (sub-)halo are all combined together into one amalgamated structure: a ``sub-halo''\footnote{We largely adopt the terminology of \textsc{Subfind} and refer to the gravitationally-bound, DM+baryonic structures identified by \textsc{Subfind}, \textsc{AHF}, and \textsc{Cantor} as ``sub-haloes'', with one exception: in isolation, we refer to the self-bound DM+baryons of a cluster's main halo (excised of embedded sub-structure) as the cluster's ``central structure'' to avoid implying this structure to be a sub-structure.}.
In the context of ICL studies, this means that there is no separation between the BCG and the ICL of a cluster imposed 
by these structure finders, 
and the implementation of any such division is left to the end-user. 
Conversely, \textsc{AdaptaHOP} 
separately identifies DM structures (``DM haloes'') and stellar structures (``galaxies'') and so does intrinsically impose an explicit distinction between galactic and intergalactic stars: the local stellar density of each stellar particle is estimated using 20 nearest neighbour particles \citep{tweed_building_2009}, and then a density threshold applied to split the two populations of stars. No unbinding procedure is employed. In \textsc{Hz-AGN}, this threshold density is $178\times$ the cosmic average total matter density \citep{dubois_dancing_2014}.
We adopt this local stellar density definition for (inter-)galactic stars, along with an arbitrary cluster border at $r_{178c}$ (centred on the central galaxy), as our fiducial ICL definition for this study -- calculating local stellar density following the methodology of \citet{tweed_building_2009} and \citet{dubois_dancing_2014}.

\subsubsection{Galaxy catalogue construction}\label{Methods-IclDef-GalDef} 

Under our fiducial ICL definition 
all stellar particles with local stellar densities below $178\times$ the cosmic average total matter density are considered intergalactic. For the purpose of then identifying ``galaxies'' among the remaining ``galactic'' stellar particles in \textsc{TNG100}, \textsc{The300 GS-7K}, and \textsc{Hydrangea} sub-haloes (referring to the mixed structures of DM and baryons identified by the structure finder), we assume each sub-halo to contain at most 
one galaxy,  
to which the sub-halo's merger tree is associated (see Section~\ref{Methods-TrackingGalaxies-MergerTrees}). 
However, we do not simply assign all stellar particles with sufficient local density in each sub-halo to this galaxy. In massive haloes, the dynamically unbound but dense stellar envelopes around infalling satellites are reassigned to the 
central structure
by the unbinding procedure (for further discussion see e.g. \citealt{rodriguez-gomez_merger_2015}, \citealt{ahvazi_progenitors_2024}). 
We seek to exclude these spatially distinct stellar envelopes when identifying the central galaxies of massive haloes (associating these envelopes instead with the satellite galaxies they surround). For this purpose, 
we perform a particle clustering search on the galactic-density stellar particles of each sub-halo with the \textsc{DBSCAN} clustering algorithm (\citealt{ester_density-based_1996}). For this search we target neighbourhood densities of $\geq178\times$ the cosmic average matter density and consider a neighbourhood size equal to the half mass radius of all the sub-halo's ``galactic'' stellar particles, centred on the density maximum as found by a ``shrinking sphere'' approach. 
The largest structure identified by this clustering search is then used as a seed for the main stellar structure in each sub-halo and -- once these searches have been carried out on all sub-haloes --
leftover galactic density stars are allocated to these structure seeds by a density walk through their 20 nearest neighbour stellar particles, emulating how stars are assigned to local density peaks to construct galaxies in \textsc{AdaptaHOP}.

\textsc{Subfind} and \textsc{Cantor} sometimes fragment individual galaxies into several distinct objects. This behaviour is undesirable
for our use case, and
our procedure for reconsolidating these fragments into a single galaxy is described in Appendix~\ref{Appendix:Methods-IclDef-Clumps}.  
\textsc{AHF} follows the inclusive particle convention (in which the particle membership lists of structures can overlap) unlike the other structure finders, which assign particles to (at most) one structure each. 
Before carrying out our galaxy identification procedure, 
we obtain exclusive membership lists for \textsc{AHF} objects by assigning each particle to the lowest particle count structure it is a member of.

In \textsc{Hz-AGN} a minimum particle count of 51 for stellar structures to 
be kept 
is also imposed.
For consistency, we impose this same minimum stellar particle count uniformly across the simulations: 
sub-threshold sub-structures of stars identified within a larger stellar structure are absorbed into this host galaxy, and the stellar particles of isolated small objects are instead recategorised as ``intergalactic''.
We confirm that imposing this minimum stellar particle count has no qualitative impact on any of our findings (though see Sections~\ref{Discussion-SimDiffs-IclMassFracs} and~\ref{Discussion-SimDiffs-FromInfallersFrac}). 

\subsubsection{BCG determination and fiducial definitions summary}\label{Methods-IclDef-BcgDetermination}

In \textsc{Hz-AGN} we identify the cluster BCG as the most massive galaxy
within $0.1\times r_{178c}$ of the cluster halo centre at $z\approx0$, and assume the main progenitor of this galaxy to always be the BCG of the cluster (progenitor structure) in all earlier snapshots (see Section~\ref{Methods-TrackingGalaxies-MergerTrees}). 
In \textsc{TNG100}, \textsc{The300 GS-7K}, and \textsc{Hydrangea} we 
assume the cluster BCG to be the galaxy identified within the cluster's central structure at $z=0$ 
and that the main progenitor of this structure always contains the central galaxy of the cluster (progenitor) in all earlier snapshots.

Our fiducial BCG definition can thus be summarised as the BCG being the stellar structure associated with the cluster's main halo, composed exclusively of stellar particles with local stellar densities greater than $178\times$ the cosmic average total matter density (with no unbinding criteria employed). Under our fiducial ICL definition, all stellar particles within an outer cluster border at $r_{178c}$ that fall below this same local stellar density threshold are considered part of the cluster's ICL component (with a small additional contribution from unresolved satellites; Section~\ref{Methods-IclDef-GalDef}).

\subsubsection{BCG-ICL separation arbitrariness}\label{Methods-IclDef-CombinedBcgIcl} 

Arguably the most disputed aspect of different ICL extraction methodologies is the separation of the BCG and the ICL: determining where to place the border between these two cluster components is non-trivial and somewhat subjective, and different selections can yield considerable differences in the determined properties of both components (e.g. \citealt{brough_preparing_2024}, \citealt{kimmig_intra-cluster_2025}, \citealt{montenegro-taborda_stellar_2025}). As a result, rather than implement an arbitrary border, several prior ICL studies have opted not to separate the BCG from the ICL and consider the combined system only (e.g. \citealt{zhang_dark_2019}, \citealt{chun_formation_2023}, \citealt{jeon_origin_2025}; see also \citealt{pillepich_first_2018} and \citealt{montenegro-taborda_stellar_2025} for further relevant discussion). 
We probe the consequences of our fiducial approach for separating the BCG and the ICL on determined ICL stellar mass fractions in Section~\ref{Results-StellarMassDistribution-AltDefIclEvo}, but also provide alternative versions of our analysis where applicable considering the combined BCG\,+\,ICL rather than the ICL alone. 

\subsection{Linking ICL stars to progenitor galaxies}\label{Methods-TrackingGalaxies}

We follow a very similar scheme for assigning $z\approx0$ ICL stars to progenitor satellite galaxies as was implemented in \citetalias{brown_assembly_2024}. We first trace the main progenitor structures of $z\approx0$ clusters back to a lookback time of approximately 12\,Gyr ($z\sim3$). 
We then track the stellar particles of satellites falling into these structures after $z\sim3$ 
and so generate lists of stellar particles associated with particular satellite galaxies, which can then be used to look up the progenitor objects of $z\approx0$ ICL stars.

\subsubsection{Snapshot spacing and merger trees}\label{Methods-TrackingGalaxies-MergerTrees}

We perform our analysis using simulation snapshots with a coarse time resolution of $\sim1$\,Gyr (henceforth referred to as ``coarsely spaced snapshots'') -- selected to limit the computational expense of our analysis while still employing a spacing smaller than one cluster crossing time, allowing each orbit of an infalling satellite to be sampled multiple times.
We follow structures between snapshots in \textsc{Hz-AGN} using merger trees generated by the 
\textsc{Treemaker} algorithm (\citealt{hatton_galics-_2003}, \citealt{tweed_building_2009}) built using a time resolution of $\sim0.05$\,Gyr between $z\approx3$ and $z\approx0$. 
In \textsc{TNG100} we use the provided merger trees output by the \textsc{Sublink} algorithm (we specifically use the baryonic-based variant trees, ``\textsc{Sublink\_gal}''; \citealt{rodriguez-gomez_merger_2015}). 
In \textsc{The300 GS-7K} we use the provided outputs of the \textsc{MergerTree} algorithm incorporated into \textsc{AHF} (we specifically use the ``skipping+stars'' variant trees first described in \citealt{Contreras-Santos_DMDG_The300_2024} that incorporate tree break patching and consider both DM and stars in the utilised merit function). In \textsc{Hydrangea} we use the provided merger trees for the \textsc{Cantor} catalogue objects as generated by the \textsc{Spiderweb} algorithm \citep{bahe_disruption_2019}. 

It is known that during a near-binary merger between a central galaxy and a massive satellite, structure finders can sometimes behave erratically -- with significant ``mass sloshing'' between the central and satellite sub-haloes, and with which structure is considered the central able to flip back-and-forth between consecutive snapshots (the so-called ``sub-halo switching problem''; for further discussion see e.g. \citealt{muldrew_accuracy_2011}, 
\citealt{behroozi_major_2015},
\citealt{chandro-gomez_accuracy_2025}). 
To limit the potential impact of this behaviour on our analysis, for some clusters 
we slightly stray (for a single snapshot) from our usual snapshot spacing of $\sim1$\,Gyr to avoid using snapshots when the cluster central is in the midst of such a merger. 

\subsubsection{Tracking infalling satellites}\label{Methods-TrackingGalaxies-ActualTracking}

We consider galaxies to become satellites of clusters once they pass $r_{178c}$, and track the stellar particles of all galaxies that become cluster satellites since a lookback time of $\sim12$\,Gyr in order to associate $z\approx0$ ICL stars with progenitor objects. 
Infalling galaxies are identified as galaxies with centres-of-mass outside $r_{178c}$ whose main descendants in the next coarsely spaced snapshot 
are cluster satellites. 
We also label as an infalling satellite any galaxy outside $r_{178c}$ which has no detected descendant in the next coarsely spaced snapshot (or that merges into the cluster's central structure) provided we find the centre-of-mass of all stellar particles formerly belonging to that galaxy within $r_{178c}$. 
To avoid double counting from ``splashback'' galaxies that resurface above $r_{178c}$ before falling back into the cluster, this infalling satellite identification is performed chronologically and each galaxy's first observed $r_{178c}$ crossing considered its original infall. 

All stellar particles belonging to an infalling galaxy when it is first identified as such 
are tagged as having that galaxy as their progenitor object. We then follow the main descendant of that progenitor galaxy through all remaining coarsely spaced snapshots, tagging to the same progenitor stellar particles formed within that galaxy after first cluster infall. 
In order to avoid double counting due to galaxy mergers, we stop tagging new stellar particles to a progenitor once the original infalling galaxy is no longer the main progenitor of the descendant galaxy those stellar particles form in. 

It is known that merger tree algorithms can sometimes struggle in the extreme environments of clusters, occasionally leading to tracking failures or inappropriate links between objects (for further discussion, see e.g. \citealt{srisawat_sussing_2013}, \citealt{poole_convergence_2017}, \citealt{chandro-gomez_accuracy_2025}). 
When a satellite either shares zero stellar particles with its supposed descendant or has no apparent descendant in the next coarsely spaced snapshot, and yet another galaxy exists containing $>50$~per cent of the stellar particles formerly belonging to the satellite, we overrule the merger tree and connect the tracked satellite to this assumed ``correct'' descendant. 

We do not attempt to separately quantify the ICL contributions from tidal stripping and violent relaxation during galaxy mergers as part of this study, and refer to the combination of these processes as the liberation of stars from satellite galaxies. Disentangling the contributions from these two processes during a major galaxy merger is non-trivial, and the proportional ICL contributions of the two channels highly sensitive to the arbitrary demarcation imposed (see \citealt{contini_moreimptime_2024} and references therein for further relevant discussion). 

\subsection{Cluster sample selection}\label{Methods-ClusterSamples}

As no structures more massive than $10^{15}\,$M\textsubscript{\sun} are present in either \textsc{Hz-AGN} or \textsc{TNG100}, 
we restrict ourselves to \textsc{The300 GS-7K} and \textsc{Hydrangea} clusters with $z=0$ halo masses less than $10^{15}\,$M\textsubscript{\sun}
to limit the mass discrepancy between our cluster samples. We also exclude from our samples any clusters with a central merger tree that cannot be followed back to a lookback time of $\sim12$\,Gyr. As described in Pearce et al. (in prep.), in some \textsc{The300 GS-7K} clusters the main merger tree branch of the cluster's DM halo clearly decouples from that of the associated $z=0$ central galaxy (due to a ``stellar core switch'' occurring during a close encounter between two massive systems). Our assumption that the main progenitor of the $z=0$ central structure always contains the main progenitor of the identified $z=0$ BCG in \textsc{TNG100}, \textsc{The300 GS-7K}, and \textsc{Hydrangea} breaks down if such a ``core switch'' occurs. We thus follow the methodology of Pearce et al. (in prep.; using our local stellar density defined central galaxies and coarsely spaced snapshots) to identify clusters that experience such a decoupling at $z\lesssim2$ and exclude these from our cluster samples as well. 
Additional details on all the simulated clusters we consider in this study are provided in Appendix~\ref{Appendix-ClusterProps}, including 
individual cluster identifiers.  

In the latest-time \textsc{Hz-AGN} simulation snapshot we use for this work (snapshot 761, $z=0.0556$), there are $14$ cluster-scale haloes
($M_{178c}\gtrsim10^{14}\,$M\textsubscript{\sun}). 
We consider only the sub-set of 10 for which we are able to follow the BCG back to a lookback time of $\sim12$\,Gyrs. 
These clusters have $M_{178c}$ values in the range $[1.00,7.52]\times10^{14}$\,M\textsubscript{\sun}. 

Our \textsc{TNG100} sample consists of the 15 most massive Friends-of-Friends (FoF) groups in the \textsc{TNG100} simulation box at $z=0$ (snapshot 99; we use $M_{200c}$ masses for this selection).
These clusters have $M_{178c}$ values in the range $[0.94,3.72]\times10^{14}$\,M\textsubscript{\sun}. Four of these clusters (snapshot 99 FoF IDs 0, 4, 5, and 10; see Appendix~\ref{Appendix-ClusterProps}) are within 5\,Mpc of the simulation boundaries at $z=0$, with two of these (IDs 0 and 10) crossing the boundaries during our period of study. Addressing concerns about edge-effects in \textsc{TNG100} raised by \citet{saulder_isolated_2020} and \citet{mitrasinovic_living_2025}, we confirm that the inclusion of these clusters does not meaningfully alter any of our results.

We restrict ourselves to \textsc{The300 GS-7K} clusters less massive than $10^{15}$\,M\textsubscript{\sun} at $z=0$ (using $M_{200c}$ masses for this selection), and also exclude clusters with central structure merger trees that either cannot be followed back to a lookback time of $\sim12$\,Gyr or that are otherwise problematic (see above).  
This yielded a \textsc{The300 GS-7K} sample consisting of 15 clusters with $M_{178c}$ values in the range $[8.56,10.71]\times10^{14}$\,M\textsubscript{\sun}. 

Of the 24 \textsc{Hydrangea} clusters, we restrict ourselves to the sub-set of 18 clusters with $M_{178c}(z=0)\lesssim10^{15}\,$M\textsubscript{\sun}. 
From this sub-set, we exclude two further clusters: 
one due to issues with the merger tree of its central structure (see above) and another for an unphysical AGN outburst at $z\sim1.5$ (identifiers CE-11 and CE-10 respectively; \citealt{bahe_hydrangea_2017}). 
This yielded a sample of 16 clusters, which have $M_{178c}$ values in the range $[1.16,6.86]\times10^{14}$\,M\textsubscript{\sun}.


\section{Stellar mass fractions}\label{Results-StellarMassDistribution}

\subsection{Distribution of cluster stellar mass at $z\approx0$}\label{Results-StellarMassDistribution-z0}

\begin{figure*}
    \includegraphics[width=2.07\columnwidth]{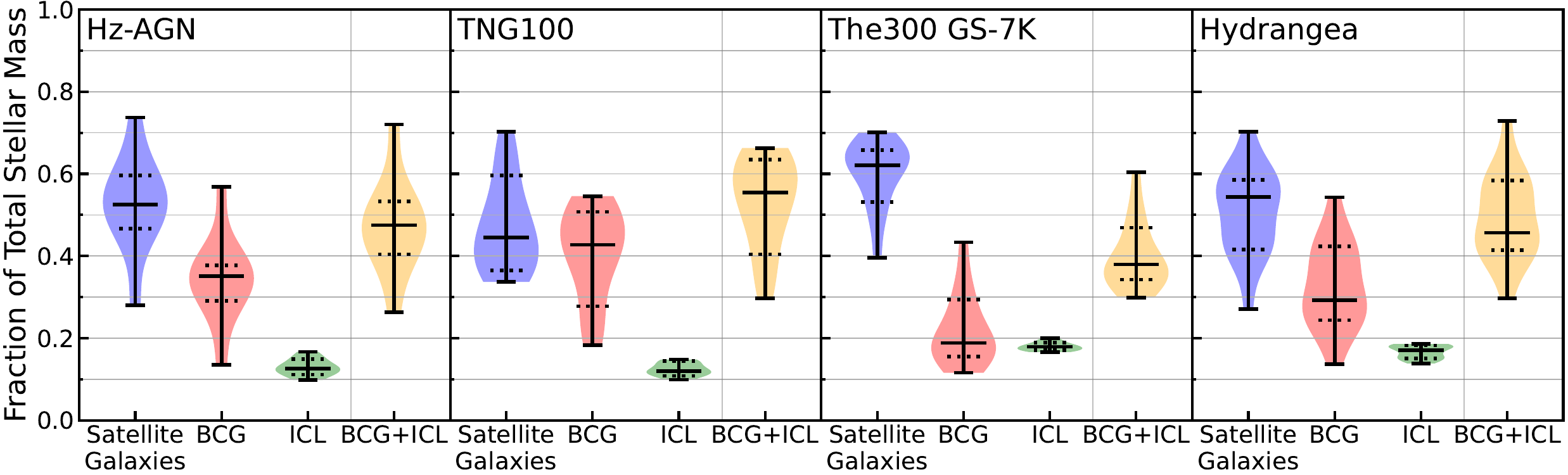}
    \caption{
    Violin plots showing the distribution of stellar mass at $z\approx0$ between satellite galaxies, the BCG, and the ICL 
    in simulated galaxy clusters 
    from \textsc{Horizon-AGN} (leftmost panel), \textsc{TNG100} (middle-left panel), \textsc{The Three Hundred Gizmo-Simba 7K} (middle-right panel), and \textsc{Hydrangea} (rightmost panel), when uniformly applying our fiducial ICL identification methodology (based on local stellar density). Combined BCG\,+\,ICL stellar mass fractions are also shown. Dotted lines indicate 16th and 84th percentiles, and solid lines denote maximum, minimum, and median values. 
    }
    \label{fig:StelFracVplot}
\end{figure*}

We begin by examining the $z\approx0$ ICL (and BCG\,+\,ICL) stellar mass fractions of the cluster samples drawn from each of the four simulations, homogenized to our fiducial ICL definition based on local stellar density (Section~\ref{Methods-IclDef}), to probe for significant intrinsic inter-simulation differences. 
We present in Figure~\ref{fig:StelFracVplot} violin plots showing the distribution of cluster stellar mass (within $r_{178c}$) between satellite galaxies, the BCG, and the ICL at $z\approx0$  separately for the cluster sample from each simulation.  
Combined BCG\,+\,ICL fractions are also included. Stellar mass fractions are given for each cluster individually in Appendix~\ref{Appendix-ClusterProps}. 

The ICL fractions of the four cluster samples are broadly consistent, all lying in the range $\sim0.1$ to $\sim0.2$, which is far less scatter than the wide range of values reported for various different ICL extraction methodologies by prior theoretical studies (between $\sim0.05$ and $>0.5$; \citealt{kluge_photometric_2021}).
We note a slight dispersion in typical ICL fraction between the four simulations, which we discuss in Section~\ref{Discussion-SimDiffs-IclMassFracs}. 
The ICL fractions we obtain 
lie towards
the lower end of the broad range of previously reported values, but are markedly similar to the findings of past studies employing comparable ICL definitions, such as \citet{rudick_quantity_2011}, who reported $z\approx0$ ICL fractions between $\sim0.09$ and $\sim0.15$ for a sample of similarly massive simulated clusters, and \citet{joo_tracing_2024}, who reported a typical ICL fraction of $\sim0.12$ for a large sample of simulated haloes (mass $>10^{13}\,$M\textsubscript{\sun}) at $z\sim0.625$. 

While the BCG\,+\,ICL stellar mass fractions of the four simulations span very similar ranges ($\sim0.3$ to $\sim0.7$), there is noticeable scatter between the average BCG\,+\,ICL fractions, 
driven chiefly by a corresponding scatter in typical BCG mass fractions. 
Most apparent is the comparatively low average BCG fraction for \textsc{The300 GS-7K} (of only $\sim0.2$), though the distribution of values in \textsc{TNG100} also appears mildly top-heavy,  
consistent with the findings of \citet{brough_preparing_2024} and \citet{kimmig_intra-cluster_2025}, who both reported slightly higher average BCG\,+\,ICL fractions in \textsc{TNG100} clusters compared to those from \textsc{Hz-AGN}. 
We discuss possible factors contributing to these differing typical BCG (and hence also BCG\,+\,ICL) mass fractions in Section~\ref{Discussion-SimDiffs-BcgIclMassFracs} (see also Appendix~\ref{Appendix-ClusterProps}) but in brief 
we largely attribute the usually lower BCG fractions of \textsc{The300 GS-7K} to these generally more massive clusters 
typically assembling more of their stellar mass closer to $z\approx0$.

\subsection{ICL mass fraction evolution}\label{Results-StellarMassDistribution-IclEvo}

\begin{figure*}
    \includegraphics[width=2.07\columnwidth]{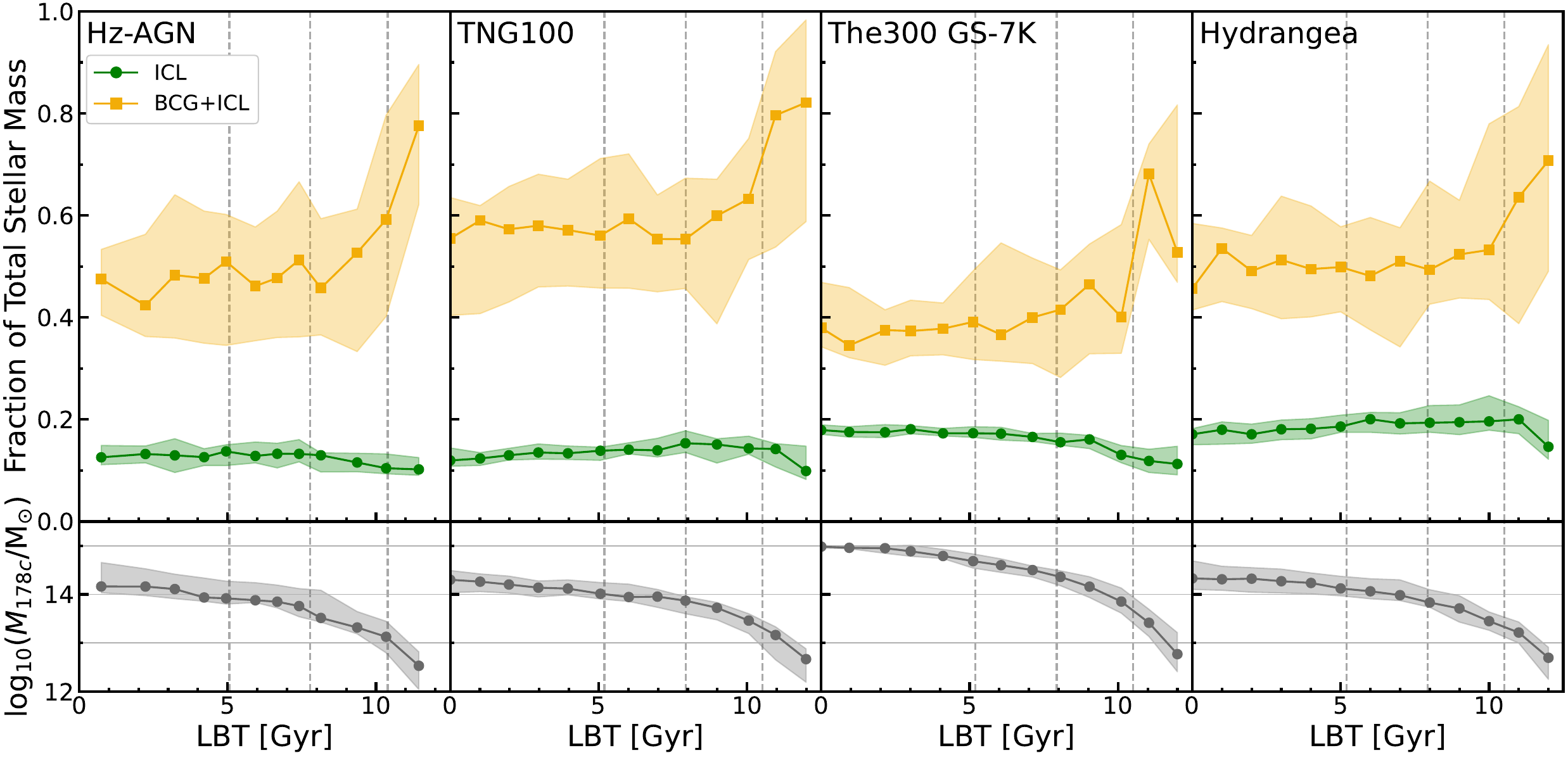}
    \caption{
    \textbf{Top Panels:} 
    Median ICL (green circles) and BCG\,+\,ICL (gold squares) stellar mass fraction as a function of lookback time (LBT) for $z\approx0$ galaxy clusters and their main progenitor structures at earlier times drawn from \textsc{Horizon-AGN} (leftmost panel), \textsc{TNG100} (middle-left panel), \textsc{The Three Hundred Gizmo-Simba 7K} (middle-right panel), and \textsc{Hydrangea} (rightmost panel). 
    \textbf{Bottom Panels:} Median mass of structures considered in the top panels as a function of lookback time. 
    The shaded regions are bounded by 16th and 84th percentiles. 
    The dashed vertical lines indicate $z=0.5,1,2$.
    }
    \label{fig:IclFracEvoPlot}
\end{figure*}

We examine next the evolution of ICL (and BCG\,+\,ICL) stellar mass fractions with redshift. Prior theoretical works have reported conflicting findings for how the ICL component evolves with redshift (e.g. \citealt{rudick_quantity_2011}, \citealt{kimmig_intra-cluster_2025}). We probe for significant inter-simulation discrepancies which might contribute to this tension, unobstructed by inconsistent ICL extraction methodologies. 
Figure~\ref{fig:IclFracEvoPlot} depicts 
median ICL stellar mass fractions as a function of lookback time for the samples of $z\approx0$ galaxy clusters and their $z>0$ progenitor structures\footnote{We continue to refer to the diffuse stellar component of even the sub-cluster scale progenitor structures as ICL.} from the four simulations, homogenized to our fiducial ICL definition (top panels; the bottom panels indicate typical progenitor structure mass as a function of lookback time).
The shaded regions are bounded by 16th and 84th percentiles as an indication of the cluster-to-cluster scatter. 
Combined BCG\,+\,ICL fractions are also included.
We present an alternative version of this analysis confined to cluster mass ($\gtrsim10^{14}\,$M\textsubscript{\sun}) structures at all times in Appendix~\ref{Appendix-MassLimitedIclFracEvo}.

Although the ICL and BCG\,+\,ICL stellar mass fractions of individual structures can fluctuate over time, we find no significant evolution in either the average ICL or BCG\,+\,ICL mass fraction 
in any simulation until at least $z\sim2$. At $z\gtrsim2$ there are hints of a common downturn in typical ICL fraction and the average BCG\,+\,ICL fractions all begin to rise, which we consider in both cases symptomatic of progenitor structures falling below the group scale as opposed to intrinsic redshift evolution.

Our finding of no evolution in the average ICL fraction (at least for $z\lesssim2$) is in agreement with the results of some prior theoretical studies -- \citet{contini_moreimptime_2024} predicted no trend in ICL fraction with redshift (for $\gtrsim10^{13}\,$M\textsubscript{\sun} haloes at $0<z<3$ based on semi-analytical models) and \citet{joo_tracing_2024} found a nearly constant median ICL fraction across various redshifts (for $\gtrsim10^{13}\,$M\textsubscript{\sun} haloes at $z=0.625$ and their progenitors back to $z\sim3$ using the \textsc{Horizon Run 5} hydrodynamical simulation) -- but disagrees with the findings of \citet{rudick_quantity_2011}, who reported ICL fractions 
that steadily fall with increasing lookback time, approaching zero by $z\sim2$. 
Although \citet{rudick_quantity_2011} employed a density-based ICL definition 
broadly similar to our fiducial ICL definition, 
their methodology differs by imposing a \textit{fixed} stellar density threshold to distinguish ICL stars (e.g. $\rho_{\textrm{thresh}} = 10^{-5}\,$M\textsubscript{\sun}\,pc\textsuperscript{-3}), whereas for this study we emulate \textsc{AdaptaHOP} and employ a stellar density threshold that scales with the \textit{redshift-dependent} cosmic average matter density (e.g. $\rho_{\textrm{thresh}}(z=0)\sim7\times10^{-6}\,$M\textsubscript{\sun}\,pc\textsuperscript{-3}, $\rho_{\textrm{thresh}}(z=2)\sim2\times10^{-4}\,$M\textsubscript{\sun}\,pc\textsuperscript{-3}) for our fiducial ICL definition.

\begin{figure*}
    \includegraphics[width=2.07\columnwidth]{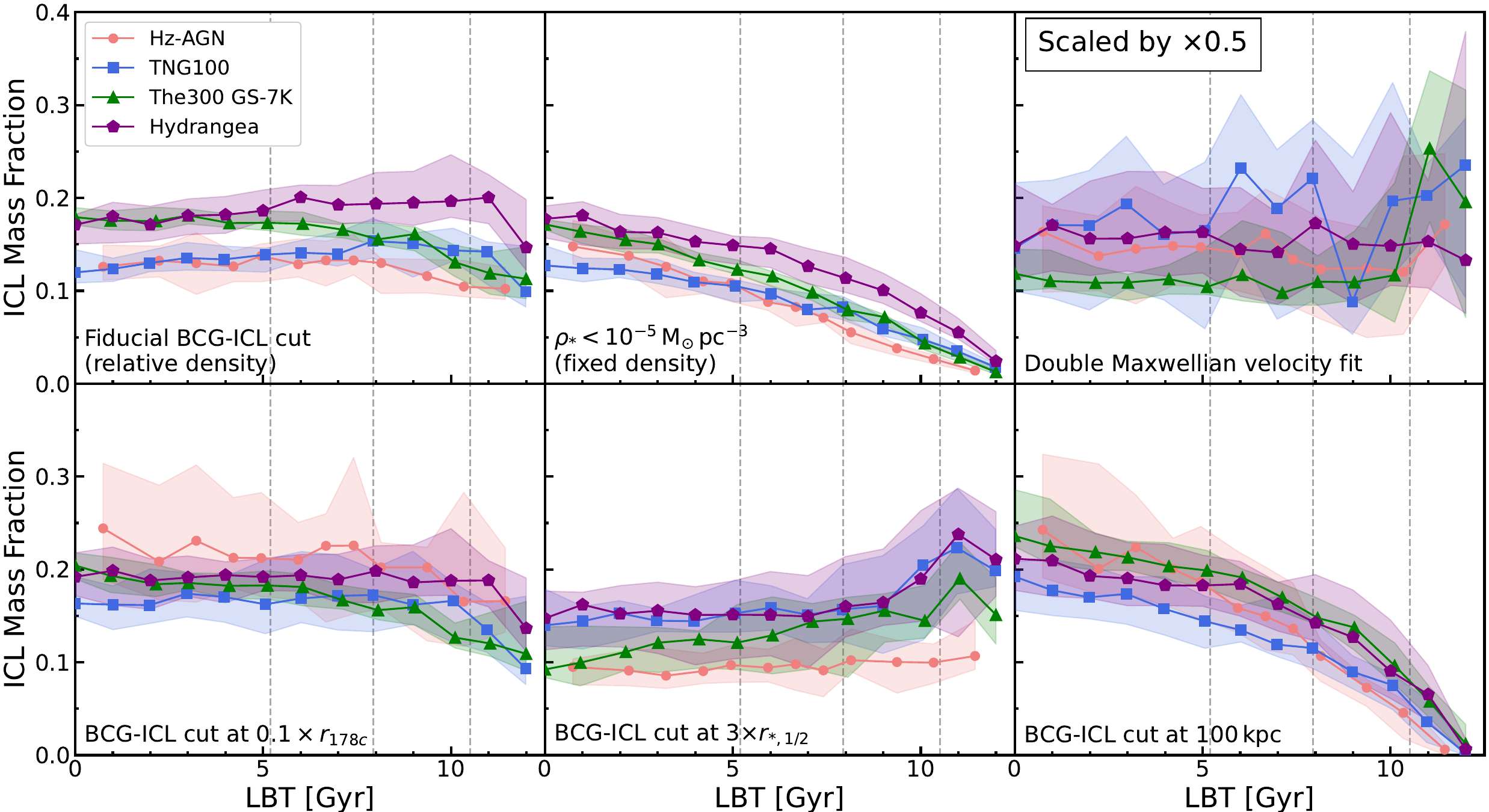}
    \caption{
    Median ICL stellar mass fraction as a function of lookback time (LBT) for a selection of different BCG-ICL separation methodologies, for galaxy clusters 
    drawn from \textsc{Horizon-AGN} (red circles), \textsc{TNG100} (blue squares), \textsc{The Three Hundred Gizmo-Simba 7K} (green triangles), and \textsc{Hydrangea} (purple pentagons). The shaded regions are bounded by the 16th and 84th percentiles of the ICL fractions from each simulation. 
    The dashed vertical lines indicate $z=0.5,1,2$ in the Planck2015 cosmology.  
    \textbf{Top Left Panel:} Our fiducial ICL extraction methodology (local stellar density cut at $\rho_{*}=178\times$ the cosmic average matter density; reproduced from Figure~\ref{fig:IclFracEvoPlot}). 
    \textbf{Top Middle Panel:} Local stellar density cut at $\rho_{*}=10^{-5}$\,M\textsubscript{\sun}\,pc$^{-3}$.
    \textbf{Top Right Panel:} Ratio between BCG and ICL masses from fitting a double Maxwellian to BCG\,+\,ICL stellar particle velocities. 
    The presented values have been scaled by $\times0.5$.
    \textbf{Bottom Left Panel:} BCG and ICL split at a galactocentric radius of $0.1\times r_{178c}$. 
    \textbf{Bottom Middle Panel:} BCG and ICL split at a galactocentric radius of three times the combined BCG\,+\,ICL stellar half mass radius. 
    \textbf{Bottom Right Panel:} BCG and ICL split at a fixed galactocentric radius of 100\,kpc. 
    }
    \label{fig:AltIclFracEvoPlot}
\end{figure*}

\subsection{Influence of BCG-ICL separation}\label{Results-StellarMassDistribution-AltDefIclEvo}

In order to probe the influence that the fiducial ICL extraction methodology we employ for this study has on our findings for the evolution of the ICL fraction with lookback time, we repeat our analysis from Section~\ref{Results-StellarMassDistribution-IclEvo} for a selection of common alternative approaches for separating the BCG from the ICL.
The first alternative BCG-ICL separation we consider is a \textit{fixed} local stellar density threshold, emulating \citet{rudick_quantity_2011} and using the same fixed stellar density threshold (of $\rho_{*}=10^{-5}$\,M\textsubscript{\sun}\,pc$^{-3}$). We additionally consider three alternative methodologies all based on different spherical aperture cuts: splitting BCG from ICL at 10 per cent of $r_{178c}$; splitting at three times the combined BCG\,+\,ICL stellar half-mass radius, $r_{*,1/2}$; and splitting at an arbitrary fixed radius. 
Lastly, we also consider a kinetic BCG-ICL separation employing a double Maxwellian fit to the velocities of BCG\,+\,ICL stellar particles (following the methodology of \citealt{dolag_dynamical_2010}, \citealt{remus_outer_2017} as implemented in \citealt{brough_preparing_2024}). 
For these repeat analyses, we retain both our arbitrary cluster border at $r_{178c}$ and our fiducial procedure for demarcating satellite galaxies from the ICL.

The results of these alternative analyses are depicted in Figure~\ref{fig:AltIclFracEvoPlot}, which shows how the median ICL fractions for each of the different BCG-ICL separation methodologies evolve as a function of lookback time for each of the four simulated cluster samples. 
We present an alternative version of this analysis restricted to cluster mass ($\gtrsim10^{14}\,$M\textsubscript{\sun}) structures at all times in Appendix~\ref{Appendix-MassLimitedIclFracEvo}. 

Fixed apertures with radii between 30 and 200\,kpc were considered, but we include only the results for a 100\,kpc aperture in Figure~\ref{fig:AltIclFracEvoPlot} as a representative example; smaller apertures result in larger characteristic ICL fractions (e.g. $\sim0.3-0.4$ at $z\approx0$ for a 30\,kpc aperture compared to $\sim0.1-0.2$ for a 200\,kpc aperture) but follow the same overall trends with lookback time 
(see \citealt{montenegro-taborda_stellar_2025} for further relevant discussion).
Similarly, separating the BCG from the ICL at $2\times r_{*,1/2}$ (as in e.g. \citealt{pillepich_first_2018}) 
rather than $3\times r_{*,1/2}$ does not alter the overall trends with lookback time shown in the bottom middle panel of Figure~\ref{fig:AltIclFracEvoPlot} and just raises the typical ICL mass fractions found slightly (such that the median ICL fractions at $z\approx0$ fall in the range $\sim0.12-0.20$). 
Projected circular apertures 
with fixed radii were also examined, which yield ICL fractions typically just below their spherical cut equivalent (e.g. within 4 percentage points at $z\approx0$ for 100\,kpc) and reproduce qualitatively the same trends with lookback time. 

Considerable scatter is apparent in Figure~\ref{fig:AltIclFracEvoPlot} between the typical ICL fractions that the different BCG-ICL split methodologies yield, even for the same sample of clusters from the same simulation (a phenomenon also reported in prior studies e.g. \citealt{brough_preparing_2024}, \citealt{montenegro-taborda_stellar_2025}). Particularly conspicuous is the kinematic methodology yielding typical ICL fractions 
significantly larger than any other BCG-ICL separation considered (with typical $z\approx0$ ICL fractions in the range $\sim0.2-0.3$; approximately twice those obtained by the other methods), qualitatively similar to the findings of \citealt{brough_preparing_2024}.

How the typical ICL fraction evolves as a function of lookback time is also not consistent between the different BCG-ICL separation methodologies. Under the fixed aperture and fixed density cut methodologies, all four simulations yield ICL fractions that decrease with lookback time, approaching zero by $z\sim3$, reproducing the findings of \citet{rudick_quantity_2011}. 
In contrast, the kinematic separation methodology as well as BCG-ICL cuts scaled by $r_{178c}$ or $r_{*,1/2}$ yield no significant evolution in the typical ICL fraction from $z\approx0$ to $z\sim2$, just as with our fiducial ICL extraction methodology. 
However, it merits highlighting that when homogenized to the same BCG-ICL split methodology, there are no significant discrepancies between the different simulations for the evolution of the ICL fraction as a function of lookback time. Essentially the same trends are reproduced by every considered simulation for each considered BCG-ICL separation methodology, and so the different trends with lookback time appear to just stem from the different ICL definitions being imposed. 

\citet{kimmig_intra-cluster_2025} previously investigated how the fraction of stellar mass in different cluster components evolved with time for galaxy clusters from the \textsc{Magneticum} simulation for various BCG-ICL splits. They used effectively the same selection of BCG-ICL separations we trial in the three bottom panels of Figure~\ref{fig:AltIclFracEvoPlot} (i.e. spherical aperture cuts based on halo over-density radii, stellar half-mass radii, and fixed apertures; see their figure~B.2) and recovered the same trends we observe: little-to-no evolution of typical ICL mass fractions with lookback time (back to $z\sim2$) except for fixed aperture approaches, which yield fractions that fall with increasing redshift. 
 
For the analysis presented in Figure~\ref{fig:AltIclFracEvoPlot}, we do not control for evolving halo or central-galaxy mass when applying the fixed aperture methodology to progenitor structures at $z>0$. This is the primary cause behind the consistent redshift evolution trends seen for this methodology, which are largely curtailed in the alternative version of this analysis limited to cluster mass objects which we present in Appendix~\ref{Appendix-MassLimitedIclFracEvo}. 
However, it also bears highlighting that for a fixed mass both galaxies and haloes are typically more centrally concentrated and physically smaller at higher redshifts (e.g. \citealt{wechsler_concentrations_2002}, \citealt{Allen_GalSizeAndMassBuildup_2025}, \citealt{McGrath_GalSizeAndColGradWithZ_2026}). 
Consequently, any aperture of fixed physical size would be expected to 
enclose more 
of a typical cluster mass halo (and combined BCG\,+\,ICL system) at $z\gtrsim2$ as compared to $z\approx0$. Not controlling for evolving typical halo or galaxy size could thus also be a secondary driver of the evolution towards lower ICL fractions at higher redshift seen with the fixed aperture methodology (traces of which are still present in the alternative analysis employing a mass-cut shown in Appendix~\ref{Appendix-MassLimitedIclFracEvo}).


\section{Assembly of ICL from cluster satellites}\label{Results2}

Previous theoretical works have occasionally disagreed on whether the main progenitor galaxies of ICL stars should be more or less massive galaxies, with differing predictions sometimes attributed to the different methodologies implemented to separate galactic and intergalactic stars (e.g. \citealt{Tang_ICLfromSats_2023}). 
In this section, we investigate which satellites are the dominant contributors of stars to the $z\approx0$ ICL as predicted by each of the four simulations (using an approach similar to that employed in \citetalias{brown_assembly_2024}) 
-- probing for significant discrepancies between the simulations for the predicted peak progenitor masses despite implementing a uniform scheme for separating galactic and intergalactic stars. For this analysis, we return to our fiducial ICL extraction methodology based on local stellar density (as compared to the cosmic average matter density; Section~\ref{Methods-IclDef}). 

\subsection{The ICL contribution from a cluster's own satellites}\label{Results2-SummarisedOrigChannels}

Our scheme for linking ICL stars to progenitor galaxies (described in Section~\ref{Methods-TrackingGalaxies}) only considers stars liberated on or after cluster infall from galaxies we track falling into our sample clusters after $z\sim3$, and is not recursive for the pre-existing diffuse stellar component of smaller clusters or massive groups accreted during cluster assembly. Though tidal stripping is generally acknowledged to be the dominant assembly channel for ICL stars by $z=0$, this usually requires considering this ``pre-processing'' of ICL stars in accreted groups to be an indirect channel (as in e.g. \citealt{contini_moreimptime_2024}) -- with prior studies often finding the contribution of this pre-processing channel to be substantial when it is considered separately. 

\begin{figure}
    \includegraphics[width=\columnwidth]{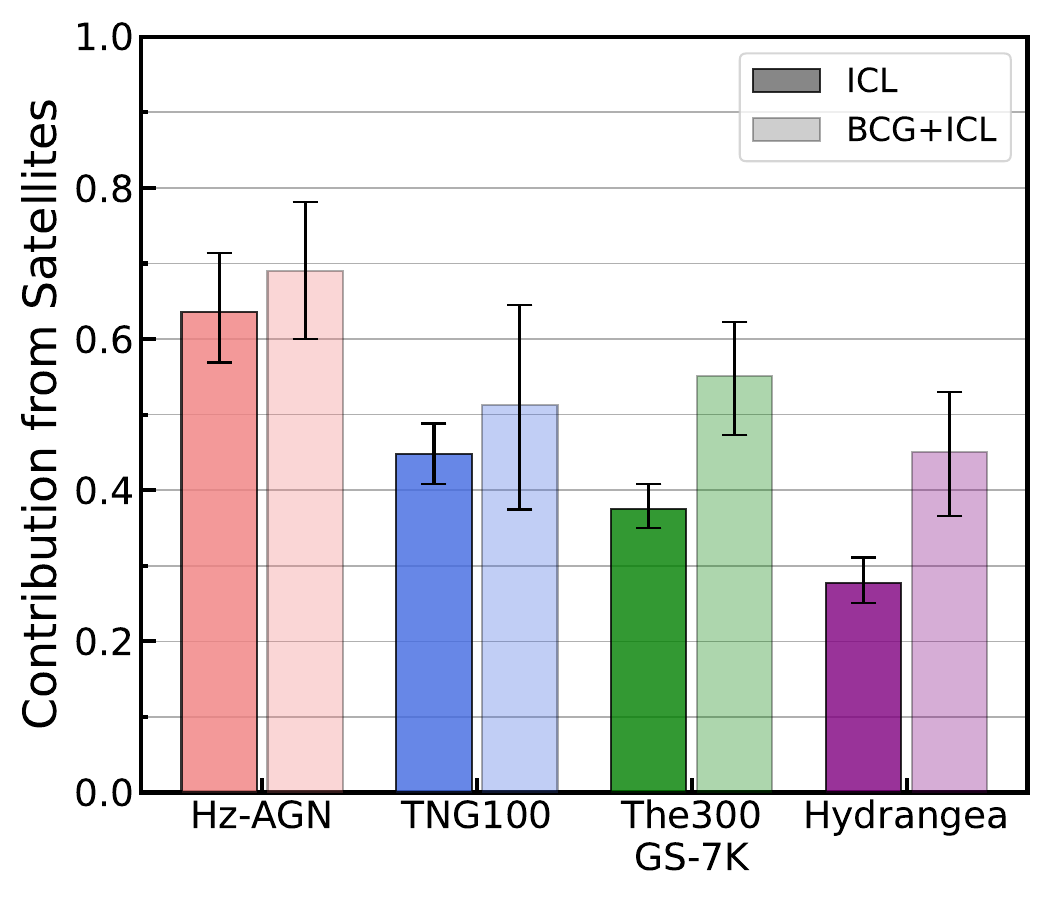}
    \caption{
    The mean fractions of $z\approx0$ ICL and BCG\,+\,ICL stars liberated within the cluster halo from a cluster's own satellite galaxies (that joined the cluster after $z\sim3$) for the galaxy cluster samples drawn from \textsc{Horizon-AGN} (red), \textsc{TNG100} (blue), \textsc{The Three Hundred Gizmo-Simba 7K} (green), and \textsc{Hydrangea} (purple). The error bars indicate the 16th-84th percentile range of each cluster sample. 
    }
    \label{fig:MeanFromSatsBarPlot}
\end{figure}

In Figure~\ref{fig:MeanFromSatsBarPlot} we present for each of the four simulated cluster samples the mean fraction of $z=0$ ICL (and BCG\,+\,ICL) stellar mass assembled by the liberation of stars within a cluster's own halo from satellite galaxies that join the cluster after $z\sim3$.  
These mean ICL mass fractions span the range $\sim0.3-0.6$ ($\sim0.4-0.7$ for BCG\,+\,ICL), always falling significantly short of unity, in qualitative agreement with the findings of prior studies: \citet{contini_moreimptime_2024} reported pre-processing to typically contribute between 20 and 40~per cent of the $z=0$ ICL stellar mass in $10^{14}-10^{15}$\,M\textsubscript{\sun} halo mass clusters (based on semi-analytic models); and \citet{jeon_origin_2025} reported only $\sim55$~per cent of the combined BCG\,+\,ICL stellar mass in a $z=0.79$ cluster (from the \textsc{NewCluster} simulation; $z=0.79$ halo mass $\sim1\times10^{14}$\,M\textsubscript{\sun}) to originate from satellite galaxies, with ICL pre-processing already contributing $>10$\,per cent of the combined BCG\,+\,ICL mass even at this early redshift. 

Though we defer a more comprehensive analysis of ICL assembly via channels other than the liberation of satellite galaxy stars to the forthcoming companion paper (Brown et al. in prep.), we confirm that in all four considered simulations most of the $z\approx0$ ICL mass not liberated within the cluster halo from a cluster's own satellite galaxies 
can be attributed to ICL pre-processing -- with stars formed within the central galaxy since $z\sim3$ also typically making a substantial contribution to the combined BCG\,+\,ICL (see Section~\ref{Discussion-SimDiffs-BcgIclMassFracs}). We defer discussion of the inter-simulation differences in Figure~\ref{fig:MeanFromSatsBarPlot} to Section~\ref{Discussion-SimDiffs-FromInfallersFrac}.
Though for the remainder of our analysis we consider only the sub-set of ICL stars 
liberated within the cluster halo from a cluster's own satellite galaxies,
we highlight that under hierarchical assembly it is broadly expected that this pre-existing diffuse stellar component of massive groups should be assembled in a self-similar fashion to the ICL of clusters (see Section~\ref{Discussion-Caveats} for further discussion). 

\subsection{Progenitor galaxies of ICL (and BCG\,+\,ICL) stars}\label{Results2-IclContribution}

\subsubsection{Liberated fraction of satellite stellar mass}\label{Results2-IclContribution-LibFrac}

\begin{figure}
    \includegraphics[width=\columnwidth]{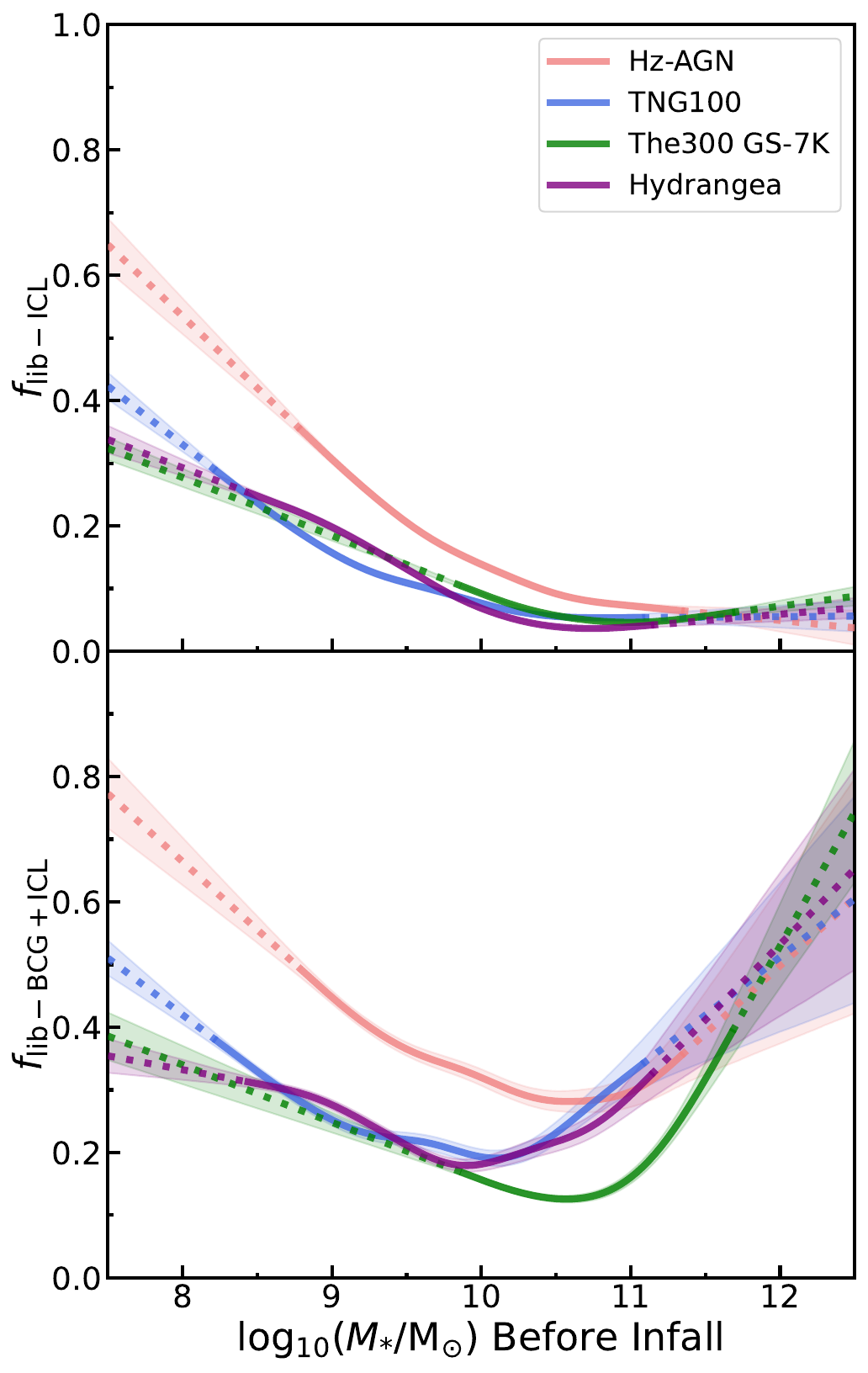}
    \caption{
    \textbf{Top Panel:} Cubic spline functions fitted to the mean fraction of associated stars liberated to the ICL ($f_\textrm{lib-ICL}$) by $z\approx0$ as a function of progenitor galaxy infall stellar mass, $M_{*}$, for 
    satellite galaxies joining \textsc{Horizon-AGN} (red), \textsc{TNG100}  (blue), \textsc{The Three Hundred Gizmo-Simba 7K} (green), and  \textsc{Hydrangea} (purple) clusters after $z\sim3$. 
    Linear extrapolation beyond the range of stellar masses used for fitting
    shown with dotted lines. The shaded regions indicate estimated uncertainties based on bootstrapping. See Appendix~\ref{Appendix-IndivSimFits} for data each fit based on. \textbf{Bottom Panel:} Same as top panel but for the combined BCG\,+\,ICL system.
    }
    \label{fig:LibFracPlot}
\end{figure}

To begin our investigation into which satellites are the largest contributors of ICL stars overall, 
we first probe how the typical ICL contribution of an individual satellite galaxy varies as a function of cluster infall stellar mass. 
We refer to the fraction of the stellar particles associated with a satellite -- meaning those stellar particles composing the galaxy just proceeding first cluster infall 
or that were born in the descendant of that galaxy thereafter (see Section~\ref{Methods-TrackingGalaxies-ActualTracking}) 
-- that become part of the $z\approx0$ ICL component as the \textit{liberated fraction}, $f_\text{lib-ICL}$, of that satellite. 

For each of the four simulations individually we calculate the mean value of $f_\text{lib-ICL}$ for 0.5\,dex wide rolling bins (bin step 0.15\,dex) in infall stellar mass, $M_{*}$, for satellite galaxies that enter any of the clusters for the first time at least $1\,$Gyr before $z\approx0$.
We then fit a cubic spline function to these mean $f_\text{lib-ICL}$ values as a function of infall stellar mass, 
ignoring both bins containing poorly-resolved galaxies (infall masses corresponding to fewer than 100 stellar particles)
and also poorly sampled bins (fewer than 5 objects per cluster). We linearly extrapolate this fit beyond the range of masses used for fitting (with the output constrained to $[0,1]$). 

We plot the resulting fitted curves for each of the four simulations in the top panel of Figure~\ref{fig:LibFracPlot}. The bottom panel of Figure~\ref{fig:LibFracPlot} shows the equivalent analysis for the combined BCG\,+\,ICL. 
The shaded regions indicate the dispersion (16th-84th percentile) of the fit lines generated in the same fashion for $10^{4}$ bootstrap resamples of the aggregate infalling satellite population of each simulation.  
The data each fit is based on are presented in Appendix~\ref{Appendix-IndivSimFits}. 
Though we include stars formed post-infall when calculating $f_{\text{lib-ICL}}$, we confirm that this has no qualitative impact on the presented findings. The fitted functions shown in Figure~\ref{fig:LibFracPlot} are not meaningfully different if reproduced as a function of total stellar mass associated to an infaller by $z=0$ rather than infall mass. 
Liberated satellite stars found at $r>r_{178c}$ at $z\approx0$ are not considered $z\approx0$ ICL stars when we calculate $f_{\text{lib-ICL}}$; we also exclude satellites just crossing $r_{178c}$ for the first time at $z\approx0$ as well as those galaxies which only ``skim'' clusters, entering then exiting $r_{178c}$ in consecutive coarsely spaced snapshots and never returning before $z\approx0$. We confirm that none of these exclusions significantly influence our findings. 

Qualitatively the same trend between expected $f_\text{lib-ICL}$ value and infall mass is reproduced by all four simulations: comparatively higher average $f_\text{lib-ICL}$ fractions at lower infall masses, which decrease with increasing infall mass until $M_{*}\sim10^{10}\,$M\textsubscript{\sun}, at which point they level out around $f_\text{lib-ICL}\sim0.05-0.1$. An initially similar general trend is also seen for the equivalent analyses with the combined BCG\,+\,ICL, differing chiefly through $f_\text{lib-BCG+ICL}$ beginning to rise again at high masses ($M_{*}\gtrsim10^{10.5}\,$M\textsubscript{\sun}; from a rapidly rising typical BCG contribution) rather than level out. Essentially the same qualitative trends were previously reported in \citetalias{brown_assembly_2024}, and we discuss the physical origins of these trends in Section~\ref{Discussion-flibTrends}. 
We defer discussion of the inter-simulation differences to Section~\ref{Discussion-SimDiffs-flib}. 

\begin{figure}
    \includegraphics[width=\columnwidth]{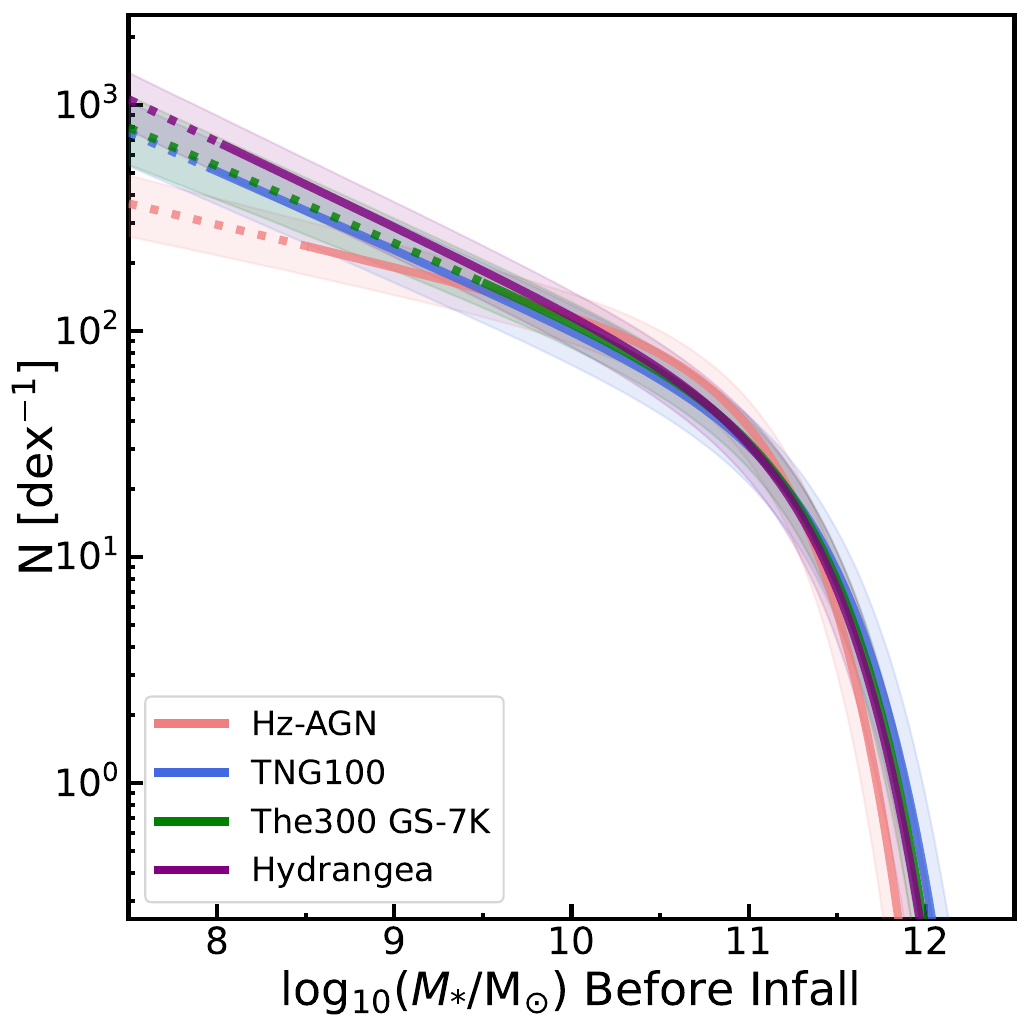}
    \caption{
    \citet{schechter_analytic_1976} function (Equation~\ref{eq:Schechter}) fits to the 
    infalling satellite stellar mass function of the cluster samples drawn from \textsc{Horizon-AGN} (red), \textsc{TNG100} (blue), \textsc{The Three Hundred Gizmo-Simba 7K} (green), and \textsc{Hydrangea} (purple) -- all renormalised to an arbitrary total accreted stellar mass of $5\times10^{12}\,$M\textsubscript{\sun}.
    Dotted lines indicate extrapolation beyond the range of masses used for fitting.
    The shaded regions indicate estimated uncertainties based on both bootstrapping and fit uncertainties. See Table~\ref{tab:Sch_Fit_Params} for fit parameters and Appendix~\ref{Appendix-IndivSimFits} for data fits based on.
    }
    \label{fig:MassFunctionPlot}
\end{figure}

\subsubsection{Infalling satellite mass function}\label{Results2-IclContribution-MassFunc}

\begin{table}
    \centering
    \caption{
    Functional parameters for the fitted infalling satellite stellar mass functions shown in Figure~\ref{fig:MassFunctionPlot}, which adopt the form of \citet{schechter_analytic_1976} functions (Equation~\ref{eq:Schechter}). 
    Parameter uncertainty estimates based on both bootstrapping and individual fit uncertainties are also given. 
    }
    \label{tab:Sch_Fit_Params}
    \begin{tabular}{|l|c|c|}
        \hline
        Simulation & $\alpha$ & $\textrm{log}_{10}(M_{k}/\textrm{M}_{\sun})$\\
        \hline
        \textsc{Horizon-AGN} & $-1.19\pm0.03$ & $11.11\pm0.08$ \\
        \textsc{TNG100} & $-1.35\pm0.02$ & $11.39\pm0.09$ \\
        \textsc{The300 GS-7K} & $-1.34\pm0.04$ & $11.33\pm0.07$ \\
        \textsc{Hydrangea} & $-1.38\pm0.02$ & $11.31\pm0.07$ \\
        \hline
    \end{tabular}
\end{table}

In addition to understanding how the typical ICL mass contribution per galaxy varies with satellite infall mass, determining which population of satellites 
contribute the bulk of ICL stars also requires quantifying the relative population sizes of differingly massive satellites falling into assembling clusters. 
We present in Figure~\ref{fig:MassFunctionPlot} infalling satellite mass function fits 
for each simulation
for galaxies that enter any of the clusters for the first time at least $1\,$Gyr before $z\approx0$. 
These fits adopt the form of the \citet{schechter_analytic_1976} function, 
\begin{equation}
    \Phi(M_{*})\cdot dM_{*} = \Phi_{n} \left( \frac{M_{*}}{M_{k}} \right)^{\alpha} \exp\left({-M_{*} / M_{k}} \right) \cdot dM_{*}
    \label{eq:Schechter}
\end{equation}
where $\alpha$ is the low-mass-end slope, $\Phi_{n}$ a normalization parameter, and $M_{k}$ a characteristic mass (corresponding to the ``knee'' of the function i.e. the mass when the function exhibits a rapid change in slope). 
The data each fit is based on are presented in Appendix~\ref{Appendix-IndivSimFits}, and fitted parameter values are given in Table~\ref{tab:Sch_Fit_Params}.
To compensate for resolution effects and enable predictions for the ICL contribution made by low-mass satellites, only stellar masses corresponding to well-resolved galaxies ($\geq100$ stellar particles) are considered for fitting and the fitted functions then extrapolated to lower masses. 
To facilitate comparison between the low-mass slopes and  ``knee'' positions of each fit, the mass functions depicted in Figure~\ref{fig:MassFunctionPlot} have all been arbitrarily renormalised to each correspond to a total accreted stellar mass of $5\times10^{12}\,$M\textsubscript{\sun}. 

We assume both parameter fit uncertainty and the dispersion in fitted parameters values across $10^{4}$ bootstrap resamples of the satellite population to be independent sources of Gaussian error, and combine these in quadrature to obtain the estimated errors on fit parameter value given in Table~\ref{tab:Sch_Fit_Params}. 
The shaded regions in Figure~\ref{fig:MassFunctionPlot} then show 16th-84th percentile confidence intervals on the fitted functions for this estimated uncertainty. 
Within estimated uncertainty the infalling satellite mass function fits for the different simulations appear in reasonable agreement (see Section~\ref{Discussion=SimDiffs-Schfit} for a brief discussion of the inter-simulation differences). 

\subsubsection{Star formation after cluster infall}\label{Results2-IclContribution-PostInfallFactor}

A complete accounting of the ICL contribution by satellite galaxies requires acknowledging the population of stars formed in some satellites after cluster infall (rather than solely the initial infall mass of these galaxies).
In \citetalias{brown_assembly_2024}, this post-infall star formation was accounted for with a constant factor equal to the mean ratio between total associated stellar mass by $z\approx0$ 
and infall mass across the entire population of tracked satellite galaxies. 
Here we instead calculate the mean ratio between total associated stellar mass by $z\approx0$ and infall stellar mass for the same rolling bins used in the analysis depicted in Figure~\ref{fig:LibFracPlot}. We fit a cubic spline function to these mean mass ratios in the same manner as our fitted $f_{\mathrm{lib}}(M_{*})$ functions, including linear extrapolation beyond the range of masses used for fitting (though with the output constrained to $[1,\infty]$ rather than $[0,1]$ and no longer disregarding low-mass satellites when fitting). 
The resulting fitted functions can be found in Appendix~\ref{Appendix-PostInfallSF} but all adopt the same essential shape, with a peak at intermediate infall masses ($10^{9.5}-10^{10}\,$M\textsubscript{\sun}), falling back to unity at lower and higher masses. 
We confirm none of our remaining findings are significantly altered by substituting this approach for that used in \citetalias{brown_assembly_2024}. 

\subsubsection{Total ICL contributions from galaxies of different mass}\label{Results2-IclContribution-ContriPlots}

\begin{figure}
    \includegraphics[width=\columnwidth]{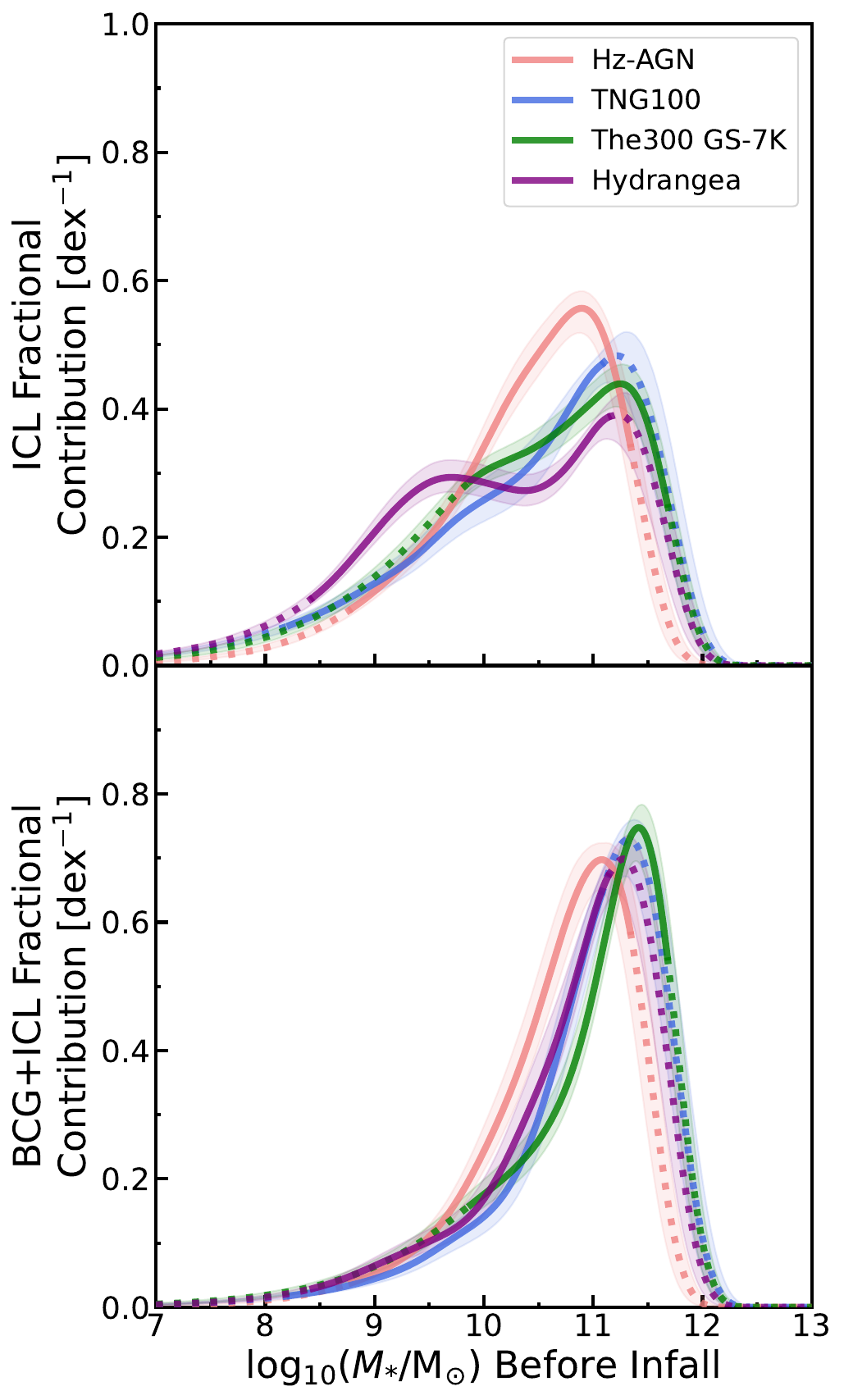}
    \caption{
    \textbf{Top Panel:} Fractional ICL mass contribution as a function of satellite galaxy infall stellar mass (per dex in infall mass; relative to the total expected ICL contribution from satellite galaxies) for the simulated cluster samples drawn from \textsc{Horizon-AGN} (red), \textsc{TNG100} (blue), \textsc{The Three Hundred Gizmo-Simba 7K} (green), and \textsc{Hydrangea} (purple). 
    Dotted lines indicate extrapolation beyond the range of masses used for fitting. The shaded regions indicate estimated uncertainties based on bootstrapping. \textbf{Bottom Panel:} Same as top panel but for the combined BCG\,+\,ICL system. 
    }
    \label{fig:IclContriPlot}
\end{figure}

\begin{figure}
    \includegraphics[width=\columnwidth]{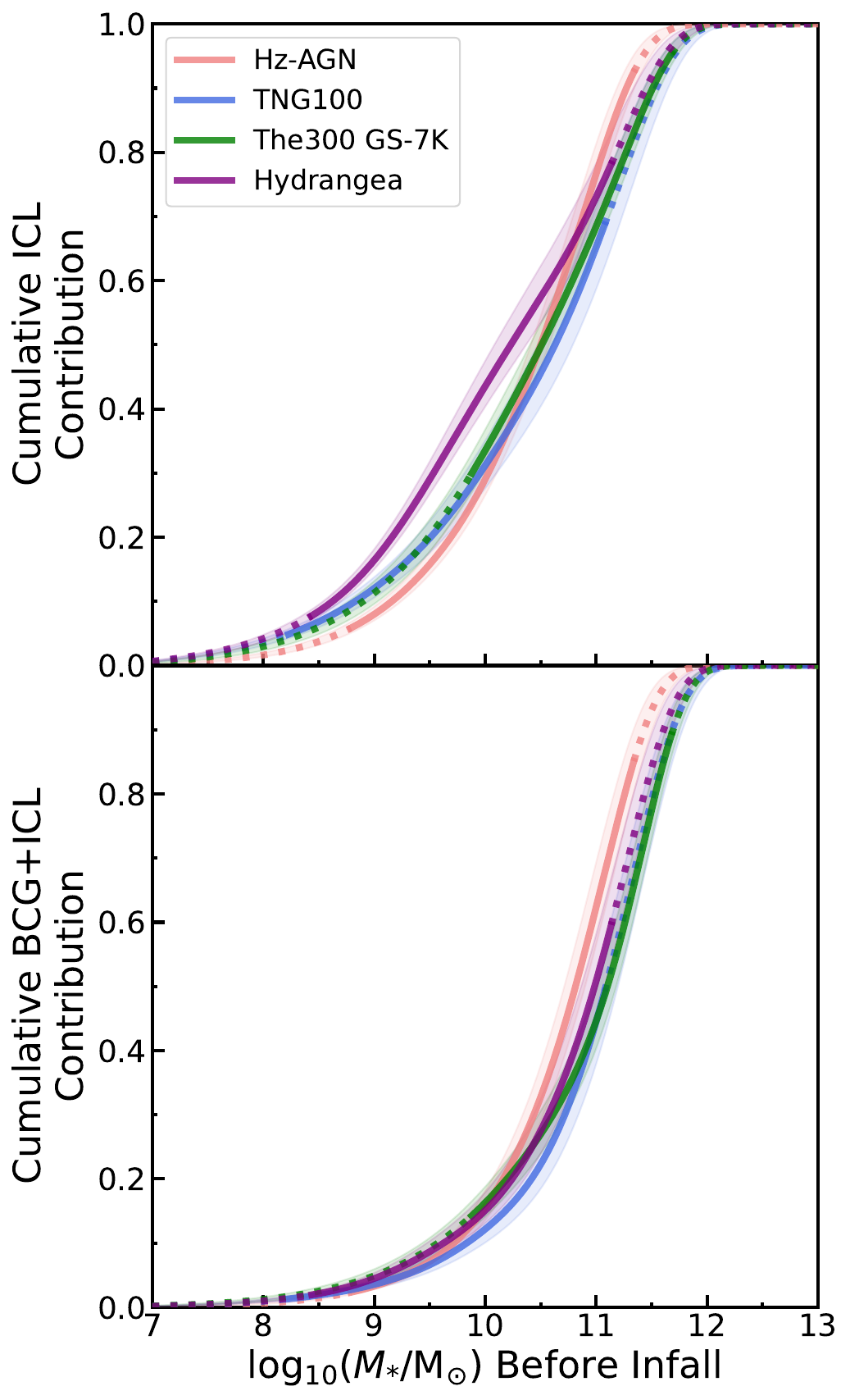}
    \caption{
    \textbf{Top Panel:} Cumulative contribution of satellite galaxies to the ICL as a function of infall stellar mass (normalized to the total predicted ICL mass contribution from satellites), 
    for the simulated cluster samples drawn from \textsc{Horizon-AGN} (red), \textsc{TNG100} (blue), \textsc{The Three Hundred Gizmo-Simba 7K} (green), and \textsc{Hydrangea} (purple). 
    Dotted lines indicate extrapolation beyond the range of masses used for fitting. The shaded regions indicate estimated uncertainties based on bootstrapping. 
    See Appendix~\ref{Appendix-FromSimFracCs} for comparisons with raw simulation data. \textbf{Bottom Panel:} Same as top panel but for the combined BCG\,+\,ICL system. 
    }
    \label{fig:IclContriCsPlot}
\end{figure}

\begin{table}
    \centering
    \caption{
    Satellite infall stellar masses corresponding to the predicted peak contributors of ICL (or BCG\,+\,ICL) stars in a $z\approx0$ cluster as per Figure~\ref{fig:IclContriPlot}. Uncertainty estimates based on repeat analyses with bootstrap resamples of the satellite populations are also given. 
    }
    \label{tab:ICL_and_BCGICL_peaks}
    \begin{tabular}{|l|c|c|}
        \hline
        Simulation & 
        $\textrm{log}_{10}(M^\text{Max ICL}_{*}/\text{M}_{\sun})$ & 
        $\textrm{log}_{10}(M^\text{Max BCG+ICL}_{*}/\text{M}_{\sun})$ \\
        \hline
        \textsc{Horizon-AGN} & $10.90^{+0.08}_{-0.12}$ & $11.08^{+0.11}_{-0.14}$ \\[1mm]
        \textsc{TNG100} & $11.22^{+0.10}_{-0.17}$ & $11.33^{+0.09}_{-0.10}$ \\[1mm]
        \textsc{The300 GS-7K} & $11.25^{+0.08}_{-0.12}$ & $11.42^{+0.05}_{-0.07}$ \\[1mm]
        \textsc{Hydrangea} & $11.23^{+0.07}_{-0.14}$ & $11.28^{+0.06}_{-0.12}$ \\
        \hline
    \end{tabular}
\end{table}

The product of our fitted functions for liberated fraction of stellar mass ($f_{\text{lib-ICL}}(M_{*})$; Figure~\ref{fig:LibFracPlot}) and total associated stellar mass by $z\approx0$ ($M^{z=0}_{*,\textrm{Tot}}(M_{*})$; Appendix~\ref{Appendix-PostInfallSF}) yield a prediction for the typical ICL mass contribution per satellite for a given infall stellar mass. Combining this with our infalling satellite mass function fit ($\Phi(M_{*})$; Figure~\ref{fig:MassFunctionPlot}) then allows us to make a prediction for the typical mass fraction of satellite galaxy sourced ICL stars at $z\approx0$ contributed by satellites with infall mass $M_{*}$, i.e. 
\begin{equation}
    f_\text{ICL}(M_{*})=\frac{f_{\mathrm{lib}}(M_{*})M^{z=0}_\text{*,\textrm{Tot}}(M_{*})\Phi(M_{*}) }{ \int f_{\mathrm{lib}}(M_{*}) M^{z=0}_\text{*,\textrm{Tot}}(M_{*})\Phi(M_{*}) \cdot dM_{*} }.
    \label{eq:final_fit}
\end{equation}
The results for each of the four simulations are shown in  Figure~\ref{fig:IclContriPlot} (top panel; equivalent analysis for the combined BCG\,+\,ICL shown in the bottom panel), 
normalized to per dex in infall mass, with dotted lines indicating any reliance on extrapolation beyond the mass ranges used to fit the constituent functions. The shaded regions indicate estimated uncertainty (16th-84th percentile confidence interval) from propagating the uncertainties of the component fitted functions.
For convenience, we state the specific infall stellar masses corresponding to the peak contributors of ICL (and BCG\,+\,ICL) stars as per Figure~\ref{fig:IclContriPlot} in Table~\ref{tab:ICL_and_BCGICL_peaks} (with uncertainties estimated from the 16th-84th percentile scatter of the peak masses from the $10^{4}$ bootstrap resample repeat analyses). 
We re-emphasise that the progenitor analysis presented in Figure~\ref{fig:IclContriPlot} considers only stars liberated within the cluster halo from a cluster's own satellite galaxies -- and does not include the ``pre-processed'' diffuse stellar component of smaller clusters and groups accreted during cluster assembly (Section~\ref{Results2-SummarisedOrigChannels}; see also Section~\ref{Discussion-Caveats} for relevant discussion).

Though there are some noticeable differences between the four predicted $f_\text{ICL}(M_{*})$ curves (which we discuss in Section~\ref{Discussion-SimDiffs-MainProgen}),
it can be seen in Figure~\ref{fig:IclContriPlot} (and Table~\ref{tab:ICL_and_BCGICL_peaks}) that all four simulations predict the dominant progenitors of ICL stars 
to be massive galaxies, with the peak contributors being
roughly Milky Way mass galaxies ($M_{*}\sim10^{11}\,$M\textsubscript{\sun}), in good agreement with the findings of several prior theoretical studies (e.g. \citealt{contini_formation_2014, contini_theoretical_2019}, \citealt{chun_formation_2023, chun_formation_2024}, \citealt{ahvazi_progenitors_2024}, \citealt{contreras-santos_origin_2025}, \citealt{mayes_coevolution_2025}). 
The peak contributors of BCG\,+\,ICL stars are then predicted by every simulation to be very slightly more massive (by $\sim0.05-0.2$\,dex) than for the ICL alone -- consistent with the findings of previous works (e.g. \citealt{bilata-woldeyes_tracing_2025}, \citealt{mayes_coevolution_2025}). 

We present the same analysis reframed as the predicted cumulative fractional contribution of satellite galaxy sourced ICL stars against satellite infall mass in Figure~\ref{fig:IclContriCsPlot}. 
A comparison between these fitted cumulative fractional ICL contribution curves and the same obtained directly from the simulations for the aggregate ICL of each cluster sample 
are presented in Appendix~\ref{Appendix-FromSimFracCs}. 

From Figure~\ref{fig:IclContriCsPlot} it can be seen that all four simulations predict $\gtrsim50$~per cent of the satellite sourced ICL mass of an average $z\approx0$ cluster to be contributed by $M_{*}\gtrsim10^{10}\,$M\textsubscript{\sun} satellites ($M_{*}\gtrsim10^{10.5}\,$M\textsubscript{\sun} for every simulation besides \textsc{Hydrangea}),  with $\sim80$~per cent from $M_{*}\gtrsim10^{9}\,$M\textsubscript{\sun} satellites ($M_{*}\gtrsim10^{9.5}\,$M\textsubscript{\sun} for every simulation besides \textsc{Hydrangea}). 
The corresponding minimum satellite infall mass thresholds for the combined BCG\,+\,ICL are $M_{*}\gtrsim10^{10.5}\,$M\textsubscript{\sun} and $M_{*}\gtrsim10^{10}\,$M\textsubscript{\sun} for $\sim50$~per cent and $\sim80$~per cent (of the satellite sourced BCG\,+\,ICL stellar mass), respectively. 

In a prior study employing \textsc{TNG100}, \citet{mayes_coevolution_2025} reported the mean ICL mass fraction attributable to progenitor objects with stellar masses $\gtrsim10^{10}\,$M\textsubscript{\sun} to be $\sim60$ per cent. This is broadly in line with our findings, with our slightly higher equivalent prediction of $\sim70$~per cent using \textsc{TNG100} thought chiefly due to \citet{mayes_coevolution_2025} also including in their sample galaxy groups (which should be assembled from on average slightly less massive progenitors than clusters).
The minimum progenitor stellar mass required to recover at least $50$~per cent of the BCG\,+\,ICL in \textsc{TNG100} clusters 
was previously reported by \citet{pillepich_first_2018} to be $\sim10^{10.5}\,$M\textsubscript{\sun} (requiring $\gtrsim10^{9.5}\,$M\textsubscript{\sun} progenitors for $\geq90$~per cent; see their figure 13), also in reasonably good agreement with our findings. 

Using the lower-resolution predecessor simulations of \textsc{The300 GS-7K}, \citet{contreras-santos_origin_2025} reported $M_{*}\geq10^{11}\,$M\textsubscript{\sun} progenitors to contribute $65-80$~per cent of the ICL in the $14.8 < \text{log}_{10}(M_{200}/\text{M}_{\sun}) < 15.6$ clusters they considered, in qualitative agreement with our result that massive galaxies appear the dominant contributors of ICL stars, though we do not find 
quite so large a contribution from these very massive galaxies
(instead predicting a typical contribution $\sim30-40$~per cent). 
This discrepancy is likely due in part to \citet{contreras-santos_origin_2025} considering typically more massive clusters, 
and that they only consider the ICL contribution from objects explicitly resolved in their (lower-resolution) simulations, whereas we extrapolate to predict the contribution from unresolved low mass galaxies as well. Additionally, the definitions they use include the pre-existing diffuse stellar component of groups that merge into their clusters as part of the ICL contribution by the former central galaxies of those groups (see Section~\ref{Discussion-Caveats} for further relevant discussion). 


\section{Discussion}\label{Discussion}

\subsection{Physical origins of liberated fraction trends}\label{Discussion-flibTrends}

All four simulations reproduce the same general trends in Figure~\ref{fig:LibFracPlot}: decreasing $f_\text{lib-ICL}$ values with increasing satellite infall mass, and $f_\text{lib-BCG+ICL}$ values that initially fall with increasing satellite mass before reversing and beginning to rapidly rise at $M_{*}\gtrsim10^{10.5}$\,M\textsubscript{\sun}. These are qualitatively the same trends noted and discussed in \citetalias{brown_assembly_2024}, and we consider these to emerge chiefly from the convolution of two phenomena:
more massive satellites generally experience more rapid orbital decay in clusters due to dynamical friction (e.g. \citealt{van_den_bosch_substructure_1999}, \citealt{boylan-kolchin_dynamical_2008}, \citealt{amorisco_contributions_2017}); and more massive galaxies also have deeper potential wells and proportionally larger DM haloes (once past the peak in the stellar-to-halo mass relation), 
so are less readily stripped of stars by gravitational interactions (e.g. \citealt{read_tidal_2006}, \citealt{smith_preferential_2016}, \citealt{martin_stellar_2024}).

The orbits of less massive satellites typically taking longer to decay provides ample opportunity for a significant fraction of their more readily stripped stellar mass to be gradually siphoned into the ICL as they spiral slowly inwards towards the cluster centre, 
yielding higher average $f_\text{lib-ICL}$ values but in general not meaningfully contributing to the BCG. By contrast, the very massive former-central galaxies accreted when a cluster subsumes a smaller cluster or massive group will not only be highly resistant to stripping, but also typically experience only a brief window for this stripping to occur between cluster infall and the inevitable merger with the cluster central galaxy relatively soon thereafter, yielding low average $f_\text{lib-ICL}$ values but high average $f_\text{lib-BCG+ICL}$ values at the highest satellite masses (see Pearce et al. in prep. for further relevant discussion).
We then consider the
levelling out of $f_\text{lib-ICL}$ at high-masses indicative of violent relaxation processes during BCG mergers becoming an increasingly significant mechanism for liberating stars to the ICL for especially massive galaxies (\citetalias{brown_assembly_2024}).  

Despite the scheme described above, it can be seen (by comparison between the two panels of Figure~\ref{fig:LibFracPlot}) that the average BCG stellar mass contribution is generally non-zero even for the lowest infall masses. 
This is driven chiefly by low mass satellites which join to-be clusters very early (first infall at $z\gtrsim2$) which, having been afforded $\gtrsim10$\,Gy of orbital decay after 
joining 
the much smaller, early progenitor of a $z\approx0$ cluster 
(and having joined with a more significant infall mass ratio compared to a similarly massive galaxy at lower redshifts; Pearce et al. in prep.) are able to eventually merge into the BCG, resulting in the underlying distribution being mildly bimodal. 
We note that earlier infall times generally correspond to higher average $f_\text{lib-ICL}$ and $f_\text{lib-BCG+ICL}$ values in all four simulations for all infall masses, but qualitatively the same trends seen in Figure~\ref{fig:LibFracPlot} persist even when considering only satellites with a specific infall time. 

\subsection{Inter-simulation differences}\label{Discussion-SimDiffs}

\subsubsection{ICL mass fractions}\label{Discussion-SimDiffs-IclMassFracs}

A slight dispersion between the typical ICL stellar mass fractions of the four simulations at $z\approx0$ 
can be seen in Figure~\ref{fig:StelFracVplot}, with the typical values for \textsc{Hz-AGN} and \textsc{TNG100} (median 0.13 and 0.12 respectively) falling slightly below those for \textsc{The300 GS-7K} and \textsc{Hydrangea} (median 0.18 and 0.17 respectively). It is initially curious that these typical fractions appear stratified by simulation approach (uniform-volume vs zoom-in) but, while it is conceivable that such a stratification could stem from some bias associated with simulation approach 
(e.g. the \textsc{Hz-AGN} and \textsc{TNG100} clusters occupying the very top of the mass hierarchy in those simulations and so being biased towards rapid recent mass growth; \citealt{onions_life_2025}), 
that this stratification is usually not retained under the various alternative BCG-ICL split methodologies trialled in Figure~\ref{fig:AltIclFracEvoPlot} prompts us to consider it somewhat coincidental.

The fiducial ICL definition we implement is based on instantaneous local stellar density, thus dynamically unbound stars transiently located within a region of sufficiently high density are in that instance regarded as galactic and not as ICL stars. 
This is pertinent for the \textsc{Hz-AGN} ICL fractions we determine, as \textsc{Hz-AGN} is known to have central galaxies with larger than expected effective radii (as compared to observations; e.g. \citealt{martin_formation_2019}, \citealt{chabanier_formation_2020}, see also figure 1 in \citetalias{brown_assembly_2024}). 
In combination with the ICL being highly centrally concentrated, these distended central galaxies may be decreasing the ICL mass fractions we find in \textsc{Hz-AGN} somewhat, with a population of stars that might have otherwise contributed to the ICL occluded beneath the extended BCG. 

We consider the comparatively high ICL fractions of the \textsc{The300 GS-7K} sample shown in Figure~\ref{fig:StelFracVplot} to at least partially be a consequence of the lower resolution of this simulation (and also of our adopted galaxy identification methodology).
Reduced simulation resolution artificially enhances the efficiency of tidal stripping (\citealt{martin_stellar_2024}, \citealt{lovell_numerical_2025}; see also \citealt{contini_formation_2014}, \citealt{van_den_bosch_dark_2018}). One might therefore anticipate that were the resolution of \textsc{The300 GS-7K} improved to be on par with the other three simulations, the typical ICL stellar mass fractions yielded may fall slightly. 
It is however also true that 
stellar stripping rates are not yet fully converged even at the higher resolutions of \textsc{Hz-AGN}, \textsc{TNG100}, and \textsc{Hydrangea}, thus it could equally be said that the ICL fractions predicted by these simulations may be slight overestimates as well. 

Additionally, we impose a minimum stellar particle count of 51 when identifying galaxies in \textsc{TNG100}, \textsc{The300 GS-7K}, and \textsc{Hydrangea} (for consistency with the implementation of \textsc{AdaptaHOP} employed in \textsc{Hz-AGN}; Section~\ref{Methods-IclDef}), corresponding to a minimum galaxy stellar mass of approximately  
$1\times10^{8}$\,M\textsubscript{\sun} in \textsc{Hz-AGN}, $5\times10^{7}$\,M\textsubscript{\sun} in \textsc{TNG100} and \textsc{Hydrangea}, and $2\times10^{9}$\,M\textsubscript{\sun} in \textsc{The300 GS-7K}. 
If this constraint is relaxed, and a minimum particle count of unity instead employed, the median ICL fraction of \textsc{The300 GS-7K} under our fiducial ICL methodology falls to $\sim0.16$ -- indicating that the ICL fractions in \textsc{The300 GS-7K} from our main analysis include a small contribution from the stellar mass of neglected, poorly-resolved galaxies (of stellar mass $\lesssim2\times10^9\,$M\textsubscript{\sun}; the median fractions of \textsc{TNG100} and \textsc{Hydrangea} are unaltered by this change). The typical ICL fractions of the \textsc{The300 GS-7K} sample under the various alternative methodologies considered in Figure~\ref{fig:AltIclFracEvoPlot} likewise fall slightly when this constraint is relaxed (usually by $\lesssim0.02$; with \textsc{TNG100} and \textsc{Hydrangea} again broadly unaffected), though none of the redshift evolution trends shown in Figure~\ref{fig:AltIclFracEvoPlot} are qualitatively altered by relaxing this constraint. 

We discuss the consequences of relaxing this minimum stellar particle threshold on the apparent fractional ICL contribution from a cluster's own satellites in Section~\ref{Discussion-SimDiffs-FromInfallersFrac}, but confirm the progenitor analysis presented in Section~\ref{Results2-IclContribution} to otherwise be insensitive to the exact threshold employed. 
The predicted peak ICL (and BCG\,+\,ICL) progenitor masses obtained by repeating the analysis presented in Figure~\ref{fig:IclContriPlot} for \textsc{TNG100}, \textsc{The300 GS-7K}, and \textsc{Hydrangea} with a minimum galaxy stellar particle threshold of unity are functionally identical to those presented in Table~\ref{tab:ICL_and_BCGICL_peaks}.

\subsubsection{BCG (and BCG\,+\,ICL) mass fractions}\label{Discussion-SimDiffs-BcgIclMassFracs}

We note some scatter between the median BCG stellar mass fractions of the four simulations in Figure~\ref{fig:StelFracVplot}, most conspicuously the comparatively low median BCG fraction found for \textsc{The300 GS-7K} (of 0.19, compared to 0.35, 0.43, and 0.29 for \textsc{Hz-AGN}, \textsc{TNG100}, and \textsc{Hydrangea} respectively). Though less prominent, the distribution of BCG mass fractions in \textsc{TNG100} also appears mildly skewed towards larger values -- with the median BCG fraction in \textsc{TNG100} coinciding with the maximum of the \textsc{The300 GS-7K} sample. 
This BCG fraction scatter is largely responsible for the similar scatter also seen in the typical BCG\,+\,ICL mass fractions, as the ICL fractions of the different simulations are broadly consistent.  
We consider this scatter chiefly the joint consequence of differing typical cluster assembly times and differing dependencies on accretion 
for assembling stellar mass between the different cluster samples. 

The \textsc{The300 GS-7K} sample clusters typically have lower DM halo assembly redshifts compared to those from the other samples (with a median $z_{50}$ value of 0.53, compared to 0.62, 0.82, and 0.77 for \textsc{Hz-AGN}, \textsc{TNG100}, and \textsc{Hydrangea} respectively; see Appendix~\ref{Appendix-ClusterProps}). 
A more recent assembly redshift suggests a larger fraction of the total cluster mass to be newly introduced -- brought in with satellite galaxies that have yet had less opportunity for disruption -- and so a larger fraction of the cluster's total stellar mass should be retained in the satellite component. 
As such, a 
correlation between assembly redshift and BCG\,+\,ICL mass fraction 
appears intuitive, and prior studies have noted such trends already (e.g. \citealt{yoo_spatial_2024}, \citealt{kimmig_intra-cluster_2025}; see also Appendix~\ref{Appendix-ClusterProps}).
Furthermore, we highlight that the \textsc{The300 GS-7K} clusters are systematically more massive at $z\approx0$ than those from the other samples 
and prior studies have also reported broad trends towards lower central stellar mass fractions in increasingly massive haloes (e.g. \citealt{pillepich_first_2018}, \citealt{montenegro-taborda_stellar_2025}). 

We also attribute part of the dispersion in typical BCG stellar mass fractions between the simulations to differing amounts of late-time central star formation. We note significantly more ``in-situ'' BCG\,+\,ICL star formation occurring at $z<3$ in \textsc{TNG100} and \textsc{Hydrangea} as compared to \textsc{Hz-AGN} or \textsc{The300 GS-7K}. 
This star formation -- occurring either within the central galaxy or directly into the ICL component -- often contributes $>20$~per cent of all $z\approx0$ BCG\,+\,ICL stellar particles in both \textsc{TNG100} and \textsc{Hydrangea} (mean fractions 0.26 and 0.25 respectively), as opposed to typically $<10$~per cent in both \textsc{Hz-AGN} and \textsc{The300 GS-7K} (mean fractions 0.08 and 0.05 respectively). These in-situ stars are also generally much more centrally concentrated by $z\approx0$ in \textsc{TNG100} as compared to \textsc{Hydrangea}. 
We leave a more thorough investigation of this in-situ star formation to a follow-up paper (though see Section~\ref{Discussion-SimDiffs-FromInfallersFrac}) -- for further relevant discussion see e.g. \citet{pillepich_first_2018}, \citet{contreras-santos_origin_2025}, \citet{montenegro-taborda_stellar_2025}, and references therein.

\subsubsection{ICL contribution from a cluster's own satellites}\label{Discussion-SimDiffs-FromInfallersFrac}

We consider the lower mean fraction of ICL stellar mass liberated 
within a cluster's own halo from satellites joining to-be clusters after $z\sim3$
in \textsc{The300 GS-7K} as compared to \textsc{Hz-AGN} (0.29 vs 0.62; Figure~\ref{fig:MeanFromSatsBarPlot}) to partly stem from the comparatively lower resolution of \textsc{The300 GS-7K}, in combination with the minimum stellar particle count of 51 we impose when identifying galaxies (Section~\ref{Methods-IclDef-GalDef}). 
The stellar particles of poorly-resolved galaxies 
below this threshold are relabelled as ``intergalactic'', which when joining clusters are therefore classified as pre-processed ICL. When this minimum stellar particle count for galaxies is relaxed 
(acknowledging poorly-resolved satellites and so slightly reducing the obtained ICL mass; Section~\ref{Discussion-SimDiffs-IclMassFracs})
the mean fraction of $z=0$ ICL stellar mass liberated from a cluster's own satellite galaxies in \textsc{The300 GS-7K} rises from 0.29 to 0.38 (and from 0.49 to 0.55 for BCG\,+\,ICL stars; the mean ICL and BCG\,+\,ICL fractions from tracked satellites in \textsc{TNG100} and \textsc{Hydrangea} are unaltered by this change). The clusters of the \textsc{The300 GS-7K} sample are also systematically more massive than those of the other samples, 
and this may also play a role in the typically smaller proportional ICL contribution by a cluster's own satellites 
in \textsc{The300 GS-7K},
as the significance of ICL pre-processing is expected to grow with cluster mass (\citealt{contini_moreimptime_2024}). 

Rather than a direct consequence of differing simulation resolution or a discrepancy in mass between the cluster samples, we  primarily attribute the lower mean fractions of $z\approx0$ ICL stellar mass liberated from a cluster's own satellite galaxies in \textsc{TNG100} and \textsc{Hydrangea} (0.44 and 0.27 respectively) to a non-negligible ICL contribution from the apparent ``in-situ'' formation of stars directly into the ICL component in these two simulations. The mean fractions of $z\approx0$ ICL stellar mass that we link to this channel (i.e. $z\approx0$ ICL stellar particles first seen within the cluster at $z<3$ but outside any resolved galaxy)
are $0.14$ and $0.24$ in \textsc{TNG100} and \textsc{Hydrangea} respectively (and $\sim0.01$ in both \textsc{Hz-AGN} and \textsc{The300 GS-7K}) – though we caution that these determined fractions are highly sensitive to the galaxy definition imposed. If this ``in-situ'' ICL component is ignored, then the mean fractions of the remaining $z\approx0$ ICL mass contributed by tracked satellites are $0.51$ and $0.36$ in \textsc{TNG100} and \textsc{Hydrangea} respectively. 
We defer a more comprehensive investigation and inter-simulation comparison of the contribution from this alternate ICL assembly channel to a forthcoming paper (Brown et al. in prep.) --
for further relevant discussion see e.g. \citet{rohr_cooler_2024}, \citet{contreras-santos_origin_2025}, \citet{mayes_coevolution_2025}, and references therein.

\subsubsection{Liberated fraction of stellar mass}\label{Discussion-SimDiffs-flib}

We find broad agreement between the fitted functions for $f_\text{lib-ICL}(M_{*})$ from \textsc{TNG100}, \textsc{The300 GS-7K}, and \textsc{Hydrangea} shown in Figure~\ref{fig:LibFracPlot}. This is despite the different models and methods employed and resolution achieved by each simulation, and also 
that we do not control for any differences 
in typical satellite infall time, infall mass ratio, satellite orbital parameters, or typical cluster assembly time between the different simulations. 

However, we do note that the fitted function for $f_\text{lib-ICL}(M_{*})$ from \textsc{Hz-AGN} diverges noticeably from the other simulations, with generally higher mean $f_\text{lib-ICL}$ values, particularly at low infall masses. 
This is likely to be a product of the high stellar-to-halo mass ratios (\citealt{dubois_dancing_2014}; see also Appendix~\ref{Appendix-ClusterProps}) and low galaxy compactnesses noted in \textsc{Hz-AGN}, with the median (and 16th-84th percentile scatter in) stellar-half mass radii for $M_{*}\in[1,5]\times10^{10}\,$M\textsubscript{\sun} satellites on first cluster infall at $z\approx0.5$ being $5.5^{+1.6}_{-0.9}$\,kpc in \textsc{Hz-AGN}, as compared to $2.6^{+2.1}_{-0.6}$, $3.6^{+0.8}_{-0.7}$, and $2.3^{+1.9}_{-1.1}$\,kpc in \textsc{TNG100}, \textsc{The300 GS-7K}, and \textsc{Hydrangea} respectively.
Reduced galaxy compactness and under-massive DM haloes correspond to shallower potentials and so increased vulnerability to tidal stripping (e.g. \citealt{pfeffer_ultra-compact_2013}, \citealt{contini_constraints_2017}, \citealt{martin_formation_2019}), from which the higher typical $f_\text{lib-ICL}$ values for \textsc{Hz-AGN} satellites logically follow. 

\subsubsection{Infalling satellite mass functions}\label{Discussion=SimDiffs-Schfit}

Within estimated uncertainty the infalling satellite mass function fits (Figure~\ref{fig:MassFunctionPlot}) for the different simulations appear in reasonable agreement, though a slight dearth of low-mass infalling satellites ($M_{*}\lesssim10^{9.5}\,$M\textsubscript{\sun}) is noted for \textsc{Hz-AGN}. We associate this with the weak supernova feedback implementation of \textsc{Hz-AGN} (\citealt{dubois_dancing_2014}, \citealt{kaviraj_horizon-agn_2017}) and so tentatively link this low-mass satellite deficit with the trace excess of $M_{*}\sim10^{10.5}\,$M\textsubscript{\sun} satellites also seen for \textsc{Hz-AGN} in Figure~\ref{fig:MassFunctionPlot}. 
This slight discrepancy at low-masses could also stem from the other three simulations somewhat over-predicting the abundance of low-mass satellites, with minor excesses of low-mass galaxies previously reported in both \textsc{TNG} and \textsc{Hydrangea} at $z\lesssim1$ (\citealt{pillepich_first_2018}, \citealt{ahad_stellar_2021}). 

\subsubsection{Main progenitors of ICL stars}\label{Discussion-SimDiffs-MainProgen}

Despite the broad agreement between the simulations for the predicted peak contributors of ICL stars, two standout discrepancies are apparent in Figures~\ref{fig:IclContriPlot} and~\ref{fig:IclContriCsPlot}: \textsc{Hz-AGN} predicts a slightly lower peak ICL contributor mass, 
and \textsc{Hydrangea} predicts a more significant ICL contribution from $M_{*}\sim10^{9}-10^{9.5}\,$M\textsubscript{\sun} satellites as compared to the other simulations. 

The prediction of less massive peak ICL (and BCG\,+\,ICL) contributors in \textsc{Hz-AGN} 
follows directly from the infalling satellite mass function fit for \textsc{Hz-AGN} having its knee 
at a lower mass than the other simulations (Figure~\ref{fig:MassFunctionPlot}; see also Table~\ref{tab:Sch_Fit_Params}). Although clear trends appeared in the obtained fits for $f_{\text{lib-ICL}}(M_{*})$ (Figure~\ref{fig:LibFracPlot}), the average value for $f_{\text{lib-ICL}}$ never strayed below $\sim0.05$ at any infall mass for any simulation and also generally did not rise above $\sim0.5$ (with the fits obtained for $M^{z=0}_{*,\textrm{Tot}}(M_{*})/M_{*}$ likewise tightly constrained; Appendix~\ref{Appendix-PostInfallSF}). This comparatively modest variation cannot significantly influence the shapes of the curves seen in Figure~\ref{fig:IclContriPlot}, which instead chiefly follow from the competition between 
rising scarcity with increasing stellar mass, and 
more massive galaxies each having more stars to potentially contribute. 
Correspondingly, it can be seen (by comparison between Tables~\ref{tab:Sch_Fit_Params} and~\ref{tab:ICL_and_BCGICL_peaks}) that both the peak ICL and BCG\,+\,ICL contributor infall masses predicted by every simulation are never more than $\sim0.2$~dex from the value of $M_{k}$ for the corresponding infalling satellite mass function fit, with 
the increased efficacy of tidal stripping at lower satellite masses 
serving only to skew these peaks towards slightly lower masses than would have been obtained by just considering the mass function alone. 

The prediction of a more significant ICL contribution from low-mass galaxies ($M_{*}\lesssim10^{10}\,$M\textsubscript{\sun}) in \textsc{Hydrangea} also follows largely from the infalling satellite mass function, which has a particularly steep low-mass slope in \textsc{Hydrangea} ($\alpha\approx-1.38$; Table~\ref{tab:Sch_Fit_Params}). A steeper mass function slope at low-masses will result in low-mass galaxies composing a higher fraction of the total stellar mass budget, yielding a prediction for $f_\text{ICL}(M_{*})$ (equation~\ref{eq:final_fit}) increasingly skewed towards lower-masses, with a heavier low-mass tail \citep{martin_intracluster_2026}\footnote{Specifically see the accompanying interactive plots they provide, available at \url{https://garrethmartin.github.io/interactive-profiles-ICL/index.html\#stripping}}. The fit to $f_\text{lib-ICL}(M_{*})$ for \textsc{Hydrangea} also levels out at a slightly lower value ($\sim0.05$) at high masses compared to the other simulations, yielding a slightly lower ICL contribution from more massive satellites, and so further enhancing the relative ICL contribution of less massive galaxies. 

\subsection{Caveats and limitations}\label{Discussion-Caveats}

The predicted ICL contributions of differingly massive satellites depicted in Figures~\ref{fig:IclContriPlot} and~\ref{fig:IclContriCsPlot} 
average over the intrinsic stochasticity of cluster assembly. 
Although we predict a significant portion of the aggregate ICL of many clusters to be associated with $M_{*}\gtrsim10^{11}\,$M\textsubscript{\sun} progenitor objects, few objects this massive would be anticipated in the assembly history of any individual cluster (Figure~\ref{fig:MassFunctionPlot}),  with each of these massive objects expected to individually make a substantial ICL mass contribution in absolute terms, even if this is only a small fraction of their infall stellar mass (Figure~\ref{fig:LibFracPlot}). 
As such, though we predict the peak contributors of ICL stars to in general be roughly Milky Way mass galaxies, individual clusters may exist in which the dominant ICL progenitor mass deviates towards higher masses, where the content of the ICL will be dominated by the contribution from only a small number of very massive objects. Alternatively, should a cluster's mass function during assembly be skewed towards lower masses (with a steep low-mass slope), similar to that we find for the \textsc{Hydrangea} sample or as was recently suggested for Hydra I by \citet{lohmann_intracluster_2026}, then for that individual cluster the apex ICL contributor mass may be pushed to lower masses.

For the analysis shown in Figures~\ref{fig:IclContriPlot} and~\ref{fig:IclContriCsPlot} we consider only satellite galaxies which fall into the studied clusters. We do not recursively link the pre-assembled diffuse stellar content of groups or smaller clusters that merge into the studied clusters with prior satellites of those groups or clusters, and instead entirely neglect this ``pre-processed'' ICL component for this analysis. As groups should be assembled from on average less massive progenitors than clusters, we anticipate that if such a recursive analysis were performed 
the predicted peak ICL (and BCG\,+\,ICL) contributor mass would fall slightly. 
As these ``pre-processed'' diffuse stars will typically be only weakly bound to the haloes in which they are delivered to clusters (as compared to still-galactic stars), they should be stripped from these halos soon after cluster infall, before the stripping of galactic stars begins (\citealt{pfeffer_ultra-compact_2013}), and so we would expect these ``pre-processed'' ICL stars to be preferentially deposited at large cluster centric radii. That these ``pre-processed'' ICL stars should also originate from typically less massive (and so typically more metal-poor; e.g. \citealt{gallazzi_ages_2005}) galaxies than the more centrally concentrated ICL assembled ``in-place'' within a cluster from its own satellite galaxies may then contribute to observed radial colour trends in the ICL (e.g. \citealt{ellien_euclid_2025}). We directly address the ICL contribution from this ``pre-processed'' channel and interrogate the origins of these radial ICL trends further in the forthcoming companion paper (Brown et al. in prep.). 


\section{Conclusions}\label{Conclusions}

In this study we have explored the ICL assembly of low-mass galaxy clusters ($M_{178c}\sim10^{14}-10^{15}$\,M\textsubscript{\sun} at $z\approx0$) using four markedly different hydrodynamical simulations: \textsc{Hz-AGN}, \textsc{TNG100}, \textsc{The300 GS-7K}, and \textsc{Hydrangea} (using samples of $10-16$ clusters per simulation). We employed a consistent approach for identifying ICL stars, allowing us to probe for significant inter-simulation discrepancies unobstructed by differing ICL extraction methodologies. We have investigated how the typical ICL stellar mass fractions of these clusters compare at $z\approx0$, how these have evolved since $z\sim3$, as well as how this evolution is influenced by the adopted methodology for separating the ICL and the BCG, and have also tracked the stars of satellite galaxies joining clusters during this same period in order to quantify the different typical ICL contributions made by galaxies with differing stellar masses at cluster infall. 
Our main findings can be summarised as follows:
\begin{enumerate}
    \vspace{-0.05cm}\item \noindent \textit{The $z\approx0$ ICL stellar mass fractions of all four simulations are broadly consistent} (Figure~\ref{fig:StelFracVplot}). When uniformly subject to our fiducial ICL definition (based on local stellar density), the ICL fractions of almost every considered cluster are confined to $[0.10,0.19]$ -- with median ICL fractions of 0.13, 0.12, 0.18, and 0.17 in \textsc{Hz-AGN}, \textsc{TNG100}, \textsc{The300 GS-7K}, and \textsc{Hydrangea}, respectively.
    The BCG\,+\,ICL stellar mass fractions of each simulation also all span the same range (between $\sim0.3$ and $\sim0.7$), with inter-simulation differences in the average BCG\,+\,ICL fraction following directly from differing typical BCG mass fractions, thought to primarily stem from differing typical cluster assembly times and masses between the four samples.\\

    \vspace{-0.05cm}\item \noindent \textit{Whether the average ICL stellar mass fraction remains constant or falls with increasing redshift is determined by the adopted ICL definition, but all four simulations reproduce the same redshift evolution trend when subject to a consistent definition} (Figure~\ref{fig:AltIclFracEvoPlot}). When considering the main progenitors of $z\approx0$ clusters, 
    our fiducial ICL definition based on local stellar density relative to the cosmic average matter density, BCG-ICL splits based on spherical aperture cuts that scale with 
    DM halo overdensity radius or BCG\,+\,ICL stellar half-mass radius, as well as kinematic BCG-ICL separations all yield no significant redshift evolution in the average ICL stellar mass fraction (at least for $z\lesssim2$). 
    Identifying ICL stars using fixed stellar density thresholds or BCG-ICL splits based on spherical apertures of fixed radius (without controlling for evolving central galaxy mass or size) instead yield typical ICL fractions that fall to approximately zero by $z\sim3$.\\

    \vspace{-0.05cm}\item \noindent \textit{All four simulations agree that a galaxy with less stellar mass on cluster infall can be expected to contribute a greater fraction of its stars to the $z\approx0$ ICL than a more massive infalling galaxy} (Figure~\ref{fig:LibFracPlot}), when considering both stars liberated by tidal stripping and those ejected during galaxy mergers (and when marginalising over cluster assembly time, satellite orbital parameters, satellite infall time, and satellite infall mass-ratio). However, 
    even the most massive galaxies that join clusters are still predicted to typically eventually contribute $\gtrsim5$~per cent of their associated stellar mass to the ICL component.\\

    \vspace{-0.05cm}\item \noindent \textit{All four simulations predict the peak contributors of ICL stars in $M_{178c}\lesssim10^{15}$\,\textup{M}\textsubscript{\sun} clusters at $z\approx0$ to be roughly Milky Way mass galaxies} (infall stellar mass $\sim10^{11}$\,M\textsubscript{\sun}; Figure~\ref{fig:IclContriPlot}), with the peak contributors of BCG\,+\,ICL stars only slightly more massive (by $\sim0.1$\,dex) than for the ICL alone. These peak contributor masses are primarily controlled by the infalling satellite mass function during cluster assembly, and are never more than $\sim0.2$~dex below the characteristic mass of the fitted mass functions (Figure~\ref{fig:MassFunctionPlot}).\\

    \vspace{-0.05cm}\item \noindent \textit{All four simulations predict $\gtrsim50$ ($\gtrsim80$)~per cent of satellite galaxy sourced ICL stars 
    to be contributed by satellites with infall stellar masses $>10^{10}$\,\textup{M}\textsubscript{\sun}} ($>10^{9}$\,M\textsubscript{\sun}; Figure~\ref{fig:IclContriCsPlot}). The corresponding satellite infall stellar mass thresholds for the combined BCG\,+\,ICL are $10^{10.5}$\,M\textsubscript{\sun} (for $>50$~per cent) and $10^{10}$\,M\textsubscript{\sun} (for $>80$~per cent). 
    
\end{enumerate}

Our results suggest that the conflicting findings of prior theoretical studies concerning both typical ICL stellar mass fractions and the evolution of these with redshift may entirely stem from the differing ICL extraction methodologies employed by these studies. Subject to a consistent ICL definition, we find no significant discrepancies between the four simulations concerning ICL mass fractions at $z\approx0$; ICL mass fraction evolution in the progenitors of $z\approx0$ low-mass clusters back to $z\sim2$; the relative ICL contributions of differingly massive satellites on cluster infall; or the main progenitor objects of ICL stars. 

What remains as yet under-explored is how the considered simulations differ in their predictions for contributions to the ICL via assembly channels beyond just the liberation of stars from cluster satellite galaxies; from the pre-processed ICL accreted along with groups and smaller clusters during cluster assembly, and from stars that form ``in-situ'' either in the central galaxy or directly into the ICL component. We intend to investigate these alternative ICL assembly channels, as well as the resulting implications for cluster-centric radial trends in stellar age and metallicity, in a forthcoming companion paper (Brown et al. in prep.). 


\section*{Acknowledgements}

The authors thank the other members of the NottICL Group -- and in particular Jesse B. Golden-Marx and Harry Gully -- for helpful discussions and comments; 
and also thank Emanuele Contini for their careful reading of the original manuscript and for their constructive comments which have helped to improve the quality and clarity of the presented work. 
H.~J.~Brown thanks Dylan Nelson for their guidance on handling \textsc{TNG100} data. 

H.~J.~Brown acknowledges support from the UK Science and Technology Facilities Council (STFC) under grant ST/Y509437/1. 
F.~.R.~Pearce and N.~A.~Hatch acknowledge support from the UK STFC under grant ST/X000982/1.
Y.~M.~Bah\'{e} acknowledges support from UK Research and Innovation through a Future Leaders Fellowship (grant agreement MR/X035166/1) and financial support from the Swiss National Science Foundation (SNSF) under project ``Galaxy evolution in the cosmic web'' (200021\_213076).
M.~Butler and N.~A.~Hatch acknowledge support from the Leverhulme Trust through a Research Leadership Award.
W.~Cui thanks Comunidad de Madrid for the Atracci\'{o}n de Talento fellowship no. 2020-T1/TIC19882 and Agencia Estatal de Investigaci\'{o}n (AEI) for the Consolidaci\'{o}n Investigadora Grant CNS2024-154838; he further acknowledges the Project PID2024-156100NB-C21 financed by MICIU/AEI /10.13039/501100011033/FEDER, EU and ERC: HORIZON-TMA-MSCA-SE for supporting the LACEGAL-III (Latin American Chinese European Galaxy Formation Network) project with grant number 101086388 and the science research grants from the China Manned Space Project.
A.~Knebe is supported by project PID2024-156100NB-C21 financed by MICIU /AEI/10.13039/501100011033 / FEDER, UE, and further thanks Oasis for supersonic.

This work made use of \textsc{NumPy} \citep{harris_array_2020}, \textsc{Matplotlib} \citep{hunter_matplotlib_2007}, \textsc{SciPy} \citep{virtanen_scipy_2020}, 
\textsc{h5py} \citep{collette_h5pyh5py_2023-1}, \textsc{pyGAM} \citep{pyGAM_ref}, \textsc{Scikit-learn} \citep{pedregosa_scikit-learn_2011}, and \textsc{Astropy} \citep{the_astropy_collaboration_astropy_2022}. 

The \textsc{Horizon-AGN} simulation was granted access to the HPC resources of CINES under allocations 2013047012, 2014047012 and 2015047012 made by GENCI, and made use of the Infinity cluster, hosted by the Institut d'Astrophysique de Paris. We warmly thank S.~Rouberol for running it smoothly. 
%
The \textsc{IllustrisTNG} simulations were undertaken with compute time awarded by the Gauss Centre for Supercomputing (GCS) under GCS Large-Scale Projects GCS-ILLU and GCS-DWAR on the GCS share of the supercomputer Hazel Hen at the High Performance Computing Center Stuttgart (HLRS), as well as on the machines of the Max Planck Computing and Data Facility (MPCDF) in Garching, Germany.
%
The authors acknowledge The Red Española de Supercomputaci\'on for granting computing time for running the hydrodynamical and DMO simulations of the \textsc{The Three Hundred} galaxy cluster project in the Marenostrum supercomputer at the Barcelona Supercomputing Center and Cibeles Supercomputers through various RES grants. The \textsc{The Three Hundred} HD hydrodynamic simulations (7K and 15K runs) were performed also on the DIaL3 -- DiRAC Data Intensive service at the University of Leicester through the RAC15 grant: dp235, and on the Niagara supercomputer at the SciNet HPC Consortium. DIaL3 is managed by the University of Leicester Research Computing Service on behalf of the STFC DiRAC HPC Facility (\url{www.dirac.ac.uk}). The DiRAC service at Leicester was funded by BEIS (ST/K000373/1), UKRI, STFC capital funding, and STFC operations grants (ST/K0003259/1). DiRAC is part of the UKRI Digital Research Infrastructure. SciNet \citep{loken_scinet_2010} is funded by Innovation, Science and Economic Development Canada; the Digital Research Alliance of Canada; the Ontario Research Fund: Research Excellence; and the University of Toronto.
%
This work used the DiRAC@Durham facility managed by the Institute for Computational Cosmology on behalf of the STFC DiRAC HPC Facility (\url{www.dirac.ac.uk}). The equipment was funded by BEIS capital funding via STFC capital grants ST/K00042X/1, ST/P002293/1, ST/R002371/1 and ST/S002502/1, Durham University and STFC operations grant ST/R000832/1. DiRAC is part of the National e-Infrastructure. 


\section*{Data Availability}

The data products of the \textsc{Horizon-AGN} simulation are available upon reasonable request through the collaboration's website: \url{https://www.horizon-simulation.org/}. The data products of the \textsc{TNG100} simulation are publicly available through the \textsc{IllustrisTNG} project's website: \url{https://tng-project.org/}. The data products of \textsc{The Three Hundred} project are available upon reasonable request through the collaboration's website: \url{https://www.the300-project.org}. The data products of the \textsc{Hydrangea} simulations are publicly available through the Leiden Observatory Science Data Repository: \url{https://ftp.strw.leidenuniv.nl/bahe/Hydrangea}. 
Further data products generated from this work are available upon reasonable request from the corresponding author. 


\bibliographystyle{mnras}
\bibliography{refs}


\appendix


\section{Recombining fragmented galaxies}\label{Appendix:Methods-IclDef-Clumps} 

Not all structures identified by \textsc{Subfind} in \textsc{TNG100} or \textsc{Cantor} in \textsc{Hydrangea} correspond to galaxies, even when considering only those containing stars. 
As these codes consider any local density peak a seed for potential sub-structure, dense clumps of stars and gas can end up segregated from the bulk of their host galaxy as their own ``sub-halo''. As these fragments isolate some of the densest, baryon-dominated regions in massive galaxies, they can not only non-negligibly lower the apparent stellar mass of their host galaxy if sub-structure is neglected (particularly if several are carved away from the same galaxy), but can themselves masquerade as extremely compact ``galaxies'' in structure catalogues. In \textsc{TNG100}, these fragments can have stellar masses as high as $\gtrsim10^{9.5}$\,M\textsubscript{\sun}, with sizes of order unity kpc, negligible DM content, and often very high specific star formation rates.
These fragments are also long-lived (commonly persisting for several Gyrs) and can multiply by fragmenting further into smaller and smaller objects, such that at times up to $\sim20$~per cent of the ``sub-haloes'' with stellar mass between $10^{8.5}-10^{9.5}$\,M\textsubscript{\sun} in and around some of the \textsc{TNG100} clusters we consider belong to this class of object, with similarly high abundances reported in \textsc{Hydrangea} previously by \citet{bahe_disruption_2019}. 
An in-depth investigation into this fragmentation is beyond the scope of this study (for further relevant discussion see e.g. \citealt{Nelson_2019_TngPublicDataRelease}, \citealt{celiz_mass-morphology_2025}, and \citealt{moreno_assessing_2025}; see also \citealt{bahe_disruption_2019}) -- but this behaviour is undesirable for our use case and we seek to correct for this fragmentation  when constructing our galaxy catalogues.

In \textsc{Hydrangea} we identify these ``anti-hierarchically'' formed fragments following the procedure described in \citet{bahe_disruption_2019}, in which these objects are labelled ``spectres''. Once our procedure for assigning stars to galaxies is nominally complete (prior to neglecting structures with insufficient particle counts; see Section~\ref{Methods-IclDef-GalDef}), if the stars of a provisional ``galaxy'' identified within a sub-halo flagged as a ``spectre'' are at least partially enclosed by the ``galaxy'' identified in the spectre's parent sub-halo, then 
the ``spectre'' galaxy is absorbed into its parent
(with relevant merger tree links either redirected to the parent or cut as appropriate; see Section~\ref{Methods-TrackingGalaxies}). In the event a spectre has no parent sub-halo, or the spectres' galaxy stars are not enclosed within the parent galaxy (and similar checks performed if applicable using the parent's parent sub-halo and so forth also fail) we retain the spectre-flagged object as a prospective galaxy. 

A very similar procedure is followed in \textsc{TNG100}, for which the \textsc{SubhaloFlag} field is provided to identify objects thought not of cosmological origin (prototypically clumps formed within galaxies through baryonic processes; \citealt{Nelson_2019_TngPublicDataRelease}). However, we note both a non-negligible number of apparent false positives 
(i.e. entire infalling satellites flagged as not of cosmological origin despite being otherwise unremarkable)
and also a considerable number of compact, DM deficient sub-structures archetypical of the galaxy fragmentation we seek to remedy that are not caught by \textsc{SubhaloFlag}. 
These false positives can stem from galaxies with tumultuous early histories, such that they are briefly divided into two sub-haloes (one of which briefly drops beneath the DM mass fraction threshold of 0.8 of \textsc{SubhaloFlag}) before recombining. The considerable number of missed fragments we note conceivably stems from our focus on galaxy clusters, which can have very large and distended FoF groups, in combination with the criteria used for \textsc{SubhaloFlag} explicitly omitting objects first detected outside their host FoF group's virial radius \citep{Nelson_2019_TngPublicDataRelease}.
We consequently do not rely on \textsc{SubhaloFlag} alone to identify these fragments in \textsc{TNG100} and append our own scheme based on that used in \citet{bahe_disruption_2019}. We flag as fragments (to be handled as described above for the ``spectres'' of \textsc{Hydrangea}) \textsc{TNG100} sub-haloes that are first seen in one of the coarsely spaced snapshots we consider (see Section~\ref{Methods-TrackingGalaxies-MergerTrees}) as a child sub-halo, with a dark-matter fraction less than 0.5, and that are either flagged by \textsc{SubhaloFlag}; are $\geq75$~per cent composed of particles formerly in their parent sub-halo's progenitor in the prior coarsely spaced snapshot; or that are sub-structure of a newly-flagged fragment.


\section{Cluster samples}\label{Appendix-ClusterProps}

\subsection{Cluster properties}\label{Appendix-ClusterProps-ActProps}

\begin{table*}
    \centering
    \caption{
    Overview of the cluster sample from \textsc{Horizon-AGN} used in this work. For each cluster we provide the maximum radius within which the mean DM density equals $178$ times the critical density ($r_{178c}$); the total mass within this radius ($M_{178c}$); the fractions of this total mass contributed by DM ($f_\text{DM}$) and by stars~($f_*$); the fractions of this stellar mass contributed by the BCG ($f_{*\text{,\,BCG}}$), satellite galaxies ($f_{*\text{,\,Sat}}$), and the ICL ($f_{*\text{,\,ICL}}$); DM halo assembly redshifts ($z_{50}$ and $z_{90}$); stellar-mass analogues to the ``magnitude gap'' ($M_{12}$ and $M_{14}$); the concentration of the cluster's DM halo ($c_\textrm{Halo-178}$); and the offset between the BCG centre of mass and the centre of mass of all DM within $r_{178c}$ (as a percentage of $r_{178c}$). See text for additional details. 
    }
    \label{tab:HAGN_cluster_props}
    \begin{tabular}{|c|c|c|c|c|c|c|c|c|c|c|c|c|c|}
        \hline
        Identifier: & $r_{178c}$ & $M_{\textrm{178c}}$ & $f_\text{DM}$ & $f_{*}$ & $f_{*\textrm{,\,BCG}}$ & $f_{*\textrm{,\,Sat}}$ & $f_{*\textrm{,\,ICL}}$ & $z_{50}$ & $z_{90}$ & $M_{12}$ & $M_{14}$ & $c_\textrm{Halo-178}$ & BCG-DM CoM\\
        BCG ID& [Mpc] & [$10^{14}$\,M\textsubscript{$\sun$}] & [$\%$] & [$\%$] & [$\%$]& [$\%$]& [$\%$]& & & & & & Offset $/r_{178c}$ [$\%$]\\
        \hline
        1& 1.04& 1.45& 84.3& 2.9& 34.0& 52.9& 13.1& 0.61& 0.26& 0.971& 1.084& $3.4$& 1.02\\
        9& 0.92& 1.00& 82.9& 2.5& 56.9& 28.0& 15.2& 1.50& 0.28& 1.350& 1.511& $6.8$& 2.62\\
        13& 1.81& 7.52& 84.2& 2.5& 13.6& 73.7& 12.7& 0.19& 0.08& 0.000& 0.474& $1.3$& 7.95\\
        19& 1.62& 5.46& 82.9& 2.4& 37.2& 46.1& 16.7& 0.69& 0.47& 0.969& 1.284& $3.9$& 1.56\\
        48& 0.96& 1.11& 84.1& 2.9& 36.2& 51.4& 12.5& 0.81& 0.12& 0.624& 0.831& $5.2$& 3.35\\
        49& 1.38& 3.35& 83.7& 2.4& 38.1& 47.3& 14.6& 1.07& 0.14& 0.907& 1.268& $5.5$& 4.19\\
        71& 1.05& 1.46& 84.0& 3.2& 32.2& 56.6& 11.1& 0.41& 0.10& 0.552& 0.767& $3.3$& 9.56\\
        132& 0.99& 1.25& 84.1& 3.1& 32.3& 57.9& 9.8& 0.62& 0.10& 0.368& 0.616& $3.5$& 12.51\\
        174& 1.18& 2.09& 84.5& 2.8& 26.6& 60.9& 12.5& 0.47& 0.27& 0.420& 0.796& $2.9$& 1.73\\
        183& 0.94& 1.06& 84.9& 3.3& 36.8& 52.0& 11.2& 0.57& 0.33& 0.376& 1.014& $5.8$& 7.37\\
        \hline
    \end{tabular}
\end{table*}
\begin{table*}
    \centering
    \caption{
    Overview of the cluster sample from \textsc{TNG100} used in this work. See Table~\ref{tab:HAGN_cluster_props} for description of headings.
    }
    \label{tab:TNG_cluster_props}
    \begin{tabular}{|c|c|c|c|c|c|c|c|c|c|c|c|c|c|}
        \hline
        Identifier: & $r_{178c}$ & $M_{\textrm{178c}}$ & $f_\text{DM}$ & $f_{*}$ & $f_{*\textrm{,\,BCG}}$ & $f_{*\textrm{,\,Sat}}$ & $f_{*\textrm{,\,ICL}}$ & $z_{50}$ & $z_{90}$ & $M_{12}$ & $M_{14}$ & $c_\textrm{Halo-178}$ & BCG-DM CoM\\
        FoF ID& [Mpc] & [$10^{14}$\,M\textsubscript{$\sun$}] & [$\%$] & [$\%$] & [$\%$]& [$\%$]& [$\%$]& & & & & & Offset $/r_{178c}$ [$\%$]\\
        \hline
        1& 1.50& 3.72& 85.6& 1.6& 22.4& 62.7& 14.8& 0.26& 0.12& 0.330& 0.826& $1.9$& 13.23\\
        0& 1.49& 3.69& 84.6& 2.2& 37.8& 51.5& 10.7& 0.63& 0.03& 0.717& 1.066& $4.2$& 11.34\\
        2& 1.43& 3.30& 84.3& 1.8& 32.8& 52.6& 14.6& 0.71& 0.13& 0.538& 1.204& $3.6$& 6.49\\
        4& 1.30& 2.49& 84.6& 1.7& 18.3& 70.3& 11.4& 0.27& 0.06& 0.210& 0.517& $2.8$& 4.83\\
        9& 1.23& 2.08& 85.0& 1.7& 48.3& 40.1& 11.2& 0.45& 0.04& 0.951& 1.256& $4.2$& 7.84\\
        6& 1.23& 2.06& 85.1& 1.9& 46.0& 42.4& 11.7& 0.91& 0.39& 0.853& 1.159& $4.8$& 3.48\\
        8& 1.22& 2.04& 85.2& 1.7& 44.5& 40.9& 14.6& 0.87& 0.11& 0.852& 1.276& $4.7$& 2.35\\
        5& 1.22& 2.01& 85.5& 1.8& 26.2& 61.8& 12.0& 0.59& 0.11& 0.304& 0.668& $3.7$& 9.35\\
        3& 1.14& 1.68& 83.3& 1.6& 42.7& 44.5& 12.8& 1.00& 0.28& 1.088& 1.204& $5.9$& 3.14\\
        10& 1.12& 1.60& 84.2& 1.7& 51.0& 37.2& 11.7& 1.10& 0.09& 1.026& 1.278& $6.4$& 2.20\\
        11& 1.05& 1.30& 85.1& 2.0& 54.6& 34.9& 10.5& 1.08& 0.37& 0.555& 1.540& $5.7$& 3.04\\
        14& 1.00& 1.12& 83.9& 1.7& 49.9& 36.3& 13.9& 1.17& 0.30& 0.977& 1.129& $5.8$& 1.60\\
        15& 0.99& 1.07& 84.6& 1.7& 41.6& 45.5& 12.9& 0.92& 0.16& 0.554& 1.042& $5.1$& 4.16\\
        17& 0.98& 1.05& 85.6& 1.9& 53.1& 33.7& 13.2& 0.82& 0.11& 1.063& 1.198& $4.6$& 4.21\\
        20& 0.94& 0.94& 84.7& 1.8& 39.6& 50.5& 9.9& 0.28& 0.10& 0.221& 1.195& $3.3$& 5.74\\
        \hline
    \end{tabular}
\end{table*}
\begin{table*}
    \centering
    \caption{
    Overview of the cluster sample from \textsc{The Three Hundred Gizmo-Simba 7K} used in this work. See Table~\ref{tab:HAGN_cluster_props} for description of headings.
    }
    \label{tab:The300_cluster_props}
    \begin{tabular}{|c|c|c|c|c|c|c|c|c|c|c|c|c|c|}
        \hline
        Identifier: & $r_{178c}$ & $M_{\textrm{178c}}$ & $f_\text{DM}$ & $f_{*}$ & $f_{*\textrm{,\,BCG}}$ & $f_{*\textrm{,\,Sat}}$ & $f_{*\textrm{,\,ICL}}$ & $z_{50}$ & $z_{90}$ & $M_{12}$ & $M_{14}$ & $c_\textrm{Halo-178}$ & BCG-DM CoM\\
        Box ID& [Mpc] & [$10^{14}$\,M\textsubscript{$\sun$}] & [$\%$] & [$\%$] & [$\%$]& [$\%$]& [$\%$]& & & & & & Offset $/r_{178c}$ [$\%$]\\
        \hline
        290& 2.06& 9.84& 84.8& 1.9& 15.9& 67.7& 16.4& 0.67& 0.38& 0.643& 0.781& $3.4$& 3.11 \\
        299& 2.12& 10.71& 84.3& 1.8& 27.9& 55.4& 16.7& 1.13& 0.36& 1.162& 1.299& $6.0$& 0.52\\
        303& 2.08& 9.99& 85.3& 1.8& 11.5& 72.1& 16.4& 0.16& 0.03& 0.198& 0.557& $1.8$& 15.37\\
        307& 2.04& 9.43& 85.2& 2.0& 18.8& 63.7& 17.6& 0.45& 0.20& 0.897& 0.932& $2.5$& 8.07 \\
        309& 2.06& 9.78& 84.9& 1.9& 14.6& 67.1& 18.3& 0.53& 0.07& 0.194& 0.757& $2.4$& 7.92 \\
        310& 1.97& 8.56& 84.5& 2.2& 23.9& 61.2& 14.9& 0.61& 0.20& 0.653& 1.116& $3.1$& 8.33 \\
        311& 2.03& 9.24& 85.4& 2.0& 21.2& 63.6& 15.2& 0.43& 0.20& 0.442& 0.888& $2.9$& 5.63 \\
        313& 2.05& 9.63& 84.5& 1.9& 29.5& 55.0& 15.5& 0.45& 0.23& 0.720& 1.210& $4.7$& 7.43 \\
        317& 2.02& 9.16& 85.5& 2.0& 16.2& 67.4& 16.4& 0.63& 0.19& 0.417& 0.970& $2.6$& 3.18 \\
        318& 2.03& 9.24& 85.5& 2.1& 15.2& 67.9& 17.0& 0.53& 0.29& 0.374& 0.668& $2.6$& 4.97 \\
        319& 2.01& 9.03& 84.8& 1.9& 34.5& 49.7& 15.8& 0.53& 0.27& 0.645& 1.364& $5.0$& 3.32 \\
        320& 2.08& 10.13& 84.7& 2.0& 43.1& 41.0& 15.9& 0.50& 0.30& 1.180& 1.590& $5.6$& 2.00 \\
        321& 2.08& 10.06& 84.8& 1.9& 17.0& 66.6& 16.4& 0.50& 0.04& 0.202& 0.891& $3.0$& 11.88\\
        323& 2.00& 8.88& 85.0& 1.8& 16.9& 66.7& 16.3& 0.56& 0.13& 0.627& 0.786& $3.9$& 6.87 \\
        324& 2.06& 9.79& 84.6& 2.0& 20.5& 63.8& 15.7& 1.02& 0.06& 0.853& 0.934& $5.0$& 7.88 \\
        \hline
    \end{tabular}
\end{table*}
\begin{table*}
    \centering
    \caption{
    Overview of the cluster sample from \textsc{Hydrangea} used in this work. See Table~\ref{tab:HAGN_cluster_props} for description of headings.
    }
    \label{tab:Hydra_cluster_props}
    \begin{tabular}{|c|c|c|c|c|c|c|c|c|c|c|c|c|c|}
        \hline
        Identifier: & $r_{178c}$ & $M_{\textrm{178c}}$ & $f_\text{DM}$ & $f_{*}$ & $f_{*\textrm{,\,BCG}}$ & $f_{*\textrm{,\,Sat}}$ & $f_{*\textrm{,\,ICL}}$ & $z_{50}$ & $z_{90}$ & $M_{12}$ & $M_{14}$ & $c_\textrm{Halo-178}$ & BCG-DM CoM\\
        Box ID& [Mpc] & [$10^{14}$\,M\textsubscript{$\sun$}] & [$\%$] & [$\%$] & [$\%$]& [$\%$]& [$\%$]& & & & & & Offset $/r_{178c}$ [$\%$]\\
        \hline
        CE-0& 1.02& 1.16& 86.0& 1.5& 43.0& 41.7& 15.3& 0.92& 0.35& 0.455& 1.295& $5.2$& 2.52\\
        CE-1& 1.00& 1.12& 86.1& 1.6& 41.3& 43.2& 15.5& 0.78& 0.12& 0.561& 0.969& $3.5$& 6.39\\
        CE-2& 1.03& 1.21& 85.8& 1.3& 54.3& 27.1& 18.6& 1.05& 0.42& 1.113& 1.293& $5.9$& 2.25\\
        CE-3& 1.07& 1.36& 85.2& 1.2& 37.6& 44.3& 18.2& 1.11& 0.41& 0.785& 1.018& $6.3$& 3.10\\
        CE-4& 1.15& 1.69& 85.7& 1.4& 28.9& 56.9& 14.2& 0.27& 0.04& 0.236& 0.916& $4.0$& 13.77\\
        CE-5& 1.07& 1.37& 84.5& 1.5& 44.6& 41.5& 13.9& 0.94& 0.18& 0.443& 1.276& $6.5$& 1.66\\
        CE-6& 1.24& 2.15& 85.0& 1.4& 29.0& 53.2& 17.9& 0.77& 0.31& 0.532& 0.667& $3.5$& 6.16\\
        CE-7& 1.23& 2.12& 84.5& 1.3& 36.5& 45.1& 17.0& 0.90& 0.30& 0.910& 1.279& $4.4$& 2.15\\
        CE-8& 1.20& 1.95& 84.9& 1.3& 29.5& 55.5& 15.0& 0.67& 0.14& 0.411& 0.903& $4.3$& 3.22\\
        CE-9& 1.36& 2.82& 84.8& 1.3& 40.9& 41.0& 18.1& 0.82& 0.29& 0.910& 1.291& $4.8$& 1.58\\
        CE-12& 1.52& 3.91& 85.0& 1.5& 24.6& 57.2& 18.2& 0.84& 0.53& 0.411& 0.911& $4.5$& 0.93\\
        CE-13& 1.54& 4.06& 84.7& 1.2& 27.6& 55.6& 16.8& 0.68& 0.04& 0.757& 1.042& $5.9$& 3.77\\
        CE-14& 1.58& 4.45& 84.6& 1.3& 21.9& 60.2& 17.9& 0.38& 0.18& 0.607& 1.006& $2.3$& 3.78\\
        CE-15& 1.67& 5.23& 85.2& 1.4& 13.7& 70.2& 16.0& 0.18& 0.03& -0.082& 0.696& $2.0$& 10.85\\
        CE-16& 1.70& 5.54& 84.7& 1.3& 25.5& 59.4& 14.5& 0.33& 0.02& 0.339& 0.874& $6.6$& 5.06\\
        CE-18& 1.83& 6.86& 84.7& 1.3& 24.3& 57.3& 18.5& 0.66& 0.09& 0.914& 0.980& $4.5$& 3.48\\
        \hline
    \end{tabular}
\end{table*}

We identify every cluster from the four samples we employ in Tables~\ref{tab:HAGN_cluster_props}, \ref{tab:TNG_cluster_props}, \ref{tab:The300_cluster_props}, and \ref{tab:Hydra_cluster_props} (for \textsc{Hz-AGN}, \textsc{TNG100}, \textsc{The300 GS-7K}, and \textsc{Hydrangea} respectively) and present for these a selection of cluster properties.  
We identify each cluster in \textsc{Hz-AGN} by the \textsc{AdaptaHOP} ID of its central galaxy in snapshot 761; in \textsc{TNG100} by the ID/index of its FoF group in the structure catalogue for snapshot 99; and in \textsc{The300 GS-7K} and \textsc{Hydrangea} by the ID of the simulation box it is centrally located in. 

In Tables~\ref{tab:HAGN_cluster_props}-\ref{tab:Hydra_cluster_props} we present for each cluster the value 
(at $z\approx0$) of $r_{178c}$,  
centring on the stellar centre of mass of the identified central galaxy (see Section~\ref{Methods-IclDef-BcgDetermination});
the total mass enclosed within this radius at $z\approx0$ ($M_{178c}$); the fractions of this total mass contributed by DM ($f_\text{DM}$) and by stellar particles ($f_{*}$); as well as the fractions of this stellar mass at $z\approx0$ contributed by the BCG, satellite galaxies, and the ICL ($f_{*\text{,\,BCG}}$, $f_{*\text{,\,Sat}}$, and $f_{*\text{,\,ICL}}$, respectively; following our fiducial ICL and galaxy identification scheme described in Section~\ref{Methods-IclDef}). 
We also give estimates for the redshifts when each cluster first assembled 50 and 90 per cent of its $z\approx0$ DM mass ($z_{50}$ and $z_{90}$ respectively) as found by fitting monotonic cubic spline functions to 
the total DM mass at $r<r_{178c}(z)$
against lookback time for the $\sim1$\,Gyr spaced snapshots we employ in this work (with these fits anchored to the $z\approx0$ masses) and interpolating. We additionally give stellar mass analogues to the ``magnitude gap'' (e.g. \citealt{golden-marx_hierarchical_2025}) defined here as $M_{1X}=\text{log}_{10}(M_{*,1}/M_{*,X})$ where $M_{*,1}$ is the BCG stellar mass and $M_{*,X}$ the stellar mass of the $(X-1)$th most massive satellite galaxy within $r_{178c}$. We present both $M_{12}$ and $M_{14}$ values for each cluster at $z\approx0$. 
To contextualise these ``stellar magnitude gap'' values, we note the median values of $M_{12}$ and $M_{14}$ across all four cluster samples to be $\sim0.6$ and $\sim1.0$ respectively. 
The DM halo concentration values we present 
(defined here as $c_\textrm{Halo-178}\equiv r_{178c}/r_s$) are computed by a similar methodology to that employed by \citet[][see also references therein]{bahe_hydrangea_2017}: 
centring on the density peak of each cluster's DM halo (as opposed to the BCG centre as elsewhere in our analysis), we fit an NFW profile with scale radius $r_{s}$ to the spherically averaged DM density in 40 equal log-width radial bins between $0.05\times r_{178c}$ and $r_{178c}$.
The offsets between the BCG stellar centre of mass and the centre of mass of all DM enclosed by $r_{178c}$ at $z\approx0$ are also given (as a percentage of $r_{178c}$). 

The \textsc{Hz-AGN} cluster sample we employ for this work has significant overlap with that used in \citetalias{brown_assembly_2024}. Slight differences between some of the values presented in Table~\ref{tab:HAGN_cluster_props} and those values given in table 1 of \citetalias{brown_assembly_2024} are due to subtle differences in methodology -- chiefly that in \citetalias{brown_assembly_2024} potential centres were used for the position of each galaxy, whereas in this work we use stellar centre of mass (slightly shifting the 
centres used when determining $r_{178c}$ and $M_{178c}$). 
Some of the values we present for the clusters of the \textsc{Hydrangea} sample in Table~\ref{tab:Hydra_cluster_props} slightly differ from those given in table A1 of \citet{bahe_hydrangea_2017} for similar reasons. 

Prior studies have reported possible ICL fraction trends with cluster mass (e.g. \citealt{joo_tracing_2024}, \citealt{canepa_dependence_2025}), with cluster dynamical state (e.g. \citealt{golden-marx_hierarchical_2025}, \citealt{kimmig_intra-cluster_2025}), and with the concentration of a cluster's DM halo (e.g. \citealt{contini_intracluster_2023}, \citealt{mayes_coevolution_2025}). Though we investigated potential trends between the ICL stellar mass fraction (under our fiducial ICL definition) and cluster mass, DM halo concentration, as well as various proxy measures of dynamical state ($z_{50}$, $z_{90}$, $M_{12}$, $M_{14}$, and the offset between the BCG and DM halo centres of mass), we could not identify any significant trends above the intra- and inter-simulation scatter for the restricted mass ranges and modest cluster sample sizes we employ in this study. However, we did note trends between these same cluster properties and the combined BCG\,+\,ICL stellar mass fractions -- with typically larger BCG\,+\,ICL stellar mass fractions seen for clusters that are less massive, that have more concentrated DM haloes, that have smaller BCG-DM centre of mass offsets, that have larger stellar-mass magnitude gap values, and that assembled earlier (consistent with the prior findings of e.g. \citealt{kimmig_intra-cluster_2025} and \citealt{golden-marx_hierarchical_2025}).  

\subsection{Extreme objects}\label{Appendix-ClusterProps-ExObjects}

We highlight cluster 13 in the \textsc{Hz-AGN} sample as a particularly extreme object -- in the early stages of two simultaneous major cluster mergers at $z\approx0$. 
Clusters 0 and 1 from \textsc{TNG100} (i.e. the two most massive FoF groups in the simulation box at $z\approx0$) are also notable for being highly disturbed, with large populations of massive satellites predominately located to one side of the identified BCG 
at $z\approx0$ such that the ``central'' galaxy in both instances appears noticeably offset from centroid of the local galaxy distribution. 
Cluster 0 from \textsc{TNG100} is also noteworthy for having non-negligible late time star formation in its central galaxy -- with the median stellar particle age at $r<0.02\times r_{178c}$ in this cluster being $<7.5$\,Gyr, as compared to $>10$\,Gyr in every other considered \textsc{TNG100} cluster. This is similar to cluster 310 from the \textsc{The300 GS-7K} sample where the median stellar particle age at $r<0.02\times r_{178c}$ is $<4$\,Gyr (with 20 per cent of this cluster's BCG\,+\,ICL stars having formed in-situ in either the BCG or ICL).
We also highlight \textsc{TNG100} cluster 3 as unusual, with $\sim50$ per cent of the $z\approx0$ BCG\,+\,ICL stars in this cluster linked to either BCG or ICL in-situ star formation, significantly higher than any other \textsc{TNG100} cluster.
Among the \textsc{Hydrangea} cluster sample, we highlight CE-4 and CE-15 as extreme objects -- 
with the $z=0$ central in CE-4 seemingly part of an in-spiralling pair along with a particularly massive satellite galaxy; and with a ``satellite'' galaxy having just joined the halo of CE-15 in the final $\sim1$\,Gyr before $z\approx0$ that actually has slightly more stellar mass than what we consider the central galaxy for our analysis (hence the negative $M_{12}$ value seen for this cluster in Table~\ref{tab:Hydra_cluster_props}) -- with this very slightly more massive object not instead ruled the BCG as it was not found within the sub-halo \textsc{Cantor} ruled the cluster's ``central structure'' at $z\approx0$ (see Section~\ref{Methods-IclDef}). 
We confirm that excluding these extreme objects from our analysis does not considerably alter any of the presented findings. 


\section{Mass fraction evolution for cluster mass progenitors}\label{Appendix-MassLimitedIclFracEvo}

\begin{figure*}
    \includegraphics[width=2.07\columnwidth]{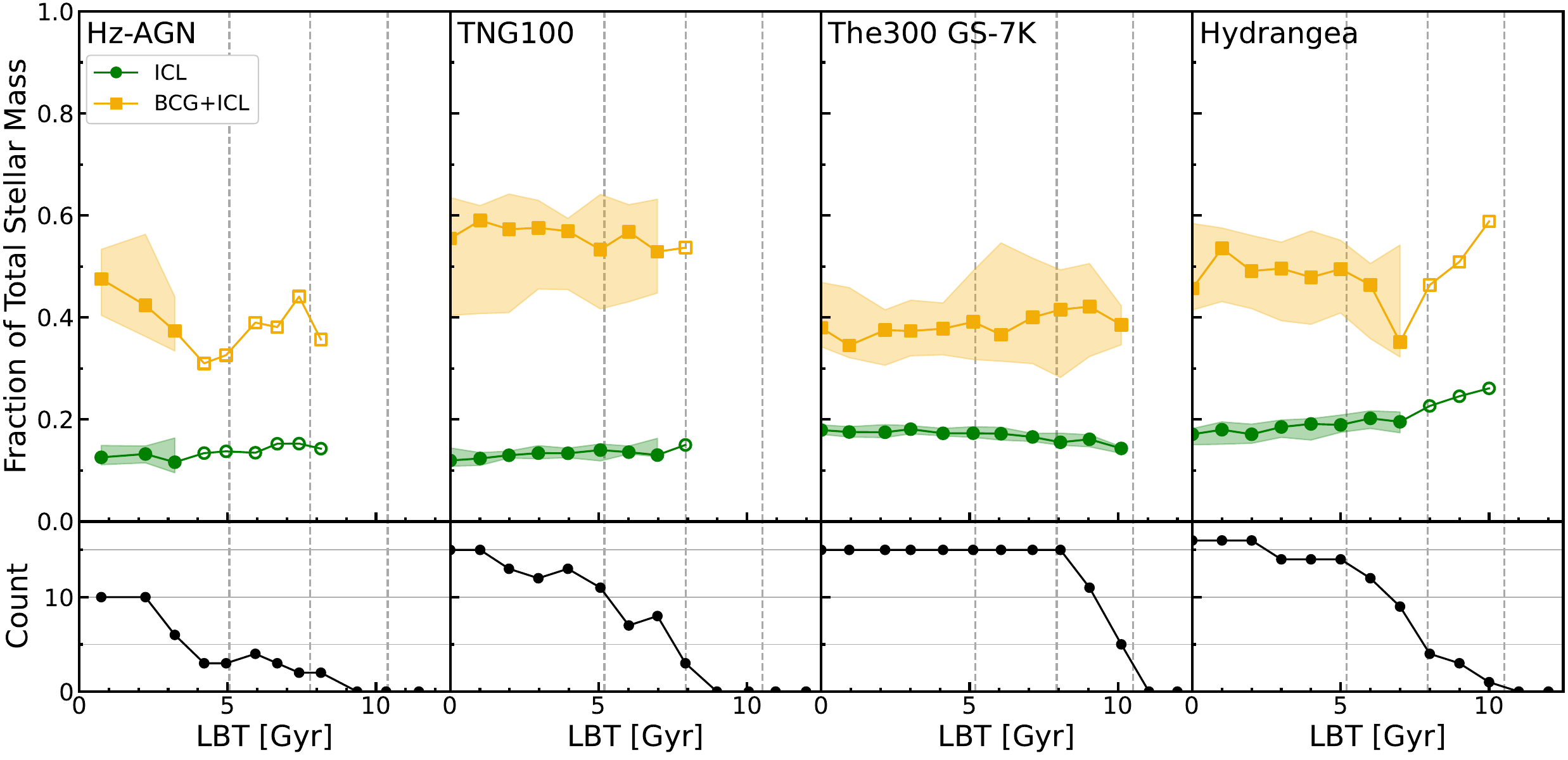}
    \caption{
    \textbf{Top Panels:} The same as Figure~\ref{fig:IclFracEvoPlot} but excluding $z>0$ progenitor structures with $M_{178c}<9\times10^{13}$\,M\textsubscript{$\sun$}. Median mass fraction values for samples of less than five structures  are indicated with unfilled markers. 
    We indicate the 16th-84th percentile dispersion via the shaded regions only for sample sizes of five or more. 
    \textbf{Bottom Panels:}  
    Number of progenitor structures above the imposed mass threshold as a function of lookback time. 
    }
    \label{fig:IclFracEvoPlot_ClusterMassOnly}
\end{figure*}

\begin{figure*}
    \includegraphics[width=2.07\columnwidth]{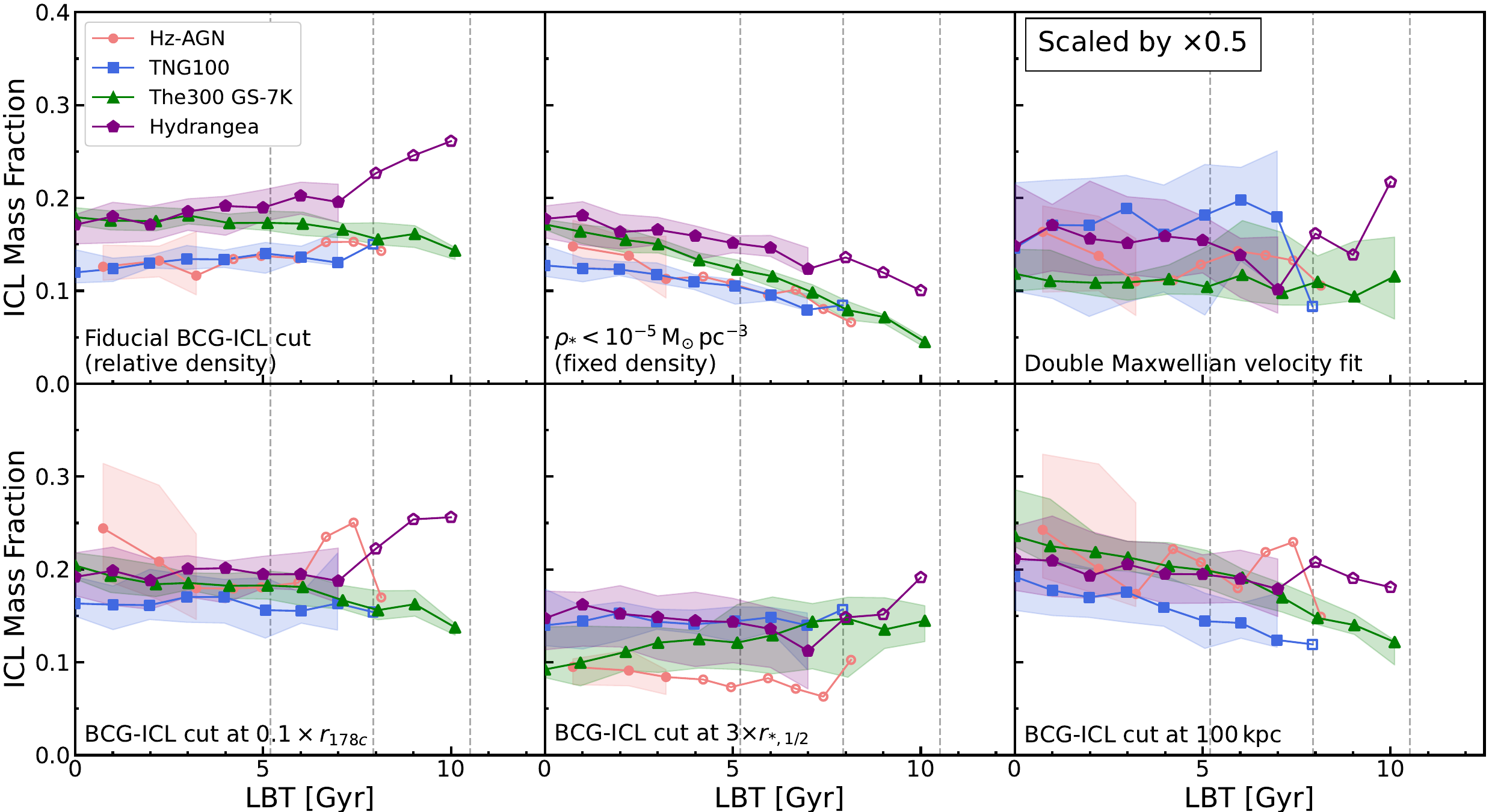}
    \caption{
    The same as Figure~\ref{fig:AltIclFracEvoPlot} but excluding $z>0$ progenitor structures with $M_{178c}<9\times10^{13}$\,M\textsubscript{$\sun$}. Median mass fraction values for samples of less than five structures are indicated with unfilled markers.
    We indicate the 16th-84th percentile dispersion via the shaded regions only for sample sizes of five or more. 
    }
    \label{fig:AltIclFracEvoPlot_ClusterMassOnly}
\end{figure*}

In Sections~\ref{Results-StellarMassDistribution-IclEvo} and~\ref{Results-StellarMassDistribution-AltDefIclEvo}, we present how the typical ICL (and BCG\,+\,ICL) stellar mass fractions of the four samples of simulated $z\approx0$ clusters and their $z>0$ progenitor structures evolve as a function of lookback time. In Figures~\ref{fig:IclFracEvoPlot_ClusterMassOnly} and~\ref{fig:AltIclFracEvoPlot_ClusterMassOnly} (counterparts to Figures~\ref{fig:IclFracEvoPlot} and~\ref{fig:AltIclFracEvoPlot} respectively) we present alterative versions of those analyses confined to cluster-mass structures ($M_{178c}\geq9\times10^{13}$\,M\textsubscript{\sun}) at all times. In Figures~\ref{fig:IclFracEvoPlot_ClusterMassOnly} and~\ref{fig:AltIclFracEvoPlot_ClusterMassOnly} unfilled markers indicate median mass fraction values found for samples of less than 5 objects, for which we also no longer indicate 16th-84th percentile dispersion via the shaded regions.

We note that in  Figure~\ref{fig:IclFracEvoPlot_ClusterMassOnly} and most panels of Figure~\ref{fig:AltIclFracEvoPlot_ClusterMassOnly} there is the suggestion 
that the typical ICL mass fraction of the (small number of) \textsc{Hydrangea} progenitors already of cluster mass begins to rise for lookback times~$\gtrsim8$\,Gyr. 
This chiefly stems from the three structures in the \textsc{Hydrangea} sample that reach cluster-scale mass first (identifiers CE-13, CE-16, and CE-18; see Table~\ref{tab:Hydra_cluster_props}) 
having some of the highest ICL mass fractions of the sample at these early lookback times, and so the median ICL fraction is pulled down as additional structures pass the imposed minimum mass threshold. 
We suspect the high early ICL fractions of these structures linked to the significant amounts of in-situ ICL star formation we note occurring in \textsc{Hydrangea} at $z>1$, such that an appreciable fraction of the $z\approx0$ ICL mass of every \textsc{Hydrangea} cluster considered is attributable to this high redshift in-situ star formation. We leave further discussion of this in-situ star formation to the forthcoming companion paper (Brown et al. in prep.). 


\section{Data and fitting for progenitor galaxies of ICL (and BCG + ICL) stars}\label{Appendix-IndivSimFits}

\begin{figure*}
    \includegraphics[width=2.07\columnwidth]{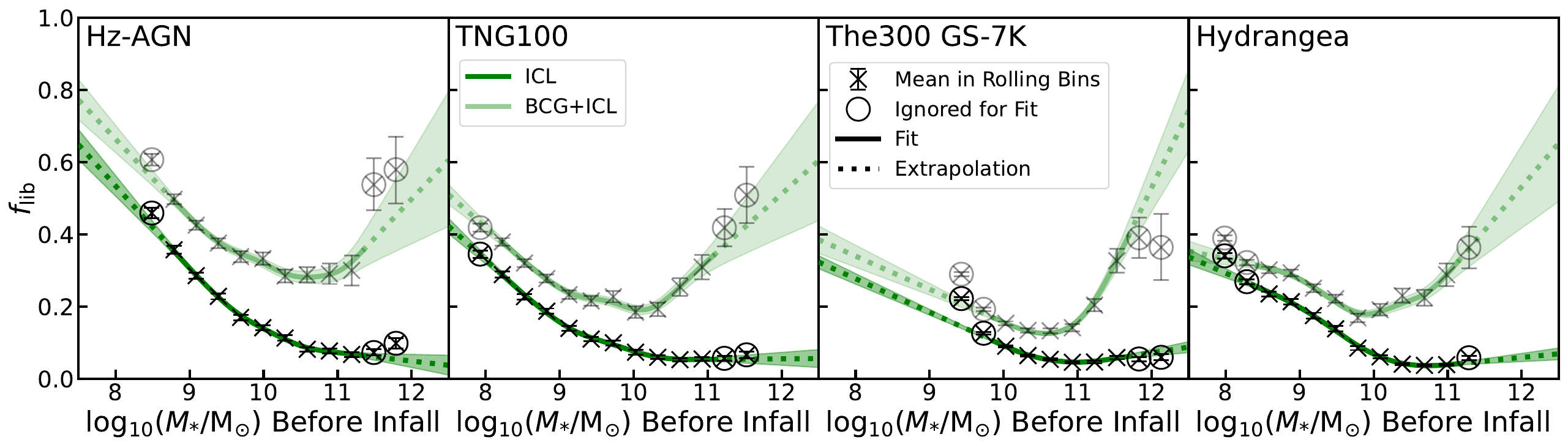}
    \caption{
    Mean fraction of stars liberated to the ICL, $f_{\textrm{lib}}$, as a function of satellite infall stellar mass, $M_{*}$, for the simulated cluster samples drawn from \textsc{Horizon-AGN} (leftmost panel), \textsc{TNG100} (middle-left panel), \textsc{The Three Hundred Gizmo-Simba 7K} (middle-right panel), and \textsc{Hydrangea} (rightmost panel). 
    For clarity only the mean value for every other bin is shown. Fitted cubic spline functions are shown in green, with bins ignored for fitting circled and linear extrapolation beyond the range of masses used for fitting indicated with dotted lines. The error bars and shaded regions indicate estimated uncertainties based on bootstrapping. The faint lines and symbols show an equivalent analysis for the combined BCG + ICL system.
    }
    \label{fig:SideBySide_flib}
\end{figure*}
\begin{figure*}
    \includegraphics[width=2.07\columnwidth]{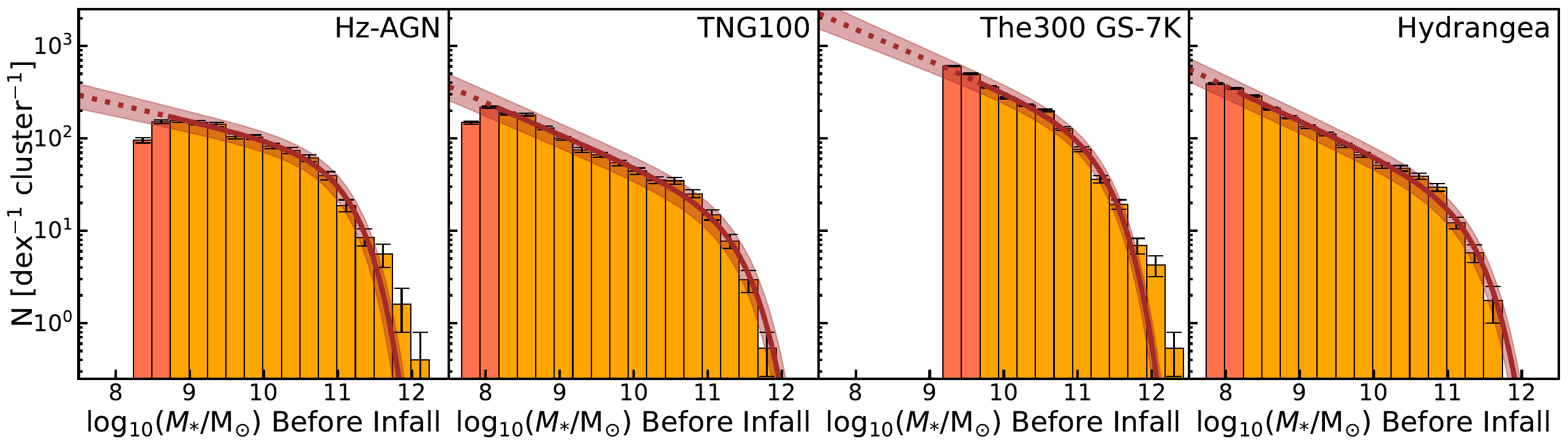}
    \caption{
    Histograms of satellite galaxy stellar masses upon cluster infall for the simulated cluster samples drawn from \textsc{Horizon-AGN} (leftmost panel), \textsc{TNG100} (middle-left panel), \textsc{The Three Hundred Gizmo-Simba 7K} (middle-right panel), and \textsc{Hydrangea} (rightmost panel) -- normalized to mean counts per cluster. Fitted \citet{schechter_analytic_1976} functions (equation~\ref{eq:Schechter}) are shown in brown, with bins ignored for fitting shown in red, and with extrapolation below the range of masses used for fitting indicated with dotted lines. The error bars indicate the dispersion in counts between bootstrap resamples (16th-84th percentile), and the shaded regions indicate estimated fit uncertainties (based on both bootstrapping and individual fit uncertainties). 
    }
    \label{fig:SideBySide_SchFit}
\end{figure*}

In 
Figures~\ref{fig:LibFracPlot} and~\ref{fig:MassFunctionPlot} we include only curves fitted to the data from each simulation while omitting the original data fitted to for clarity of presentation. We present those data here. 

In Figure~\ref{fig:SideBySide_flib} (counterpart to Figure~\ref{fig:LibFracPlot}) we present for each simulation individually the mean fraction of associated stellar mass liberated to the ICL (or BCG\,+\,ICL) by $z\approx0$ as a function of satellite infall stellar mass, as found for 0.5\,dex wide rolling infall stellar mass bins (bin step 0.15\,dex) and to which we fit a cubic spline function, linearly extrapolated beyond the range of masses used for fitting (see Section~\ref{Results2-IclContribution-LibFrac} for further details). For clarity of presentation, only every other bin is shown in Figure~\ref{fig:SideBySide_flib}. The error bars and shaded regions indicate the dispersion (16th-84th percentile) in mean values and functions fitted among the $10^{4}$ bootstrap resamples. 

In Figure~\ref{fig:SideBySide_SchFit} (counterpart to Figure~\ref{fig:MassFunctionPlot}) we present for each simulation individually a histogram of satellite infall stellar masses (0.25\,dex bin width; counts normalised to mean counts per cluster) to which we fit a \citet{schechter_analytic_1976} function (Equation~\ref{eq:Schechter}). 
Before fitting, we merge any bins containing fewer than five objects across the entire cluster sample for each simulation with the adjacent lower-mass bin, and when fitting exclude bins including masses that corresponding to poorly resolved galaxies ($<100$ stellar particles; see Section~\ref{Results2-IclContribution-MassFunc} for further details).
The error bars indicate the dispersion (16th-84th percentile) in bin counts among the $10^{4}$ bootstrap resamples, and the shaded regions show 16th-84th percentile confidence intervals for the fitted curves (based both on parameter fit uncertainty and the dispersion in fit parameters across the bootstrap resamples). 


\section{Post-infall satellite star formation}\label{Appendix-PostInfallSF}

\begin{figure*}
    \includegraphics[width=2.07\columnwidth]{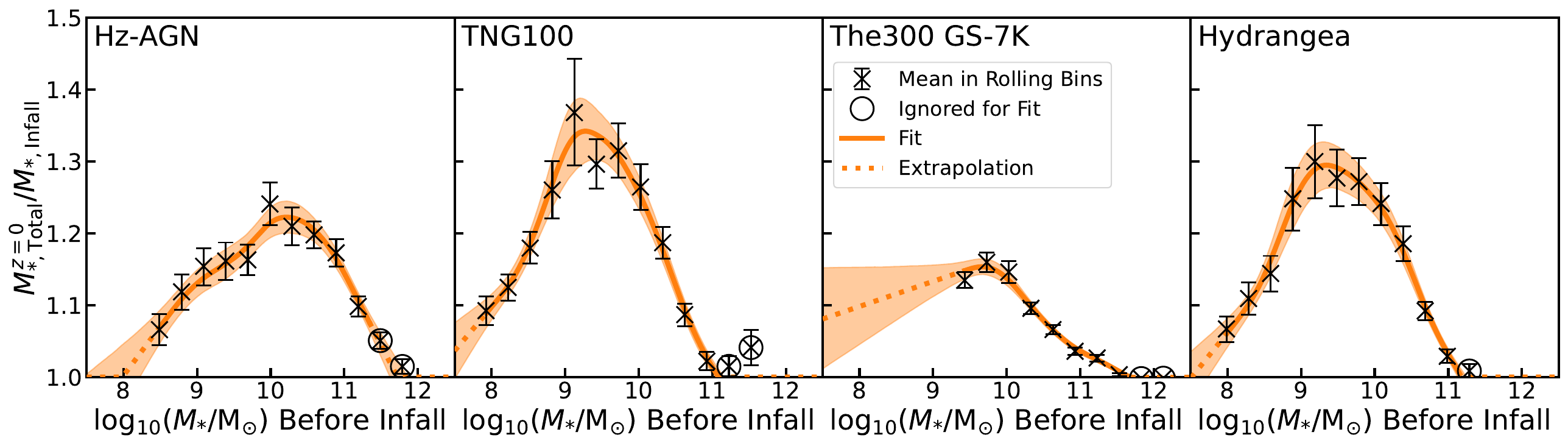}
    \caption{
    Mean ratio between total stellar mass associated with a satellite by $z\approx0$ and satellite infall stellar mass as a function of infall stellar mass for the simulated cluster samples drawn from \textsc{Horizon-AGN} (leftmost panel), \textsc{TNG100} (middle-left panel), \textsc{The Three Hundred Gizmo-Simba 7K} (middle-right panel), and \textsc{Hydrangea} (rightmost panel). For clarity only the mean value for every other bin is shown. Fitted cubic spline functions are shown in orange, with bins ignored for fitting circled and linear extrapolation beyond the range of masses used for fitting indicated with dotted lines. The error bars and shaded regions indicate estimated uncertainties based on bootstrapping.
    }
    \label{fig:SideBySide_all2infall}
\end{figure*}

In Figure~\ref{fig:SideBySide_all2infall} we present for each simulation individually the mean ratio between the total stellar mass associated with a satellite galaxy by $z\approx0$ 
(see Section~\ref{Methods-TrackingGalaxies-ActualTracking}) 
and satellite infall stellar mass as a function of infall mass, considering the same population of galaxies and following the same binning procedure as the analyses depicted in Figures~\ref{fig:LibFracPlot} and~\ref{fig:SideBySide_flib}. 
We fit to these mean mass ratios a cubic spline function in the same manner as our fitted $f_{\mathrm{lib}}(M_{*})$ functions (see Section~\ref{Results2-IclContribution-LibFrac} for details), including linear extrapolation beyond the range of masses used for fitting, but with the output constrained to $[1,\infty]$ rather than $[0,1]$ and no longer disregarding low-mass satellites when fitting.
For clarity, only every other bin is shown in Figure~\ref{fig:SideBySide_all2infall}. The error bars and shaded regions indicate the dispersion (16th-84th percentile) in mean values and functions fitted among the $10^{4}$ bootstrap resamples.

The fitted functions for each of the simulations all have the same fundamental shape -- with a peak at intermediate infall masses ($10^{9.5}-10^{10}\,$M\textsubscript{\sun}), falling back to unity at both lower and higher masses. 
We attribute the differing heights and precise positions of this peak in each panel to the differing sub-grid physics prescriptions and implementations of the different simulations as well as factors we ignore for this analysis (most pertinently differing typical cluster assembly times). 


\section{Cumulative fractional ICL (and BCG\,+\,ICL) contributions direct from simulations}\label{Appendix-FromSimFracCs}

\begin{figure*}
    \includegraphics[width=2.07\columnwidth]{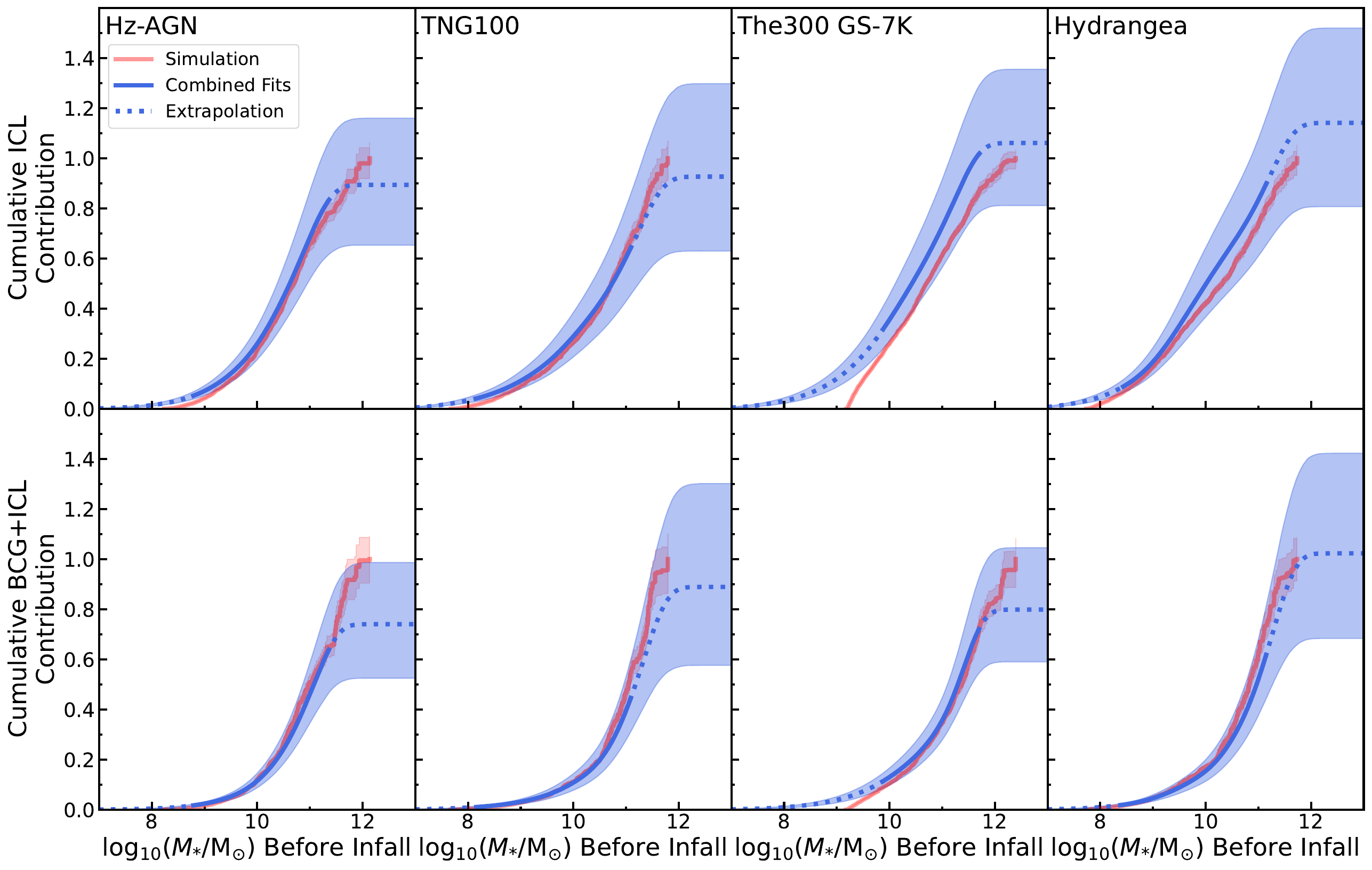}
    \caption{
    \textbf{Top Panels:} Cumulative ICL mass contribution as a function of satellite infall stellar mass for the cluster samples drawn from \textsc{Horizon-AGN} (leftmost panel), \textsc{TNG100} (middle-left panel), \textsc{The Three Hundred Gizmo-Simba 7K} (middle-right panel), and \textsc{Hydrangea} (rightmost panel) -- taken directly from each simulation (without fitting) and normalised relative to the total ICL contribution from satellite galaxies (shown in red). Predicted equivalent curves  
    (reproduced from Figure~\ref{fig:IclContriCsPlot}) are shown in blue, normalised relative to the raw data.
    Extrapolation beyond the range of masses used for fitting is indicated by dotted lines. The shaded regions indicate the dispersion (16th-84th percentile) among equivalent analyses performed for $10^{4}$ bootstrap resamples of the infalling satellite population (also normalized relative to raw data from main analysis). 
    \textbf{Bottom Panels:} Same as top panels but for the combined BCG\,+\,ICL system.
    }
    \label{fig:SideBySide_fracContriCs}
\end{figure*}

In 
Figure~\ref{fig:IclContriCsPlot} we present predictions for the cumulative contribution to the ICL mass  sourced from satellite galaxies 
as a function of satellite infall stellar mass. 
These predictions are based on fitted functions (as per Equation~\ref{eq:final_fit}) and include extrapolation to unresolved low satellite masses.
An equivalent analysis can be obtained directly from the simulations without fitting or extrapolation -- which we give in Figure~\ref{fig:SideBySide_fracContriCs}, 
where the cumulative fraction of ICL mass from satellite galaxies (for the aggregate ICL of each simulation's entire cluster sample) is shown as a function of satellite infall stellar mass (in red). The equivalent predicted curves (based on fitting and extrapolation) are reproduced from Figure~\ref{fig:IclContriCsPlot} for comparison (in blue), normalized relative to the raw data so over-/under-prediction of total contributed ICL mass can be assessed. The shaded regions indicate the dispersion (16th-84th percentile) of equivalent analyses for $10^{4}$ bootstrap resamples of the infalling satellite population, all also normalized relative to the main analysis of the raw data. 
The analysis presented in Figure~\ref{fig:SideBySide_fracContriCs} is similar to that presented for \textsc{Hz-AGN} alone in figure~7 of \citetalias{brown_assembly_2024}. 

Significant broadening of the blue shaded regions can be observed in every panel of Figure~\ref{fig:SideBySide_fracContriCs} for stellar masses beyond $\sim10^{11}$\,M\textsubscript{\sun} -- a consequence of the significant ICL (and BCG\,+\,ICL) mass contribution each individual object this massive is expected to make. As such, even small changes to the infalling satellite mass function (encompassed by the shaded regions shown in Figures~\ref{fig:MassFunctionPlot} and~\ref{fig:SideBySide_SchFit}) can alter the total predicted ICL mass by as much as $\gtrsim50$~per cent. 


\bsp	
\label{lastpage}
\end{document}